\documentclass[11pt]{article}

\usepackage[usenames,dvipsnames,table]{xcolor}
\usepackage[utf8]{inputenc}

\pdfoutput=1 
\usepackage{jheppub} 

\usepackage{graphicx}
\usepackage{amsmath, amssymb}
\usepackage{subcaption}

\usepackage{ytableau}

\newcommand{\be}{\begin{eqnarray}}
\newcommand{\ee}{\end{eqnarray}}
\newcommand{\bea}{\begin{eqnarray}}
\newcommand{\eea}{\end{eqnarray}}

\newcommand{\nn}{\nonumber}
\newcommand{\bn}{\begin{enumerate}}
\newcommand{\en}{\end{enumerate}}

\usepackage{tikz,bm}
\usetikzlibrary{arrows}
\usetikzlibrary{patterns}
\usetikzlibrary{decorations.pathmorphing}
\usepackage{xcolor}
\usepackage{amsmath,amssymb,amsthm,mathdots}

\usetikzlibrary{positioning}
\usetikzlibrary{chains}
\usetikzlibrary{arrows, arrows.meta ,fit,decorations.pathreplacing}
\tikzstyle{every picture}+=[remember picture]
\tikzstyle{na} = [baseline]

\usetikzlibrary{arrows, decorations.markings, calc, fadings, decorations.pathreplacing, patterns, decorations.pathmorphing, positioning}
\tikzset{>={Latex[width=1.5mm,length=1.5mm]}}

\def\CF{{\cal F}}

\def\CN{{\cal N}}
\def\CO{{\cal O}}

\def\CS{{\cal S}}

\def\Z{\mathbb{Z}}

\def\half{\frac{1}{2}}

\def\fsu{\mathfrak{su}}
\def\fso{\mathfrak{so}}

\def\fu{\mathfrak{u}}

\title{A 5d perspective on the compactifications of\\6d SCFTs to 4d $\mathcal{N}=1$ SCFTs}

\preprint{}

\author[]{Evyatar Sabag}
\author[]{and Matteo Sacchi}

\affiliation[]{Mathematical Institute, University of Oxford, Andrew-Wiles Building,\\Woodstock Road, Oxford, OX2 6GG, UK}

\abstract{Compactifying 6d superconformal field theories (SCFTs) to 4d $\mathcal{N}=1$ theories on two-punctured spheres (tubes) and tori with flux is realized using duality domain walls in 5d $\mathcal{N}=1$ Kaluza--Klein (KK) theories, which are usually denoted by \emph{flux domain walls}. We revisit this construction and study it in detail from the 5d perspective, specifically rephrasing it using the box graph description of the extended Coulomb branch phases of 5d theories. This perspective could be helpful in understanding how to equivalently realize the 4d $\mathcal{N}=1$ models from geometric engineering in M-theory. Along the way, we show how to recover various properties of the 4d theories from the 5d perspective, such as the flux associated to the domain wall configurations and the presence of a $\fu(1)$ global symmetry in the 4d theory descending from the KK symmetry on the tube, which is broken to a discrete subgroup on a flux torus. We demonstrate all of these ideas using the rank 1 E-string theory.}

\begin{document} 

\maketitle
\flushbottom

\section{Introduction}

Many properties of supersymmetric quantum field theories can be found by compactification from higher dimensional string-, M-, F- or field theories on manifolds preserving some supersymmetry. Such efforts have led to understanding several features and even classifications of strongly coupled theories and superconformal field theories (SCFTs) in 6d \cite{Heckman:2013pva,DelZotto:2014hpa,Heckman:2015bfa,Bhardwaj:2015xxa,Bhardwaj:2015oru} and in 5d \cite{Jefferson:2017ahm,Jefferson:2018irk,Bhardwaj:2018yhy,Bhardwaj:2018vuu,Bhardwaj:2019jtr,Bhardwaj:2019fzv,Apruzzi:2019vpe,Apruzzi:2019opn,Apruzzi:2019enx,Apruzzi:2019kgb,Bhardwaj:2020gyu}. Moving down to 4d, many properties of 4d $\CN =2$ Lagrangian as well as strongly coupled non-Lagrangian theories and relations to 6d $(2,0)$ SCFTs compactified on Riemann surfaces were found in the seminal work of Gaiotto \cite{Gaiotto:2009we}, and a classification was suggested in \cite{Bhardwaj:2013qia}. In recent years a similar effort was focused on understanding properties of less understood 4d $\CN =1$ models using relations to compactifications of 6d $(1,0)$ SCFTs on a Riemann surface with flux \cite{Gaiotto:2015usa,Razamat:2016dpl,Bah:2017gph,Kim:2017toz,Kim:2018bpg,Kim:2018lfo,Razamat:2018gro,Razamat:2019mdt,Pasquetti:2019hxf,Razamat:2019ukg,Razamat:2020bix,Sabag:2020elc,Hwang:2021xyw,Bah:2021iaa} (see also \cite{Razamat:2022gpm} for a recent review).\footnote{Recently in \cite{Sacchi:2021afk,Sacchi:2021wvg} preliminary steps were made to investigating a similar problem, but starting from 5d $\mathcal{N}=1$ SCFTs and leading to 3d $\mathcal{N}=2$ theories upon compactification on Riemann surfaces with flux. See also \cite{Naka:2002jz,Bah:2018lyv,Hosseini:2018usu,Legramandi:2021aqv} for some results from the holographic perspective.} These efforts have revealed many dualities and symmetry enhancements of the underlying 4d $\CN =1$ theories understood from geometry, yet there are many models that are not well understood including many strongly coupled theories and SCFTs. One may hope that framing these known relations in terms of geometric engineering from M- or F-theory, as it was done in 5d and 6d, may lead to a better understanding of 4d $\CN =1$ theories as this will allow the use of many powerful tools of geometric engineering.

In this work we focus on the constructions of 4d $\CN =1$ models related to compactifications of 6d $(1,0)$ SCFTs on a tube or torus with flux for the global symmetry, as it was considered in \cite{Kim:2017toz,Kim:2018bpg,Kim:2018lfo}. In these constructions one starts by compactifying a 6d $(1,0)$ SCFT on a circle with some choices of holonomies for continuous abelian subgroups of the 6d flavor symmetry. These result in 5d Kaluza-Klein (KK) theories, which in some favourable cases have low energy effective field theory descriptions in terms of weakly coupled gauge theories. In the cases in which the KK theory admits a gauge theory description, one can then build duality domain walls with half-BPS boundary conditions relating two such 5d effective field theories. In this construction the two theories will be associated with two different values of the holonomy, and will have the same UV fixed point SCFT \cite{Gaiotto:2015una}. With some abuse of terminology this phenomenon is sometimes referred to as "UV duality", hence the name duality domain wall. Note that in general such duality domain walls will contain 4d $\CN =1$ degrees of freedom which are not always easy to predict and in general only half of the 5d supersymmetry will be preserved. When the two theories related by the duality domain wall have a UV completion in 6d and their KK theories differ by an holonomy for abelian factors of the 6d flavor symmetry, this set-up is denoted as a \textit{flux domain wall}. 

\begin{figure}[h]
\center
\includegraphics[width=1\textwidth]{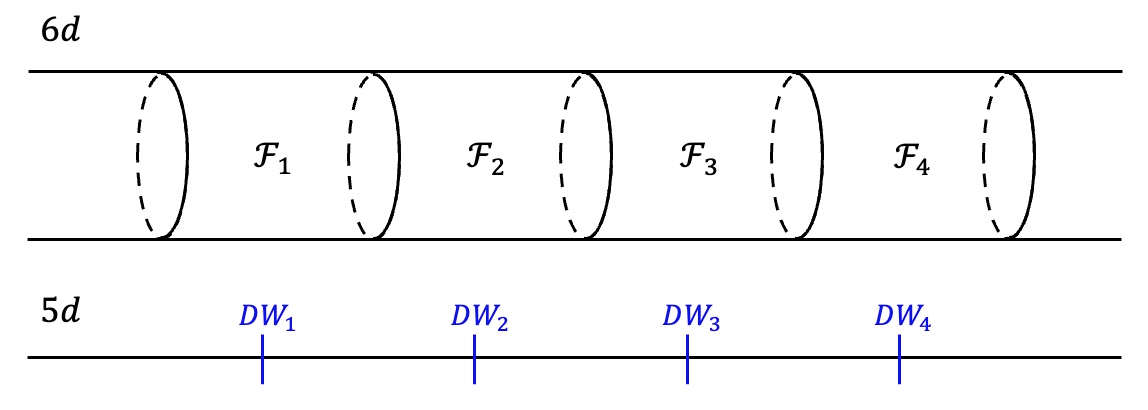} 
\caption{In 6d the compactification on a flux tube or torus is constructed by gluing several basic flux tubes with flux $\mathcal{F}_i$, where the type of gluing determines the total flux. Each basic flux tube is realized as a flux domain wall in 5d, and these are concatenated along an interval or circle to generate a tube or a torus.}
\label{F:ConDW}
\end{figure}

These flux domain walls will be the basic building blocks to generate more general flux tube and torus compactifications.
Indeed, one can concatenate several such flux domain walls to generate others with different values of the flux.\footnote{Similar 4d $\mathcal{N}=1$ models which arise from concatenating on a circle several domain walls but between 5d gauge theories that are UV completed by 5d SCFTs have been studied in \cite{Garozzo:2020pmz}.} Such a concatenation will include non-trivial identifications in the gluing region that will determine the total domain wall flux. One can also concatenate flux domain walls in this way on a circle to generate flux tori. Generating a flux tube will require to cap the two sides of the domain wall with half-BPS boundary conditions preserving the same half supersymmetry of the domain wall BPS boundary conditions (see Figure \ref{F:ConDW}).

The above constructions were mostly used in past works to construct examples of such reductions for various 6d SCFTs. In this work we aim to study the construction itself in a top-down approach in which we work out its details starting from 6d using field theory analysis. Specifically we will study this construction more in terms of the 5d KK theories and show why some results found for the 4d theories using 4d consistency conditions and anomaly arguments, can be seen to be required from the 5d perspective and the consistency of the construction of the domain walls. Our analysis is indeed completely independent of the study of anomalies and other protected observables in 4d, such as the supersymmetric index.

Another motivation for this work is that the aforementioned 5d theories can be geometrically engineered from M-theory on non-compact Calabi-Yau three-folds \cite{Intriligator:1997pq,Douglas:1996xp,Jefferson:2017ahm,Jefferson:2018irk,Bhardwaj:2018yhy,Bhardwaj:2018vuu,Closset:2018bjz,Apruzzi:2018nre,Apruzzi:2019vpe,Apruzzi:2019opn,Apruzzi:2019enx,Bhardwaj:2019jtr,Bhardwaj:2019fzv,Saxena:2020ltf,Apruzzi:2019kgb,Bhardwaj:2020gyu,Closset:2020afy,Closset:2021lwy,Closset:2020scj}. This together with the exact details of the field theory construction we present, which include the 5d extended Coulomb branch (ECB) phase on each side of each domain wall, can be used in future work to generate $G_2$ manifolds that will produce the resulting 4d $\CN =1$ models upon compactification from M-theory \cite{Acharya:1998pm,Acharya:1996ci,Acharya:2000gb,Atiyah:2000zz,Atiyah:2001qf,Witten:2001uq,Acharya:2001gy,Acharya:2004qe,Halverson:2014tya,Halverson:2015vta,Braun:2016igl,daCGuio:2017ifs,Braun:2017ryx,Braun:2017uku,Braun:2017csz,Braun:2018fdp,Fiset:2018huv,Acharya:2018nbo,Pantev:2009de,Braun:2018vhk,Hubner:2020yde,Acharya:2020vmg}.

The rest of the paper is organized as follows. In the first part of Section \ref{S:6dto5d} we review 6d $(1,0)$ and 5d $\CN =1$ theories and how to relate them via circle reduction, while in the second part we review the 5d extended Coulomb Branch (ECB) and its box graph description. In Section \ref{S:5dFDW} we describe the construction of flux domain walls as a variation of a real mass parameter in the 5d theory across one direction. In Section \ref{S:HFDW} we show how to construct higher flux domain wall and flux tori by concatenating several basic flux domain walls. In Section \ref{sec:4dDWtheory} we review the 4d domain wall theories and give arguments on how to conjecture Lagrangians for them. In Section \ref{S:DW4dT} we review some of the properties of the 4d theories arising from the flux tube compactification, showing how they match those predicted from the 5d construction of the previous sections. We conclude the paper in Section \ref{S:Conc} and give some future directions of research.

\newpage

\section{6d SCFTs to 5d KK theories}
\label{S:6dto5d}

\subsection{6d $(1,0)$ and 5d $\mathcal{N}=1$ field theories and their relations}

We start by reviewing 6d $(1,0)$ theories. The 6d $(1,0)$ and $(2,0)$ supersymmetry algebras can be extended to superconformal algebras, thus admitting the existence of 6d SCFTs \cite{Minwalla:1997ka}. 6d $(1,0)$ theories have eight real supercharges and their bosonic symmetries are the Lorentz symmetry $\mathfrak{so}(5,1)$ and an $\mathfrak{su}(2)_R$ R-symmetry. Such 6d field theories admit the following multiplets: hypermultiplets containing four real scalars parametrizing the Higgs branch of vacua and a spinor, vector multiplets containing a vector and a spinor (co-spinor of $\mathfrak{so}(5,1)$) in the doublet of $\mathfrak{su}(2)_R$, and tensor multiplets containing a two-form with a self-dual field strength, a real scalar parametrizing the tensor branch of vacua and a spinor in the doublet of $\mathfrak{su}(2)_R$.

Considering a theory with a simple gauge group and hypermultiplets, the gauge kinetic term $\frac{1}{g^2}\int \text{tr} \left(F \wedge \star F\right)$ implies that $g$ has dimension one making the theory IR free, but it requires some UV completion. If we add a tensor to the theory we can't write a well-defined Lagrangian for the theory, as the self-dual field strength of the tensor makes the kinetic term vanish $H\wedge \star H=H\wedge H=-H\wedge H=-H\wedge \star H$. Nevertheless, one can consider such a theory as having a Lagrangian with the added self-duality constraint. In such a description the tensor multiplet scalar $\phi$ couples to the gauge field with the term
\be
c\int \phi\, \text{tr} \left(F \wedge \star F\right)\,,
\ee
where $c$ is a constant. Thus, we can redefine $\phi$ to absorb the gauge coupling, leading to an effective coupling
\be
\frac{1}{g_{eff}^2}=c\langle \phi \rangle\,.
\ee
This means that the gauge coupling is non-vanishing on the tensor branch. The theory also has instantons (strings in 6d) and from the kinetic term we can see that they have tension $T_{BPS} \propto \frac{1}{g_{eff}^2}$. In addition, $T_{BPS} (\propto \frac{1}{g_{eff}^2})$ is the only dimensionful parameter of the theory and it vanishes in the origin of the tensor branch, implying the theory could have a UV completion that is conformally invariant giving an SCFT.

Before we continue by compactifying the 6d theory on a circle to 5d, we wish to shortly review 5d $\mathcal{N}=1$ theories. The 5d $\mathcal{N}=1$ supersymmetry algebra can be extended to a superconformal algebra,\footnote{The 5d $\mathcal{N}=2$ algebra can't be extended to a superconformal algebra on the other hand, see \cite{Minwalla:1997ka}.} thus admitting the existence of SCFTs with this amount of supersymmetry. 5d $\mathcal{N}=1$ theories have eight real supercharges and their bosonic symmetries are the Lorentz symmetry $\mathfrak{so}(4,1)$ and an $\mathfrak{su}(2)_R$ R-symmetry. Such 5d field theories admit the following multiplets: hypermultiplets including four real scalars and a spinor, and vector multiplets including a gauge field, a real scalar and a spinor in the doublet of $\mathfrak{su}(2)_R$. One can also consider a tensor multiplet dual to a vector multiplet containing a two-form, a real scalar and a spinor in the doublet of $\mathfrak{su}(2)_R$. Due to this, both the tensor and vector multiplets from 6d will reduce to 5d vector multiplets. The moduli space of vacua of these theories may be composed by two types of branches: the Coulomb branch parametrized by the scalars in the vector multiplets and the Higgs branch parametrized by the scalars in the hypermultiplets.

Next, we wish to reduce a 6d $(1,0)$ SCFT and its effective tensor branch field theory parametrized by scalars $\phi_i^{6d}$ on a circle in the $x^5$ direction with radius $R_5$. The SCFT reduction leads to the so-called 5d Kaluza--Klein (KK) theory, which has a Coulomb branch of vacua parametrized by real scalars $\phi_i^{5d}$. These scalars are the zero modes of the tower of massive KK modes produced by the reduction of $\phi_i^{6d}$ on the circle. 

The $(1,0)$ 6d SCFTs unlike the $(2,0)$ SCFTs in many occasions have a non-vanishing flavor symmetry $G_F^{6d}$ and one can also consider coupling abelian sub-algebras $\fu(1)_h\subset \mathfrak{g}_F^{6d}$ to background gauge fields $A_{h,\mu}^{6d}$. Turning on a non-trivial holonomy for $A_{h,\mu}^{6d}$ around the circle and compactifying to $5d$ will then produce real mass parameters $m_h$ which will parametrize the space of relevant deformations of the 5d KK theory. In addition, the circle radius $R_5$ we compactify on gives us an additional mass parameter $m_{KK}=R_5^{-1}$. This mass parameter can be related to the $\fu(1)_{KK}$ symmetry of translations along the compactification circle. These mass parameters can naturally couple to the gauge kinetic term through $m_h\int \text{tr} \left(F \wedge \star F\right)$; therefore the effective gauge coupling of each simple gauge factor will be determined by a combination of the bare coupling coming from the scalar in the associated vector multiplet and the real mass parameters. In total the vector multiplet scalars $\phi_i^{5d}$ and the mass parameters $m_h$ parametrize the so-called \emph{extended Coulomb branch} (ECB) of the 5d theory. 

On a generic point of the Coulomb branch, which is parametrized only by the $\phi_i^{5d}$, the gauge group is broken to its Cartan $U(1)^r$ where $r$ is the dimension of the Coulomb branch and is called the rank of the 5d theory. The effective low energy Lagrangian is \cite{Intriligator:1997pq}
\be
\mathcal{L}_{eff}=G_{ij}\left(d\phi^i \wedge \star d\phi^j + F^i \wedge \star F^j \right) + \frac{c_{ij\ell}}{24 \pi^2} A^i \wedge F^j \wedge F^\ell + \dots\,,
\ee
where we dropped the superscript $5d$ for brevity, as all the fields in the Lagrangian are of the 5d theory. In this Lagrangian the couplings are determined by the prepotential $\mathcal{F}$, which is a cubic polynomial of the vevs of the scalars $\phi^i$ and the real masses $m_h$. The effective gauge coupling which is the metric on the ECB $G_{ij}$ and the Chern--Simons coefficient $c_{ij\ell}$ are given by
\be
G_{ij}=\frac{\partial^2\mathcal{F}}{\partial\phi^i\partial\phi^j}\,,\qquad c_{ij\ell} = \frac{\partial^3\mathcal{F}}{\partial\phi^i\partial\phi^j\partial\phi^\ell}\,.
\ee
Note that the Chern--Simons coefficient $c_{ij\ell}$ must be an integer in order to get a well-defined theory. The prepotential is given by
\be
\mathcal{F} = \left(\frac{1}{2g_0^2} f_{ij}\phi^i\phi^j + \frac{k}{6}d_{ij\ell}\phi^i\phi^j\phi^\ell\right) + \frac{1}{12} \left( \sum_{\alpha\in roots} \left| \alpha_i \phi^i \right|^3 - \sum_h \sum_{w\in R_h} \left| w_i \phi^i + m_h\right|^3\right)\,,
\ee
where the first parenthesis contain the classical contribution, while the second parenthesis contain the one-loop contribution. In the expression $g_0$ denotes the bare gauge coupling, $k$ is the Chern--Simons level, $f_{ij}=\text{Tr}\left(T_iT_j\right),\,d_{ij\ell}=\half\text{tr}_F\left(T_i\{T_j,T_\ell\}\right)$ where $T_i$ are the Cartan generators of the Lie algebra $\mathfrak{g}$ associated to the gauge group $G$, $\alpha$ denote the roots of $\mathfrak{g}$ and $w$ denote the weights of the representation $R_h$ of $\mathfrak{g}$ under which the hypermultiplet $h$ transforms. In the above expression repeated upper and lower indices are implicitly summed, with $i,j,\ell=1,...,rk(G)$.

The spectrum at each point of the ECB contains massive BPS particles and strings. The particles (instantons) are electrically charged and their masses can be inferred from  the gauge kinetic term. Their charges are given by
\be
m/\sqrt{2}= Z_e = n_e^i \phi_i + f^h m_h\,,
\ee
where $n_e^i$ denotes the electric charge under a low energy gauge symmetry $\mathfrak{u}(1)_i$ factor on the Coulomb branch, while $f^h$ denotes the charge of a flavor $\mathfrak{u}(1)_h$ related to $m_h$. The strings are magnetically charged and descend from the 6d BPS strings that don't wind the compactification circle. Their tension is given by
\be
T/\sqrt{2} = Z_m = n_m^i \phi_{D,i}\,,
\ee
where $n_m^i$ are the magnetic charges under $\mathfrak{u}(1)_i$ and $\phi_{D,i}$ are derived from the prepotential $\mathcal{F}$ as $\phi_{D,i}=\frac{\partial \mathcal{F}}{\partial \phi_i}$.

Finally, in order to reach an SCFT point we need to take the compactification circle radius to zero $R_5 \to 0$ (or equivalently take $m_{KK}\to \infty$). This will make the entire tower of KK modes infinitely massive and integrate them out leaving us only with the zero modes. If we take the limit at the ECB origin we will find a 6d SCFT point, while taking some of the real masses to infinity first, effectively integrating out some of the hypermultiplets, and then taking $R_5 \to 0$ will lead to a genuine 5d SCFT.

\subsubsection*{Free hypermultiplet reduction example}

Here we will give an example for the compactification of a simple 6d $
\mathcal{N}=(1,0)$ field theory containing a free hypermultiplet transforming under an $\mathfrak{su}(2)_F$ flavor symmetry along with the Lorentz and $R$-symmetry $\mathfrak{so}(5,1)\oplus \mathfrak{su}(2)_R$. This analysis was initially done in \cite{Chan:2000qc} and we will summarize some of the relevant results. We will use the free hypermultiplet example to exemplify some of the features and properties of the compactification of 6d SCFTs on a flux tube to 4d. Going back to the example, under $\mathfrak{so}(5,1)\times \mathfrak{su}(2)_R\times \mathfrak{su}(2)_F$ the supersymmetry generators $Q^i_\alpha$ transform as $\left(\textbf{4},\textbf{2},\textbf{1}\right)$, the fermions $\psi_\alpha^a$ transform as $\left(\textbf{4},\textbf{1},\textbf{2}\right)$, and the bosons $\phi^{ia}$ transform as $\left(\textbf{1},\textbf{2},\textbf{2}\right)$. In addition we will choose anti-symmetric 6d Dirac matrices $\Gamma_{\alpha\beta}^\mu$.
Throughout this example $\alpha,\beta,\gamma=1,...,4$ are spinor indices of $\mathfrak{so}(5,1)$, $\mu,\nu=0,...,5$ are vector indices of $\mathfrak{so}(5,1)$, $i,j,k=1,2$ and $a,b,c=1,2$ are indices of the $\mathfrak{su}(2)_R$ and $\mathfrak{su}(2)_F$ doublets, respectively.

The hypermultiplet action is
\be
S=\int d^6 x\left[-\frac{1}{4}\epsilon_{ij}\epsilon_{ab}\partial_\mu \phi^{ia} \partial^\mu \phi^{jb} + \half \epsilon_{ab} \psi_\alpha^a \Gamma^{\mu\alpha\beta}\partial_\mu \psi^b_\beta\right]\,,
\ee
with equations of motion
\be
\partial_\mu \partial^\mu \phi^{ia} = 0\,,\quad \Gamma^{\mu\alpha\beta}\partial_\mu \psi^a_\alpha = 0\,.
\ee
The supersymmetry transformations are given by
\be
\delta\phi^{ia} = 2\eta^{\alpha i}\psi^a_\alpha\,,\quad \delta\psi^a_\alpha =  \eta^{\beta i} \epsilon_{ij} \Gamma^\mu_{\alpha\beta} \partial_\mu \phi^{ja}\,,
\ee
where $\eta^{\alpha i}$ is the supersymmetry parameter.

When compactifying the above theory on a circle to 5d, a KK tower of states is generated with the lowest state being massless and constant on the circle. Looking at $\mathfrak{u}(1)_F\subset \mathfrak{su}(2)_F$ corresponding to the $T^3$ generator of $\mathfrak{su}(2)_F$ (the third Pauli matrix), it has an associated current $J_\mu$ which we can couple to a background gauge field $A_\mu=A_{3,\mu} T^3$. Creating a Wilson line for $A_\mu$ around the circle is equivalent to changing the periodicity of the 6d hypermultiplet fields, which will be identified up to a $\mathfrak{u}(1)_F$ rotation with themselves. Setting $A_{3,5}=m$ gives the hypermultiplet fields the following $x^5$ dependence
\be
\phi^a(x^
{\widetilde{\mu}},x^5) & = & \phi^b(x^{\widetilde{\mu}})\left(e^{imx^5 T^3}\right)^a_{\ b}\,,\nonumber\\
\psi^a(x^
{\widetilde{\mu}},x^5) & = & \psi^b(x^{\widetilde{\mu}})\left(e^{imx^5 T^3}\right)^a_{\ b}\,,
\ee
where $\widetilde{\mu}=0,...,4$ is a vector index of $\mathfrak{so}(4,1)$ and
\be
\label{E:GenSU2}
T^3 = 
\begin{pmatrix}
	1 & 0\\
	0 & -1
\end{pmatrix}\,.
\ee

The action in  5d  is found by inserting the above field assignments into the 6d action
\be
\label{E:MassiveHyperAction}
S & = & \int d^5x \left[-\frac{1}{4}\epsilon_{ij}\epsilon_{ab}\left(\partial_{\widetilde{\mu}} \phi^{ia} \partial^{\widetilde{\mu}} \phi^{jb} + m^2 \phi^{ia} \phi^{jb}\right)\right.\nonumber\\ 
&& \qquad \qquad + \left.\half \epsilon_{ab} \psi_\alpha^a \Gamma^{\widetilde{\mu}\alpha\beta}\partial_{\widetilde{\mu}} \psi^b_\beta +\half im\epsilon_{ab}(T^3)_c^{\ b} \psi^a_\alpha \Gamma^{5\alpha\beta} \psi^c_\beta\right]\,.
\ee
The associated equations of motion are the equations of a massive 5d hypermultiplet
\be\label{5dhypeom}
(\partial_{\widetilde{\mu}} \partial^{\widetilde{\mu}} + m^2) \phi^{ia} = 0\,,\quad \Gamma^{\widetilde{\mu}\alpha\beta}\partial_{\widetilde{\mu}} \psi^a_\alpha + im (T^3)_b^{\ a} \Gamma^{5\alpha\beta} \psi^b_\alpha = 0\,.
\ee
The supersymmetry transformations derived form the 6d ones are
\be
\label{E:susyTrans}
\delta\phi^{ia} = 2\eta^{\alpha i}\psi^a_\alpha\,,\quad \delta\psi^a_\alpha = \epsilon_{ij} \eta^{\beta i} \Gamma^{\widetilde{\mu}}_{\alpha\beta} \partial_{\widetilde{\mu}} \phi^{ja} + im\epsilon_{ij}\eta^{\beta i}\Gamma_{\alpha\beta}^5 (T^3)^{\ a}_b \phi^{jb}\,.
\ee

\subsection{5d gauge theories ECBs and the box graph description}

We will now discuss the Coulomb branch and extended Coulomb branch (ECB) phases of 5d $\mathcal{N}=1$ gauge theories and review a compact and visual way to present the ECB phases known as the (decorated) box graph description \cite{Hayashi:2014kca,Braun:2014kla,Lawrie:2015hia,Braun:2015hkv} (see also \cite{Apruzzi:2019enx} for specific applications to 5d SCFTs).

Consider a gauge theory with gauge algebra $\mathfrak{g}_{guage} = \bigoplus_i \mathfrak{g}_i\oplus \mathfrak{u}(1)_1 \oplus \cdots \oplus \mathfrak{u}(1)_{r_A}$, where $
\mathfrak{g}_i$ are simple algebras of rank $r_i$. The Coulomb branch of the theory is isomorphic to 
\be
\mathbb{R}^{r_A} \times \prod_i \mathcal{C}_i\,,
\ee
where $\mathcal{C}_i=\mathbb{R}^{r_i}/W_{\mathfrak{g}_i}$ is the Weyl chamber cone and $W_{\mathfrak{g}_i}$ is the Weyl group of $\mathfrak{g}_i$. It is natural to choose a root basis $\vec{\alpha}_j^{(i)}$ of positive simple roots of $\mathfrak{g}_i$,\footnote{In the following sections we will need to change this basis to a less natural basis when we glue flux tubes.} such that $\mathcal{C}_i$ are the fundamental Weyl chambers of $\mathfrak{g}_i$
\be
\mathcal{C}_i = \left\{ \vec{\phi} \in \mathbb{R}^{r_i} \,\,\,|\,\,\, \langle \vec{\phi}, \vec{\alpha}_j^{(i)} \rangle>0\,\,\,\, \forall\, j \right\}\,.
\ee 

Suppose now that we have a hypermultplet transforming in representation $R_h$ of the gauge symmetry $\mathfrak{g}$. Under the $r$ Cartan $\mathfrak{u}(1)$'s of $\mathfrak{g}$ it carries charges in accordance with the weights $\vec{w}_I$ of the representation $R_h$. Some hypermultiplet becomes massless on points of the Coulomb branch where $\langle \vec{\phi},\vec{w}_I\rangle=0$ for some $\vec{w}_I\in R_h$. These loci represent walls separating the Coulomb branch to subchambers or phases where $\langle \vec{\phi},\vec{w}_I\rangle$ has a non vanishing value for each $\vec{w}_I \in R_h$. The signs of these $\langle \vec{\phi},\vec{w}_I\rangle$ determine a unique phase of the Coulomb branch and the relations between these phases describe the Coulomb branch structure. 

We will use the decorated box graph to represent the above Coulomb branch phases. The box graphs are used to represent the weight diagram of an irreducible representation, where each box represents a weight and its neighbouring boxes represent weights related to it by an addition or subtraction of a simple positive root $\vec{\alpha}_j$. The decorated box graph is a box graph with additional $\pm$ signs, where in each box related to a weight $\vec{w}$ the sign corresponds to the sign of $\langle \vec{\phi},\vec{w}\rangle$. We will use a convention where weights represented by boxes are related to other boxes lying above or to the left by additions of positive simple root. Thus, above and to the left of a positive sign box one can only assign positive signs. Vice versa below and to the right of a negative sign box one can only assign negative signs.

We can further use the (decorated) box graphs to describe the ECB. This is done by promoting the hypermultiplet masses $m_h$ to parameters of the Coulomb branch, effectively weakly gauging part of the flavor symmetry. The box graph describing the ECB will be the one for matter in representations $(R_{gauge},R_{BG})$ of $\mathfrak{g}_{gauge}\oplus \mathfrak{g}_{BG}$, where $\mathfrak{g}_{BG}$ is the flavor symmetry. Each box of the graph now corresponds to a weight $\vec{w}_{I,J}=(\vec{w}_{gauge,I},\vec{w}_{BG,J})$ of $(R_{gauge},R_{BG})$ and the $\pm$ sign associated to it corresponds to the sign of $\langle (\vec{\phi},\vec{m}),\vec{w}_{I,J}\rangle$, where $(\vec{\phi},\vec{m})$ is a vector collecting all the vevs for the vector scalars $\phi_i$ and all the real masses $m_h$. The ECB phase will determine which hypermultiplets can become massive and thus be decoupled. Specifically, each hypermultiplet corresponds to a fixed weight $\vec{w}_{BG,J}$\footnote{Or pairs of opposite weights if the flavor symmetry is a real group.} of the flavor symmetry and transforms in a representation $R_{gauge}$ of the gauge symmetry, and we will be able to give mass to it only if all of its weights under the gauge symmetry $\vec{w}_{I,gauge}$ are such that they have the same sign of $\langle (\vec{\phi},\vec{m}),\vec{w}_{I,J}\rangle$.

\subsubsection{Example: ECB phases for the 5d reduction of the rank 1 E-string}

As an example, we will consider the simplest class of 5d SCFT: the rank one 5d theories arising from the 6d rank one E-string theory reduced on a circle possibly with holonomies for the flavor symmetry which will be mapped to mass deformations in 5d. Reducing without holonomies for flavor symmetries we find the marginal theory \cite{Apruzzi:2019opn}, which at weak coupling admits a low energy effective $\mathfrak{sp}(1)=\mathfrak{su}(2)_{gauge}$ gauge theory description with 8 fundamental hypermultiplets \cite{Seiberg:1996bd,Ganor:1996pc} (semotimes we will compactly refer to this gauge theory as $\mathfrak{su}(2)+8F$). Due to the pseudo reality of $\mathfrak{su}(2)_{gauge}$ the flavor symmetry is $\mathfrak{so}(16)$. Thus the ECB phases will be those of matter in the representation $(\textbf{2},\textbf{16})$ of $\mathfrak{su}(2)_{gauge}\oplus \mathfrak{so}(16)$.

\begin{figure}[t]
\center
\includegraphics[width=1\textwidth]{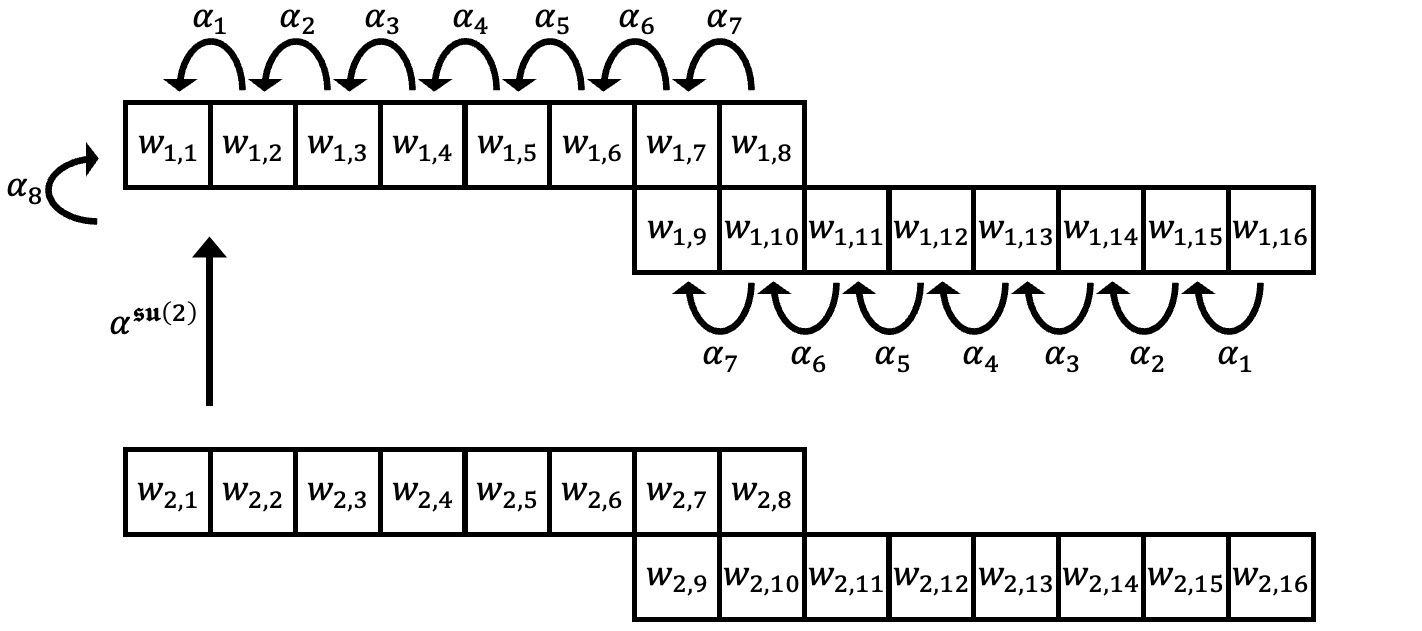} 
\caption{The rank one 5d SCFT undecorated box graph presenting the $(\textbf{2},\textbf{16})$ representation of $\mathfrak{su}(2)_{gauge}\oplus \mathfrak{so}(16)$. Each box represents a weight $\vec{w}_{i,j}$ in the representation, where $i=1,2$ labels a weight in the doublet representation of $\mathfrak{su}(2)_{gauge}$ and $j=1,...,16$ labels a weight in the vector representation of $\mathfrak{so}(16)$. The weights are related to one another by the simple positive root $\vec{\alpha}^{\mathfrak{su}(2)}$ of $\mathfrak{su}(2)_{gauge}$ and $\vec{\alpha}^{\mathfrak{so}(16)}_j$ of $\mathfrak{so}(16)$.}
\label{F:Rank1BG}
\end{figure}

In order to represent the ECB phases we will first define a basis for the relevant roots and weights. We will use a basis where the $\mathfrak{so}(16)$ vector representation $\textbf{16}$ weights are given by $(\pm 1,0,0,0,0,0,0,0)$ $+$ permutations (16 total), while the roots of $\mathfrak{so}(16)$ are given by $(\pm 1,\pm1,0,0,0,0,0,0)$ $+$ permutations (112 total) and the spinor representation weights are given by $(\pm\half,\pm\half,\pm\half,\pm\half,\pm\half,\pm\half,\pm\half,\pm\half)$ with even number of minus signs for the spinor and odd for the cospinor (128 each).\footnote{This basis will prove useful for describing the eventual 4d flux tube theories.} In this basis the positive simple roots are given by
\be
\label{SU2_8Froots}
\mathfrak{so}(16): & \qquad &
\begin{cases}
\vec{\alpha}_1^{\mathfrak{so}(16)}=(0;1,-1,0,0,0,0,0,0), & \vec{\alpha}_2^{\mathfrak{so}(16)}=(0;0,1,-1,0,0,0,0,0),\\
\vec{\alpha}_3^{\mathfrak{so}(16)}=(0;0,0,1,-1,0,0,0,0), & \vec{\alpha}_4^{\mathfrak{so}(16)}=(0;0,0,0,1,-1,0,0,0),\\
\vec{\alpha}_5^{\mathfrak{so}(16)}=(0;0,0,0,0,1,-1,0,0), & \vec{\alpha}_6^{\mathfrak{so}(16)}=(0;0,0,0,0,0,1,-1,0),\\
\vec{\alpha}_7^{\mathfrak{so}(16)}=(0;0,0,0,0,0,0,1,-1), & \vec{\alpha}_8^{\mathfrak{so}(16)}=(0;0,0,0,0,0,0,1,1),\\
\end{cases}\nonumber\\
\mathfrak{su}(2): & \qquad & \vec{\alpha}^{\mathfrak{su}(2)} =  (2;0,0,0,0,0,0,0,0)\,,
\ee
while the highest weight of $(\textbf{2},\textbf{16})$ is $\vec{w}_{1,1}=(1;1,0,0,0,0,0,0,0)$. We present the undecorated box graph for this representation in Figure \ref{F:Rank1BG}. 

\begin{figure}[t]
\center
\includegraphics[width=1\textwidth]{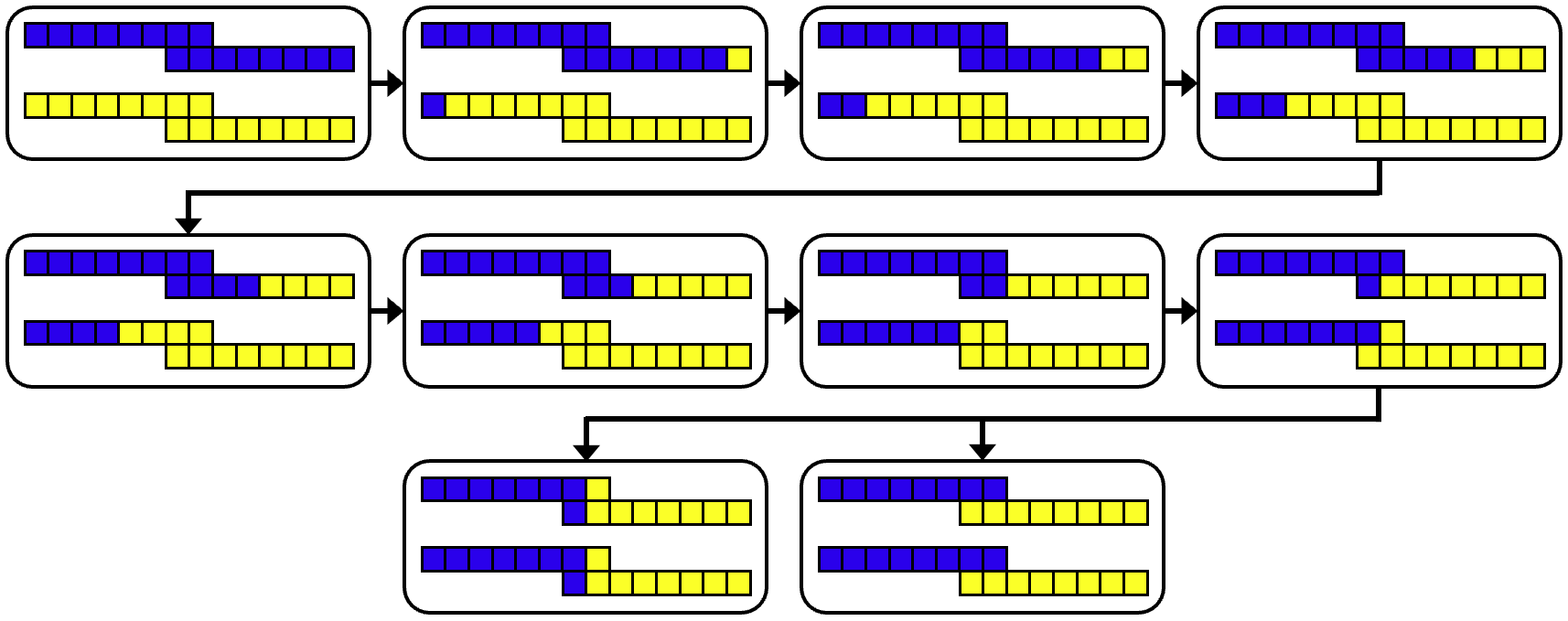} 
\caption{All the possible consistent ECB phases of the rank one 5d $\fsu(2)+8F$ gauge theory described by decorated box graphs. On the top left corner we have the phase matching the marginal theory. Each subsequent arrow leads to a phase with one flipped box sign in each of the $\textbf{16}$ representations allowing an additional hypermultiplet mass to be turned on.}
\label{F:Rank1ECB_BG}
\end{figure}

Adding the decoration to the box graphs one can write all the consistent ECB phases. Turning on no holonomies for 6d flavor symmetries corresponds to no masses for the 5d hypermultiplets giving the marginal theory. Thus, the consistent decorated box graph for this phase is the one where each of the two $\textbf{16}$ representation graphs have opposite uniform coloring, where we assign blue/yellow to the $+/-$ signs, respectively. In addition recall that this representation is self-conjugate requiring the total number of $+$ boxes to equal the number of $-$ boxes, see the top left corner of Figure \ref{F:Rank1ECB_BG}. One can then start flipping signs of boxes in a consistent manner to reach phases where one can give masses to some or even all the hypermultiplets. The full depiction of the consistent ECB phases of the rank one theory is given in Figure \ref{F:Rank1ECB_BG} \cite{Apruzzi:2019enx}.

Finally, note that the holonomies for the abelian subgroup of the 6d flavor symmetry that we can turn on correspond to generators of this symmetry. In the case of the rank one 5d theories these arise from the 6d rank one E-string SCFT compactification. Thus, the abelian symmetries we turn on holonomies for correspond to generators of $\mathfrak{e}_8$, whose non-vanishing generators in our chosen basis are given by $(\pm 1,\pm1,0,0,0,0,0,0)$ $+$ permutations (112 total) and $(\pm\half,\pm\half,\pm\half,\pm\half,\pm\half,\pm\half,\pm\half,\pm\half)$ with even number of minus signs (128 total). These together with the eight singlets are exactly the weights of the adjoint $\textbf{120}$ representation and spinor $\textbf{128}$ representation of $\mathfrak{so}(16)$, respectively. One can see that in the free hypermultiplet example \eqref{E:GenSU2} we also chose a generator of $\mathfrak{su}(2)$ for the 6d flavor symmetry. On the other hand, the masses we turn on in the 5d theory correspond to individual hypermultiplets, which in our chosen basis correspond to each placing in the basis vector. This means that in order to get a specific number of hypermultiplets to be massive could require more than a single generator whose associated holonomy is turned on. In fact the only two cases of this example that only require a single generator are giving the same mass (up to a sign) to all the hypermultiplets corresponding to the generators $(\pm\half,\pm\half,\pm\half,\pm\half,\pm\half,\pm\half,\pm\half,\pm\half)$ again with even number of minuses and giving the same mass (up to a sign) to two hypermultiplets corresponding to the generators $(\pm 1,\pm 1,0,0,0,0,0,0)$ $+$ permutations. In Section \ref{S:DW4dT} we will show how this fact translates to the simplicity of the matching flux domain wall 5d/4d coupled system theories compared to others.

\section{5d flux domain walls}
\label{S:5dFDW}

\subsection{General discussion}

After having reviewed some facts about circle compactifications of 6d theories to 5d, we will now move to discuss the so-called 5d \emph{flux domain walls}. A 5d domain wall is a configuration in which two distinct 5d theories are separated by an interface at a position, say, $x_4=0$ in the $x_4$ direction. On the wall it will typically leave some 4d degrees of freedom (d.o.f.'s) that interact in some way with the 5d bulk theories. Such a set-up was studied in \cite{Gaiotto:2015una}, where in particular \emph{duality domain walls} were considered. This refers to the situation in which the two 5d bulk theories are actually different gauge theory phases of the same UV SCFT. The typical situation considered in \cite{Gaiotto:2015una} is the one in which the two 5d bulk theories are the same gauge theory, but with a non-trivial identification of their flavor symmetry mass parameters which corresponds to a Weyl reflection of the UV flavor symmetry which is not part of the Weyl group of the IR flavor symmetry. Hence, from the low energy gauge theory perspective the deformations are different, but they are instead equivalent from the point of view of the UV SCFT so in this sense they are "dual".\footnote{In some cases, the symmetry relating the two 5d bulk theories is not a Weyl element of the continuous global symmetry at the UV point, but it is still a discrete symmetry of the SCFT.} In this configuration, one doesn't turn on vevs for the vector multiplet scalars $\phi_i$, while turning on mass parameters that the Weyl reflection acts on non-trivially. In the simplest situations that we will consider only a single mass parameter will be turned on, even though the corresponding Weyl transformation may be associated with a combination of several positive roots and not just a single one.

Although this construction applies to arbitrary 5d theories, we will consider the case in which the theories descend from some 6d SCFT, like the $\mathfrak{su}(2)+8F$ theory that is UV completed by the 6d rank 1 E-string theory which we reviewed in the previous section. In this case the 5d duality domain wall induced by the non-trivial identification of the flavor mass in the gauge theories, can also be understood as the compactification of the 6d theory on an infinitely long tube with flux for part of its flavor symmetry $\mathfrak{g}^{6d}_F$, which is why this set-up is called a \emph{flux domain wall}. To see this, we should imagine that the aforementioned mass parameter, call it $m$, has a non-trivial profile $m(x_4)$ in the $x_4$ direction such that its value will change between the left and the right 5d bulk theories. Remember from the discussion of the previous section that mass parameters in 5d KK theories descend from the $x_5$ direction component of the background gauge field $A^{6d}_\mu$ of the corresponding symmetry. Hence, a variable mass parameter in the $x_4$ direction will naturally lead to a non-vanishing $F^{6d}_{45}$ which is the source of the flux
\begin{align}
\frac{1}{2\pi}\int_{\Sigma}F_{45}^{6d}\neq 0\,.
\end{align}
Notice that at this stage $\Sigma$ is an infinitely long tube, so the flux does not need to be integrally quantized. We will come back to the quantization condition in the next section when considering the case of the torus compactification.

It turns out the precise profile of the variable mass is not important, while the only relevant thing is that its value changes sign between $x_4\to-\infty$ and $x_4\to +\infty$. In particular, the mass has to cross the zero value at the location of the wall $x_4=0$ in order for some non-trivial 4d theory to live on the wall. Notice that this means that the Weyl reflection relating the mass parameter between the left and the right 5d bulk theories corresponds to complex conjugation of the $\mathfrak{u}(1)$ inside the flavor symmetry that is associated to such a mass, which is the main situation we will consider.
The sign change of the mass can be understood considering again the simple example of the free hypermultiplet in 6d. Even though the analysis can't be applied in the same way, one can expect the same conclusion to hold also for flux tube compactifications of 6d SCFTs. Again, the discussion for the free hypermultiplet can be found in \cite{Chan:2000qc} and here we shall review its main points.

\subsubsection*{Back to the free hypermultiplet example}

The  6d  free hypermultiplet theory has a Noether current $J_\mu^A$ in the adjoint of $\mathfrak{su}(2)_F$ with $A=1,2,3$. Using the supersymmetry transformations on the current one can find its full supermultiplet
\be
J^A_\mu & = & \frac{1}{4}i\epsilon_{ij}\epsilon_{ab} (T^A)_c^{\ b}(\phi^{jc}\partial_\mu \phi^{ia} - \partial_\mu \phi^{jc} \phi^{ia}) - \half i \epsilon_{ab}(T^A)_c^{\ b} \Gamma_\mu^{\alpha\beta} \psi_\alpha^a \psi_\beta^c\,,\nonumber\\
S_\alpha^{j,A} & = & i\epsilon_{ba}(T^A)_c^{\ b} \phi^{ja}\psi_\alpha^c\,,\nonumber\\
\Delta^{ij,A} & = & \half i\epsilon_{ba}(T^A)_c^{\ b}\phi^{ic}\phi^{ja}\,,
\ee
where $T^A$ denote the Pauli matrices. The  supersymmetry transformations are
\be
\delta J^{\mu A} & = & \epsilon_{ij} \eta^{\alpha i} (\Gamma^{\mu\nu})_\alpha^{\ \beta}\partial_\nu S_\beta^{j,A}\,,\nonumber\\
\delta S_\beta^{j,A} & = & \eta^{j\gamma} \Gamma^\mu_{\beta\gamma} J_\mu^A + \epsilon_{ki}\eta^{\gamma k} \Gamma^\mu_{\beta\gamma} \partial_\mu \Delta^{ij,A}\,,\nonumber\\
\delta \Delta^{ij,A} & = & \eta^{\alpha i} S_\alpha^{j,A} + \eta^{\alpha j} S_\alpha^{i,A}\,,
\ee
where we used the equation of motion \eqref{5dhypeom} for $\psi_\alpha$ in the transformation of $J_\mu^A$.

Next we wish to consider the case where the $A_5$ component varies along the $x^4$ direction, meaning the field strength $F_{45}$ is non-vanishing. This will generate a  5d  hypermultiplet with a varying mass $m(x^4)$. Coupling the background gauge field $A_\mu$ to the theory requires adding to the action
\be
\label{E:CouplingCurrent}
\int d^6 x J_\mu A^\mu
\ee
plus a term proportional to $A^2$ to preserve gauge invariance. This coupling gives the terms proportional to $m$ and $m^2$ in the action \eqref{E:MassiveHyperAction} when considering only the $J_\mu^3$ component related to the subgroup $\fu(1)_F\subset \fsu(2)_F$. The variation of this action with a varying mass is given by
\be
\delta S = \int d^5 x\, m'(x^4)\epsilon_{ij}\epsilon_{ab} (T^3)_c^{\ b} \eta^{\gamma i} (i\Gamma^4\Gamma^5)_\gamma^\alpha \psi_\alpha^a \phi^{jc}\,,
\ee
with $m'(x^4)\equiv dm/dx^4$. In order to preserve supersymmetry we demand that the supersymmetric variation of the full action vanishes. Let's consider adding the following term to the Lagrangian:
\be
L_{new} = \frac{1}{4}m'(x^4)\epsilon_{ab} (T^3)_c^{\ b}\epsilon_{ij} (T^3)_k^{\ j}\phi^{ia}\phi^{kc}\,.
\ee
The supersymmetric variation of this term is
\be
\delta L_{new} = \frac{1}{2}m'(x^4)\epsilon_{ab} (T^3)_c^{\ b}\epsilon_{ij} (T^3)_k^{\ j} 2\eta^{\alpha i}\psi^{a}_\alpha\phi^{kc}\,.
\ee
Thus the supersymmetric variation vanishes if
\be
\label{E:susyCond}
(T^3)_j^{\ i} \eta^{\alpha j} = \eta^{\gamma i}(i\Gamma^4\Gamma^5)_\gamma^\alpha\,.
\ee
This equation breaks half of the supersymmetry, leaving  4d  $\mathcal{N}=1$ supersymmetry in the directions $x^0,\cdots,x^3$. The term added to preserve supersymmetry is proportional to $\Delta^{12,3}$, and since $m'(x^4)$ is $\int dx^5 F_{45}$ we see that the extra term it is proportional to
\be
\int d^6 x F_{45} \Delta^{12,3}\,,
\ee
which is required to be added together with \eqref{E:CouplingCurrent} due to supersymmetry.

The full action for a hypermultiplet with a varying mass is thus
\be
S & = & \int d^5x\left[-\frac{1}{4}\epsilon_{ij}\epsilon_{ab} (\partial_{\widetilde{\mu}} \phi^{ia} \partial^{\widetilde{\mu}} \phi^{jb} + m(x^4)^2 \phi^{ia}\phi^{jb} - m'(x^4)(T^3)_c^{\ b}(T^3)_k^{\ j} \phi^{ia}\phi^{kc})\right.\nonumber\\
 & & \left. +\half \epsilon_{ab}\psi^a_\alpha \Gamma^{\widetilde{\mu}\alpha\beta} \partial_{\widetilde{\mu}} \psi^b_\beta + \half im(x^4)\epsilon_{ab} (T^3)_c^{\ b} \psi^a_\alpha \Gamma^{5\alpha\beta} \psi^c_\beta \right]\,.
\ee
It preserves the supersymmetry transformations \eqref{E:susyTrans} when $\eta$ solves \eqref{E:susyCond}. The equations of motion are
\be
\label{E:massiveEOM}
\left(\partial_{\widetilde{\mu}} \partial^{\widetilde{\mu}} + m(x^4)^2\right)\phi^{ia} - m'(x^4)(T^3)^i_j(T^3)^a_b\phi^{jb} & = & 0\,,\nonumber\\
\Gamma^{\widetilde{\mu}\alpha\beta}\partial_{\widetilde{\mu}} \psi^a_\beta + im(x^4)(T^3)^a_b \Gamma^{5\alpha\beta}\psi^b_\beta & = & 0\,.
\ee

We want to determine the 4d modes obtained from this construction. The bosons and fermions will have a 4d massless mode for every solution of the equations of motion \eqref{E:massiveEOM}. The solution to the fermionic equation is
\be
\psi(x^4)=e^{-i\Gamma^4\Gamma^5 T^3 \int_0^{x^4} m(y)dy} \psi_0\,,
\ee
where we suppressed the indices. Both matrices $i\Gamma^4\Gamma^5$ and $T^3$ have eigenvalues $+1$ and $-1$. Thus, for the solution to be normalizable we require
\be\label{cond0modes}
\int^{x^4}_0 m(y)dy \xrightarrow{x^4 \to \pm \infty} \infty \qquad or \qquad \int^{x^4}_0 m(y)dy \xrightarrow{x^4 \to \pm \infty} -\infty\,.
\ee
In the former case the solution is normalizable if $\psi_0$ has the same eigenvalue under $i\Gamma^4\Gamma^5$ and $T^3$ and in the latter case the eigenvalues must be opposite. In both cases we get two chiral spinors related by a reality condition leaving one independent chiral spinor.

For the bosonic equation in a similar way the solution is
\be
\phi(x^4)=e^{-iT^3_R T^3_F \int_0^{x^4} m(y)dy} \phi_0\,,
\ee
where $T^3_R$ and $T^3_F$ are the third pauli matrix of the $\mathfrak{su}(2)_R$ and $\mathfrak{su}(2)_F$ flavor symmetry, respectively. There is a normalizable solution in the same two cases from before, both with two solutions related by the reality condition. Thus, in total we get a single massless $\mathcal{N}=1$ chiral multiplet.

The condition \eqref{cond0modes} for the existence of the zero modes implies that $m(x^4)$ must cross zero at some point, as mentioned before. Notice that in the case of a 5d gauge theory that is UV completed by an SCFT tuning the mass parameter to zero, which is the only ECB deformation we are turning on, corresponds in general to going back to the strongly coupled point. In the case of 5d gauge theories descending from 6d SCFTs, this corresponds to the 5d KK theory obtained from circle compactification with a codimension one locus where the holonomies of the 6d theory vanish. Hence, we generically expect our flux domain wall configuration to be characterized by strong coupling effects localized on the wall where the 4d theory lives. This makes the determination of the 4d d.o.f.'s in the case of a 6d SCFT a difficult task, unlike the case of the free hyper that we just reviewed. We will come back to this point in Section \ref{S:DW4dT}. 

\subsection{Example: basic flux domain wall for the rank 1 E-string}

We will now apply these ideas to study flux domain walls associated with the rank 1 E-string theory.
The simplest domain wall we can consider is the one associated with the following flux for the $\mathfrak{e}_8$ global symmetry:\footnote{Notice that this flux violates Dirac's quantization condition for a closed Riemann surface, but as we commented previously this is admissible since we are compactifying the 6d E-string theory on an infinitely long tube. Nonetheless, one can use such flux for a closed Riemann surface with the addition of discrete flux. We will discuss flux quantization more thoroughly in the next section.}
\begin{align}
\vec{\mathcal{F}}=\left(-\frac{1}{2},-\frac{1}{2},-\frac{1}{2},-\frac{1}{2},-\frac{1}{2},-\frac{1}{2},-\frac{1}{2},-\frac{1}{2}\right)\,,
\end{align}
where we are parametrizing the flux vector in terms of the Cartan of $\mathfrak{e}_8$ given by the embedding $\mathfrak{so}(2)^8\subset \mathfrak{so}(16)\subset \mathfrak{e}_8$ and the normalization is consistent with our definitions in \eqref{SU2_8Froots}. In a different parametrization, this corresponds to flux $-\frac{1}{2}$ for the $\mathfrak{u}(1)$ that is the Cartan of the $\mathfrak{su}(2)$ in the decomposition $\mathfrak{e}_7\oplus \mathfrak{su}(2)\subset \mathfrak{e}_8$. Hence, the associated flux domain wall is characterized by two copies of the 5d $\mathfrak{su}(2)+8F$ gauge theory with a variable mass parameter $m_{\mathfrak{u}(1)}$ for the $\mathfrak{u}(1)$ in the decomposition $\mathfrak{su}(8)\oplus \mathfrak{u}(1)\subset \mathfrak{so}(16)$. In particular, as we discussed, the mass should invert its sign between the two sides of the wall, which corresponds to complex conjugation for such $\mathfrak{u}(1)$ symmetry. Remember that $\mathfrak{so}(16)$ is the flavor symmetry of the gauge theory, which is enhanced to the full $\mathfrak{e}_8$ only in the UV. The sign flip of the mass parameter corresponds to a Weyl reflection of $\mathfrak{e}_8$ which is not in the Weyl group of $\mathfrak{so}(16)$, specifically it is the Weyl reflection of the $\mathfrak{su}(2)$ in the decomposition $\mathfrak{e}_7\oplus \mathfrak{su}(2)\subset \mathfrak{e}_8$. This means that we are really considering a duality domain wall in the sense of \cite{Gaiotto:2015una}.

We would like now to determine the ECB phases of the two 5d bulk theories that are consistent with such a configuration. For this purpose, it is useful to look at the branching rule of the vector representation of $\mathfrak{so}(16)$, under which the 8 hypers or equivalently the 16 half-hypers transform, with respect to its $\mathfrak{su}(8)\oplus \mathfrak{u}(1)$ subgroup, as
\begin{align}\label{eq:SO16vectoSU8U1}
{\bf 16}\to{\bf 8}^{+1}\oplus\overline{\bf 8}^{-1}\,.
\end{align}
We then see that turning on the mass parameter $m_{\mathfrak{u}(1)}$ for the $\mathfrak{u}(1)$ corresponds to giving mass to all the flavors. Hence, we must be in the ECB phase where all the 8 flavors can be integrated out, which is the following:\footnote{This is not the only phase which is consistent with integrating out all the flavors. There is also another phase, that the reader can find depicted in the bottom left of Figure \ref{F:Rank1ECB_BG} or in the second to last row of Table 6 of \cite{Apruzzi:2019enx}. The theory resulting from integrating out the flavors is again a pure $SU(2)$ gauge theory, but with a theta-angle $\theta=\pi$. Upon reduction to 4d this is expected to lead to a Witten anomaly \cite{Witten:1982fp} for the corresponding global symmetry of the puncture of the tube, since both the theta-angle in 5d and the Witten anomaly in 4d are controlled by $\pi_4(SU(2))=\mathbb{Z}_2$. All the 4d models we consider are free of such an anomaly, so we will not consider this phase.}
\begin{align}\label{eq:BGE7U1}
\ytableausetup{centertableaux,boxsize=12pt}
\begin{ytableau}
*(blue) & *(blue) & *(blue) & *(blue) & *(blue) & *(blue) & *(blue) & *(blue) \\
\none[] & \none[] & \none[] & \none[] & \none[] & \none[] & *(yellow) & *(yellow) & *(yellow) & *(yellow) & *(yellow) & *(yellow) & *(yellow) & *(yellow) \\ \none[] \\
*(blue) & *(blue) & *(blue) & *(blue) & *(blue) & *(blue) & *(blue) & *(blue) \\
\none[] & \none[] & \none[] & \none[] & \none[] & \none[] & *(yellow) & *(yellow) & *(yellow) & *(yellow) & *(yellow) & *(yellow) & *(yellow) & *(yellow) \\
\end{ytableau}
\end{align}

If we fix the above box graph to describe the ECB phase for one of the two 5d bulk theories, say the left one, then the box graph of the other theory is in principle not well-defined. This is because the inner products of the left theory $\langle (\phi,\vec{m}),\vec{w}_{I,J}\rangle$ will have definite signs specified by the box graph in \eqref{eq:BGE7U1}, while those of the right theory $\langle (\phi,\vec{\widetilde{m}}),\vec{w}_{I,J}\rangle$ do not have a definite sign, since the map of the mass parameters between the two sides in the $\mathfrak{su}(8)\oplus \mathfrak{u}(1)\subset \mathfrak{so}(16)$ parametrization is
\begin{align}
(m_{\mathfrak{u}(1)}+m_{\mathfrak{su}(8)}^1,\cdots,m_{\mathfrak{u}(1)}+m_{\mathfrak{su}(8)}^8)\,\,\rightarrow\,\,(-m_{\mathfrak{u}(1)}+m_{\mathfrak{su}(8)}^1,\cdots,-m_{\mathfrak{u}(1)}+m_{\mathfrak{su}(8)}^8)\,,
\end{align}
where $\sum_{a=1}^8m_{\mathfrak{su}(8)}^a=0$. Nevertheless, we should remember that the only parameter that we want to turn on is the mass $m_{\mathfrak{u}(1)}$, while the vector scalar vevs $\phi$ and the other masses $m_{\mathfrak{su}(8)}^a$ are set to zero. In this situation, the sign flip of $m_{\mathfrak{u}(1)}$ is effectively equivalent to the sign flip of the entire vector of masses $\vec{m}$, so the box graph associated with the 5d bulk theory on the right of the wall is
\begin{align}\label{eq:BGE7U1bis}
\ytableausetup{centertableaux,boxsize=12pt}
\begin{ytableau}
*(yellow) & *(yellow) & *(yellow) & *(yellow) & *(yellow) & *(yellow) & *(yellow) & *(yellow) \\
\none[] & \none[] & \none[] & \none[] & \none[] & \none[] & *(blue) & *(blue) & *(blue) & *(blue) & *(blue) & *(blue) & *(blue) & *(blue) \\ \none[] \\
*(yellow) & *(yellow) & *(yellow) & *(yellow) & *(yellow) & *(yellow) & *(yellow) & *(yellow) \\
\none[] & \none[] & \none[] & \none[] & \none[] & \none[] & *(blue) & *(blue) & *(blue) & *(blue) & *(blue) & *(blue) & *(blue) & *(blue) \\
\end{ytableau}
\end{align}

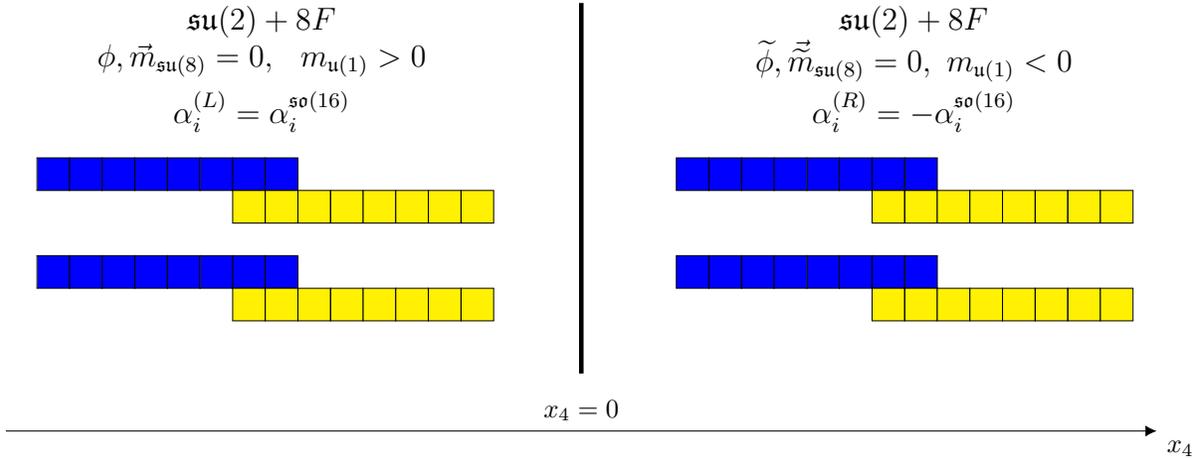
\begin{figure}[t]
\center
\begin{tikzpicture}[baseline,scale=0.85]
\tikzstyle{every node}=[font=\scriptsize]
\node[] (5dL) at (-5,2.4) {\fontsize{12pt}{12pt}$\mathfrak{su}(2)+8F$};
\node[] (Lpar) at (-5,1.8) {\fontsize{12pt}{12pt}$\phi,\vec{m}_{\mathfrak{su}(8)}=0,\quad m_{\mathfrak{u}(1)}> 0$};
\node[] (Lroot) at (-5,1) {\fontsize{12pt}{12pt}$\alpha^{(L)}_i=\alpha_i^{\mathfrak{so}(16)}$};

\node[] (5dLyt) at (-5,-1) {\ytableausetup{centertableaux,boxsize=12pt}
\begin{ytableau}
*(blue) & *(blue) & *(blue) & *(blue) & *(blue) & *(blue) & *(blue) & *(blue) \\
\none[] & \none[] & \none[] & \none[] & \none[] & \none[] & *(yellow) & *(yellow) & *(yellow) & *(yellow) & *(yellow) & *(yellow) & *(yellow) & *(yellow) \\ \none[] \\
*(blue) & *(blue) & *(blue) & *(blue) & *(blue) & *(blue) & *(blue) & *(blue) \\
\none[] & \none[] & \none[] & \none[] & \none[] & \none[] & *(yellow) & *(yellow) & *(yellow) & *(yellow) & *(yellow) & *(yellow) & *(yellow) & *(yellow) \\
\end{ytableau}};

\draw[black,line width=0.6mm] (0,-3.1) -- (0,2.7);

\draw[black,solid,->] (-9,-4) -- (9,-4);

\node[below right] at (9,-4) {\fontsize{10pt}{10pt}$x_4$};
\node[above] at (0,-4) {\fontsize{10pt}{10pt}$x_4=0$};

\node[] (5dR) at (5.2,2.4) {\fontsize{12pt}{12pt}$\mathfrak{su}(2)+8F$};
\node[] (Rpar) at (5.2,1.8) {\fontsize{12pt}{12pt}$\widetilde{\phi},\vec{\widetilde{m}}_{\mathfrak{su}(8)}=0,\,\,m_{\mathfrak{u}(1)}< 0$};
\node[] (Rroot) at (5.2,1) {\fontsize{12pt}{12pt}$\alpha^{(R)}_i=-\alpha_i^{\mathfrak{so}(16)}$};

\node[] (5dRyt) at (5,-1) {\ytableausetup{centertableaux,boxsize=12pt}
\begin{ytableau}
*(blue) & *(blue) & *(blue) & *(blue) & *(blue) & *(blue) & *(blue) & *(blue) \\
\none[] & \none[] & \none[] & \none[] & \none[] & \none[] & *(yellow) & *(yellow) & *(yellow) & *(yellow) & *(yellow) & *(yellow) & *(yellow) & *(yellow) \\ \none[] \\
*(blue) & *(blue) & *(blue) & *(blue) & *(blue) & *(blue) & *(blue) & *(blue) \\
\none[] & \none[] & \none[] & \none[] & \none[] & \none[] & *(yellow) & *(yellow) & *(yellow) & *(yellow) & *(yellow) & *(yellow) & *(yellow) & *(yellow) \\
\end{ytableau}};
\end{tikzpicture}
\caption{Structure of the ECB phases associated with the domain wall of flux $\vec{\mathcal{F}}=\left(-\frac{1}{2},\cdots,-\frac{1}{2}\right)$. The symbols $\widetilde{\phi}$, $\widetilde{m}_{\mathfrak{su}(8)}$ denote the parameters in the right theory.}
\label{basicFDWestring}
\end{figure}

At first sight, this box graph might seem inconsistent, especially because it violates the rule that above and to the left of a positive sign box one can only assign positive signs, while vice versa below and to the right of a negative sign box one can only assign negative signs. We should remember, though, that this is just a matter of conventions depending on the subset of roots we choose to define the box graph. Specifically, the conventional choice is to pick the positive simple roots, which we wrote in \eqref{SU2_8Froots} for the case under consideration. The point is that if we select the positive simple roots for the left theory, then necessarily we have to pick the negative simple roots\footnote{The negative simple roots are the positive simple roots with the sign flipped.} for the right theory. This is because, in our set-up where only $m_{\mathfrak{u}(1)}$ is turned on, its sign flip is effectively equivalent at the level of the box graph to the flip of all the $\mathfrak{so}(16)$ roots.\footnote{Notice that this is a Weyl operation of $\mathfrak{so}(16)$, which seems to be in contrast with our previous statement that the sign flip of $m_{\mathfrak{u}(1)}$ is equivalent to a Weyl reflection of $\mathfrak{e}_8$ which is not in the Weyl group of $\mathfrak{so}(16)$. The crucial point is that we can equivalently use the $\mathfrak{so}(16)$ Weyl transformation once we have set to zero all the masses $m_{\mathfrak{su}(8)}^a$.} Another way to say this is that the ECB is defined modulo the Weyl action of the group, that is we consider only one Weyl chamber representative. If we choose the usual Weyl chamber for the left theory, then we have to choose a different Weyl chamber for the right one because of the non-trivial map of the parameters between the two. In fact, we can equivalently represent the ECB phase of the theory on the right side of the wall with the usual box graph provided that we use a different convention for the roots, namely if the roots on the left of the wall are $\alpha^{(L)}_i=\alpha_i^{\mathfrak{so}(16)}$ then those on the right are $\alpha_i^{(R)}=-\alpha_i^{\mathfrak{so}(16)}$. The configuration that we just described associated with the domain wall of flux $\vec{\mathcal{F}}=\left(-\frac{1}{2},\cdots,-\frac{1}{2}\right)$ is summarized in Figure \ref{basicFDWestring}.

We will present more complicated flux domain walls in the next section, after having discussed how to glue several copies of this basic one in order to generate configurations with multiple domain walls concatenated. We will also come back to this basic flux domain wall in Section \ref{S:DW4dT}, where we will review the field content of the 4d model resulting from the configuration in Figure \ref{basicFDWestring} after further compactifying the $x_4$ direction, which was first proposed in \cite{Kim:2017toz}.

\section{Higher flux domain walls, tubes and tori}
\label{S:HFDW}

\subsection{Flux quantization, hypermultiplet masses, and the ECB phases}

The set-up offered so far of 5d flux domain wall corresponds to a compactification of a 6d $(1,0)$ SCFT on an infinite flux tube. In this section we will generalize this construction to a torus compactification that will enforce a quantization of the flux supported on it. The flux torus can be thought of as a concatenation of several flux domain walls, which are our building blocks, on a circle. The quantization condition on the torus thus allows us to associate a flux to each fundamental flux domain wall. We will also consider flux domain walls associated with different values of the flux.

In order to construct a 4d theory generated by a compactification of a 6d $(1,0)$ SCFT on $T^2$, we will consider a compact $x^4$ direction in addition to the $x^5$ direction. As before we will turn on a background gauge field $A_{h,\mu}$ associated to a single $U(1)_h\subset G_F^{6d}$,\footnote{In this section we pay particular attention to the precise global structure of all the groups and do not just focus on their Lie algebras.} such that it satisfies the quantization condition 
\be
\label{E:QuanCond}
\frac{1}{2\pi}\int_{T^2} F_{h,45} = \frac{n}{q_{min}^{6d}},\qquad n \in \mathbb{Z}\,,
\ee
where $q_{min}^{6d}$ is the minimal charge under $U(1)_h$ that appears in the spectrum of the 6d theory.
Let us consider a regime where $R_5\ll R_4$ with $R_\mu$ the radius of the $S^1$ in the  $x^\mu$ direction. We can then first reduce along the $x^5$ direction with holonomy
\be
W(x^4)=\int_0^{2\pi R_5} A_{h,5}(x^4,x^5)dx^5,
\ee
that varies from $0$ to $2\pi n/q_{min}^{6d}$ as $x^4$ varies from $0$ to $2\pi R_4$ such that \eqref{E:QuanCond} is satisfied.

Next, we want to look at weakly coupled regimes of the 5d KK theory. Assuming we don't turn on vevs for the vector multiplet scalars and we only turn on an holonomy for $U(1)_h \subset G_F^{6d}$, the parameters that will control the gauge coupling will be $m_h$, the hypermultiplet mass parameter associated to the holonomy $W(x^4)$, and $m_{KK}=R_5^{-1}$, the compactification circle length scale. Compactifying a 6d hypermultiplet on a circle  generates a tower of KK states. As we saw in \eqref{E:MassiveHyperAction}-\eqref{5dhypeom}, if the 5d hypermultiplet coming from a 6d circle compactification has charge $q_{h}^{5d}$ under $U(1)_h$, then the holonomy we turn on shifts the masses of the tower of states such that they are given by
\be\label{eq:mass}
m_{h,\ell}(x^4) = \left|\frac{q_{h}^{5d} W(x^4)}{2\pi R_5} - \frac{\ell}{R_5}\right|\,,
\ee
with $\ell \in \mathbb{Z}$. In the low energy limit only the lowest mass state will be relevant. Since the masses depend on $x^4$ the lowest mass state will change with $x^4$, thus giving us an effective low energy description near each $x^4$ value where $W(x^4)$ is an integer multiple of $2\pi /q_{h}^{5d}$ with a specific mass parameter. Note that if we have matter with different charges under $U(1)_h$, the strongest condition will come from the one with minimal charge $q_{min}^{5d}$. Thus, the masses of all the 5d states will approach zero near each $x^4$ value where $W(x^4)$ is an integer multiple of $2\pi /q_{min}^{5d}$. 

It is then important to determine what are the minimal charges $q_{min}^{6d}$ and $q_{min}^{5d}$. In the general case the 6d theory won't necessarily include hypermultiplets and may include BPS instantons (strings). To determine $q_{min}^{6d}$ we can simply decompose the representations of all the hypermultiplets and BPS strings under 
\be
G_F^{6d} \to \frac{H_F^{6d} \times U(1)_h}{\mathcal{Z}^{6d}}\,,
\ee
where $H_F^{6d}$ is the simply connected group of the algebra which is the commutant of $U(1)_h$ in $G_F^{6d}$ and $\mathcal{Z}^{6d}$ is some discrete group which is a subgroup of the center of $H_F^{6d}$ such that none of the states in the 6d spectrum are charged under it. Note that in this decomposition $G_F^{6d}$ is the flavor symmetry group of the 6d theory and not the flavor symmetry algebra, so the precise global structure of the symmetry is crucial in order to understand what $\mathcal{Z}^{6d}$ is. In the case in which $\mathcal{Z}^{6d}=\Z_k$, the minimal charge under $U(1)_h$ will be $q_{min}^{6d}=k$. 

In the above circle reduction from 6d to 5d we intentionally distinguished between the minimal charge in 6d $q_{min}^{6d}$ and 5d $q_{min}^{5d}$ as these can differ. The 5d minimal charge is determined in a similar way to the 6d minimal charge, but now we have to consider the decomposition of the 5d flavor symmetry
\be
G_F^{5d} \to \frac{H_F^{5d} \times U(1)_h}{\mathcal{Z}^{5d}}\,.
\ee
Again, $H_F^{5d}$ is the simply connected group of the algebra which is the commutant of $U(1)_h$ in $G_F^{5d}$ and $\mathcal{Z}^{5d}$ is some discrete subgroup which will determine the minimal charge $q_{min}^{5d}$. For $\mathcal{Z}^{5d}=\mathbb{Z}_m$ we will have $q_{min}^{5d}=m$. $G_F^{5d}$ is the flavor symmetry of the 5d KK theory not including $U(1)_{KK}$, which can differ from the flavor symmetry $G_F^{6d}$ at the SCFT point. This difference can lead to a difference between $\mathcal{Z}^{6d}$ and $\mathcal{Z}^{5d}$ and thus also to $q_{min}^{6d}\neq q_{min}^{5d}$. In general $q_{min}^{5d}\ge q_{min}^{6d}$ and in the case of inequality this will lead to $n\cdot q_{min}^{5d}/q_{min}^{6d}$ instead of $n$ domain walls required to build a torus with flux $n/q_{min}^{6d}$. This implies that the flux associated to a single domain wall is $1/q_{min}^{5d}$. Note that $q_{min}^{5d}/ q_{min}^{6d} \in \mathbb{Z}$ and we will usually work in a normalization where $q_{min}^{6d}=1$ for simplicity. Below we will show using the rank 1 E-string example how this comes into play.

Going back to the mass parameters in \eqref{eq:mass}, we can relate the lowest mass parameters around each point in the $x_4$ direction where $W(x_4)$ crosses a multiple of $2\pi /q_{min}^{5d}$ by setting $m^{(0)}_h=m_{h,\ell=0}(x^4)$ around $x^4=0$, while the other mass parameters are set to be 
\be\label{eq:massmixKK}
m_h^{(p)}=m_{h,\ell=p}(x^4)=m_h^{(0)}-p\, m_{KK}
\ee
around $x^4$ satisfying $W(x^4)=2\pi p/q_{min}^{5d}$, where $p=0,\cdots,\widetilde{n}$ and $\widetilde{n}=nq_{min}^{5d}/q_{min}^{6d}$. In the 5d theory this can be thought of as turning on a mass parameter corresponding to the symmetry\footnote{This notation refers to the charges under the various abelian symmetries.}
\be\label{eq:mixKK}
U(1)_{h}^{(p)}=\frac{1}{2}\left(U(1)_h^{(0)} -\frac{1}{p}\, U(1)_{KK}\right)\,.
\ee
This shows explicitly that in the case of a flux torus as described above we have a theory with a $U(1)_{h}^{(0)}$ symmetry while the $U(1)_{KK}$ symmetry is actually broken to a $\mathbb{Z}_{\widetilde{n}}$ symmetry. Indeed, on the torus the points $x_4=0$ where $W(x_4)=0$ and $x_4=2\pi R_4$ where $W(x_4)=2\pi n/q_{min}^{6d}$ are actually identified because of the periodic boundary conditions and imposing that $U(1)_h^{(0)}=U(1)_{h}^{(\widetilde{n})}$ or equivalently $m_h^{(0)}=m_h^{(\widetilde{n})}$ implies that all the charges under the KK symmetry are identified modulo $\widetilde{n}$.\footnote{This can be easily seen at the level of fugacities. Denoting by $h_{(p)}$ the $U(1)_h^{(p)}$ fugacity for $p=0,\cdots,\widetilde{n}$ and by $\kappa$ the $U(1)_{KK}$ fugacity, the relation \eqref{eq:mixKK} implies that $h_{(p)}=h_{(0)}\kappa^{-p}$. Requiring $h_{(\widetilde{n})}=h_{(0)}$ indicates that $\kappa$ is a $\mathbb{Z}_{\widetilde{n}}$ fugacity satisfying $\kappa^{\widetilde{n}}=1$.}

With the above picture in mind we can now see that for all the $x^4$ values where $W(x^4)$ is an integer multiple of $2\pi/q_{min}^{5d}$ the effective gauge coupling goes to infinity indicating a strongly coupled theory located on the 4d domain wall. These interfaces are also where the mass parameter crosses zero. Since $R_5 \ll R_4$ we can think of the region near these $x^4$ values as approximately being a flux domain wall or an infinite flux tube as we discussed in the previous section. Thus, we can expect as in the free hypermultiplet example that there are chiral zero modes localized on these interfaces. Note that in order to get a higher flux with $n>1$ the mass parameters need to cross zero in the same direction $\widetilde{n}$ times. In our convention the crossing is from a negative sign on the left of the wall to a positive sign on the right of the wall for a positive flux. This is possible thanks to the fact that the mass parameter we are considering near the $p$-th domain wall is $m_h^{(p)}=m_h^{(0)}-p\, m_{KK}=m^{(p-1)}-m_{KK}$, so that even if the mass $m^{(p-1)}$ of the previous domain wall already changed from negative to positive, the shift by $-m_{KK}$ allows the mass $m_h^{(p)}$ to be negative before the $p$-th domain wall, so that it can again cross zero from negative to positive value. On the other hand if the crossing of the mass is in an opposite direction for two flux domain walls, concatenating them will lead to a cancellation of the flux. In this case the mass parameter on both walls remains the same one and the associated $U(1)_{h}^{(p)}$ symmetry is not shifted by $U(1)_{KK}$ between the two walls.

The final knot to tie in the above picture is how to relate the 5d ECB phases of neighbouring flux domain walls\footnote{The gluing points in the $x^4$ direction will be where $W(x^4)$ is an odd integer multiple of $\pi/q_{min}^{5d}$.} when concatenating them to create a higher flux domain wall as part of a higher flux tube or torus. As it was discussed in the former section the two ECB phases on the two sides of the domain wall can be related by a Weyl reflection. Thus, in the same manner we can identify the ECB phase on the right side of one domain wall with the ECB phase on the left side of its neighbouring domain wall up to a Weyl element. Denoting by $\vec{\mathcal{F}}_1$ and $\vec{\mathcal{F}}_2$ the fluxes of the glued domain walls under the Cartans of $G_F^{5d}$ and by $w\in W_G$ a Weyl element of $G_F^{5d}$, then the resulting flux will be
\be
\vec{\mathcal{F}}_{tot}=\vec{\mathcal{F}}_1+w\left(\vec{\mathcal{F}}_2\right)\,.
\ee
The simplest situation is the one in which $w$ is the trivial Weyl element. If we glue several copies of the same flux domain wall in such a way, then the resulting flux will just be that of the single domain wall multiplied by the number of glued domain walls. On the other hand, we can consider a Weyl reflection that flips the sign of all the entries of the flux vector. Relating the ECB phases of two neighbouring domain walls in such a way will lead to a cancellation of the flux and give in total a trivial domain wall. This gluing will accordingly lead to a non-trivial identification of the flavor symmetries of the 5d theories, in particular we will have that $U(1)_{h}^{(0)}=-U(1)_{h}^{(1)}$.

In general when the 5d effective low energy description has more then one hypermultiplet and thus more then one possible mass parameter, one can choose to glue domain walls in a way that adds the flux for some of the $U(1)$'s associated to some hypers and deduct the flux for others. This is done by identifying the glued domain walls with a Weyl element that flips the sign of only some of the entries of the flux vector. In this case each $U(1)$ we wish to add its flux in the gluing will mix with $U(1)_{KK}$ while the others won't as prescribed above. The mapping of the ECB phases between the two domain walls will depend on the choice of Weyl transformation. This will also imply a non-trivial identification of the $U(1)$'s of the glued walls on top of the aforementioned shift by $U(1)_{KK}$. In what comes next we will continue with the rank 1 E-string theory flux domain wall example and show several instances of gluings.

\subsection{Example: higher flux domain walls and tori for the rank 1 E-string}

In the following examples we will show how one can combine several copies of the simple rank 1 E-string domain wall discussed in the former section of flux $\vec{\mathcal{F}}=\left(-\half,\cdots,-\half\right)$, the building block of our construction, to generate various other flux domain walls and tori. Here we will consider flux domain walls of the form $\vec{\mathcal{F}}=\left(-n,...,-n,0,...,0\right)$ with $k$ non-vanishing entries and $8-k$ zeros. Let us start describing some general properties of these configurations and later we will explain the gluing needed to generate them in some examples. We will mostly focus on the case of even $k$ as the odd $k$ case is more involved on the one hand and adds no qualitative value on the other hand.

The above flux breaks the 6d global symmetry as follows:
\be
E_8 \to \frac{H \times U(1)_h}{\Z_2}\,,
\ee
where the commutant $H$ of the $U(1)_h$ that gets flux depends on the value of $k$, specifically it is $E_7$ for $k=2,8$, $Spin(14)$ for $k=4$ and $E_6\times SU(2)$ for $k=6$ \cite{Kim:2017toz}. The global structure of the residual group can be understood by looking at the branching rules (see for example Appendix A of \cite{Kim:2017toz}) and it tells us that in this case $\mathcal{Z}^{6d}=\Z_2$. Hence, in a canonical normalization of the $U(1)_h$ symmetry, the minimal charge in 6d is $q_{min}^{6d}=2$.

On the other hand, in the 5d KK theory the $E_8$ flavor symmetry is broken to a $Spin(16)/\Z_2$ flavor symmetry.\footnote{The $\Z_2$ quotient is due to the decomposition of the representation $\bf 248$ of $\mathfrak{e}_8$ to representations of $\mathfrak{so}(16)$, where we find the spinor charged under one of the $\Z_2$'s of $Spin(16)$, allowing us to mod out by the other. Note that this group is different from the flavor symmetry group of the low energy effective field theory which is $SO(16)/\Z_2$ due to the matter in the vector representation and where now the $\Z_2$ quotient is thanks to a reabsorbtion by a gauge transformation associated with the $\Z_2$ center of the $SU(2)$ gauge group.} Then the flux breaks it further as follows:
\be\label{eq:genSO16decomp}
\frac{Spin(16)}{\Z_2} \to \frac{SU(8)\times U(1)_a}{\Z_{4}} \to \frac{SU(k)\times SU(8-k) \times U(1)_a \times U(1)_b}{\Z_{4}}\,,
\ee
where $k$ is even and the flux is associated to a combination of $U(1)_a$ and $U(1)_b$ that we will pin down in the examples below.\footnote{For the case of odd $k<7$ one can expect the decomposition
\be
SO(16) \to \frac{SU(8)\times U(1)_a}{\Z_{4}} \to \frac{SU(k)\times SU(6-k) \times U(1)_a \times U(1)^3}{\Z_{4}}\,.
\ee
For $k=7$ one should expect the same decomposition as $k=1$.
} Accordingly, the vector representation $\textbf{16}$ which the hypermultiplets transform under will follow the branching rules
\be
\textbf{16} \to \textbf{8}^{+1} \oplus \overline{\textbf{8}}^{-1} \to (\textbf{k},\textbf{1})^{+1,+(4-\frac{k}{2})} \oplus (\textbf{1},\textbf{8-k})^{+1,-\frac{k}{2}} \oplus (\overline{\textbf{k}},\textbf{1})^{-1,-(4-\frac{k}{2})} \oplus (\textbf{1},\overline{\textbf{8-k}})^{-1,+\frac{k}{2}}\,.\nn\\
\ee

The above global structure of the flavor symmetry breaking under the flux can be deduced by the branching rules of the E-string theory BPS string in the $\textbf{248}$ of $\mathfrak{e}_8$ under $\mathfrak{e}_8 \to \mathfrak{so}(16)$ and the subsequent above decomposition. After the first decomposition in \eqref{eq:genSO16decomp} we have
\be
\textbf{248}\to \textbf{120} \oplus \textbf{128} & \to & \left[\textbf{63}^{0} \oplus \textbf{1}^{0} \oplus \textbf{28}^{+2} \oplus \overline{\textbf{28}}^{-2}\right] \oplus \left[\textbf{70}^0 \oplus \textbf{28}^{-2} \oplus \overline{\textbf{28}}^{+2} \oplus \textbf{1}^{\pm 4} \right]\,,
\ee
where we define $\textbf{R}^{\pm q}\triangleq \textbf{R}^{+q}\oplus \textbf{R}^{-q}$ for a representation $\textbf{R}$. One can see these representations aren't charged under a $\Z_4$ subgroup of the $\Z_8$ center. For example, the $\textbf{28}$ has charge $+2$ under the $\Z_8$ center and gets an additional contribution of $+2$ from the $U(1)$; thus, the $\textbf{28}^{+2}$ has charge $+4$ under the $\Z_8$ center preserving a $\Z_4$ subgroup.\footnote{Another way to say this is that the representation $\textbf{28}^{+2}$ is invariant under a transformation of $SU(8)\times U(1)$ which is of the form $\textrm{diag}(\textrm{e}^{\frac{2\pi i}{4}},\cdots,\textrm{e}^{\frac{2\pi i}{4}})$ for the $SU(8)$ part and $\textrm{e}^{\frac{2\pi i}{4}}$ for the $U(1)$ part and thus preserves a $\Z_4$ diagonal subgroup of the $SU(8)$ center $\Z_8$ and of $U(1)$.} The last decomposition gives
\be
\textbf{248} & \to & [(\textbf{A}_{\fsu(k)},\textbf{1})^{0,0} \oplus(\textbf{1},\textbf{A}_{\fsu(8-k)})^{0,0} \oplus (\textbf{k},\overline{\textbf{8-k}})^{0,+4} \oplus (\overline{\textbf{k}},\textbf{8-k})^{0,-4}]\nonumber\\ 
&& \oplus [(\boldsymbol{\Lambda}^2_{\fsu(k)},\textbf{1})^{\pm 2,(8-k)} \oplus (\textbf{k},\textbf{8-k})^{\pm 2,(4-k)} \oplus (\textbf{1},\overline{\boldsymbol{\Lambda}^2}_{\fsu(8-k)})^{\pm 2,-k}]\nonumber\\ 
&& \oplus [(\overline{\boldsymbol{\Lambda}^2}_{\fsu(k)},\textbf{1})^{\pm 2,-(8-k)} \oplus (\overline{\textbf{k}},\overline{\textbf{8-k}})^{\pm 2,-(4-k)} \oplus (\textbf{1},\boldsymbol{\Lambda}^2_{\fsu(8-k)})^{\pm 2,k}] \nonumber\\
&& \bigoplus_{\ell =0}^{8-k}(\boldsymbol{\Lambda}^{k-4+\ell}_{\fsu(k)},\overline{\boldsymbol{\Lambda}^{\ell}}_{\fsu(8-k)})^{0,2(2\ell -8+k)} \oplus 2\times (\textbf{1},\textbf{1})^{0,0} \oplus (\textbf{1},\textbf{1})^{\pm 4,0}
\ee
where $\textbf{A}$ and $\boldsymbol{\Lambda}^m,\overline{\boldsymbol{\Lambda}^m}$ correspond to the adjoint, and the irreducible $m$ index antisymmetric and its complex conjugate representations, respectively.\footnote{Note that $\boldsymbol{\Lambda}^{m}_{\fsu(n)}$ is empty if $m<0$ or $m>n$ in the above notation.} Note that all the representations under the decomposition $\fso(16)\to \fsu(k) \oplus \fsu(8-k) \oplus \fu(1)_a \oplus \fu(1)_b$ are not charged under a $\Z_4$ subgroup of the center remembering that $k$ is even.\footnote{In the case of odd $k$ the decomposition would be different but we expect to still have a $\Z_4$ subgroup of the center under which no representation is charged.}

This leads to the relation $q_{min}^{5d}= 2q_{min}^{6d}$ for the above decompositions. Therefore, we should expect a flux $n$ domain wall or torus to be composed of $2n$ minimal flux domain walls with flux to the same $\fu(1)$. In particular, we can see that the flux associated to the domain wall we considered in Figure \ref{basicFDWestring}, which corresponds to the case $k=8$, is $-\frac{1}{2}$.

\begin{figure}[t]
\center
\begin{tikzpicture}[baseline,scale=0.85]
\tikzstyle{every node}=[font=\scriptsize]
\node[] (5dL) at (-4.5,3.8) {\fontsize{10pt}{10pt}$\mathfrak{su}(2)+8F$};
\node[] (Lpar) at (-4.5,3) {\fontsize{10pt}{10pt}$\phi,m_{{1\le i\le k}}>0,\, m_{k< i\le 8}=0$};
\node[] (Lroot) at (-4.5,2.2) {\fontsize{10pt}{10pt}$\alpha^{(L)}_i=\alpha_i^{\mathfrak{so}(16)}$};

\node[] (Lpar) at (-6.95,-1.15) {\fontsize{10pt}{10pt}$k$};
\node[] (Lpar) at (-5,-1.15) {\fontsize{10pt}{10pt}$8-k$};

\node[] (Lpar) at (-1.9,-0.85) {\fontsize{10pt}{10pt}$k$};
\node[] (Lpar) at (-3.85,-0.85) {\fontsize{10pt}{10pt}$8-k$};

\node[] (5dLyt) at (-4.5,-1) {\ytableausetup{centertableaux,boxsize=12pt}
\begin{ytableau}
*(blue) & *(blue) & *(blue) & *(blue) & *(blue) & *(blue) & *(blue) & *(blue) \\
\none[] & \none[] & \none[] & \none[] & \none[] & \none[] & *(blue) & *(blue) & *(blue) & *(blue) & *(yellow) & *(yellow) & *(yellow) & *(yellow) \\ \none[] \\ \none[] \\
*(blue) & *(blue) & *(blue) & *(blue) & *(yellow) & *(yellow) & *(yellow) & *(yellow) \\
\none[] & \none[] & \none[] & \none[] & \none[] & \none[] & *(yellow) & *(yellow) & *(yellow) & *(yellow) & *(yellow) & *(yellow) & *(yellow) & *(yellow) \\
\end{ytableau}};

\draw[black,line width=0.6mm] (0,-3.1) -- (0,3.8);

\draw[black,solid,->] (-9,-4) -- (9,-4);

\draw [decorate, decoration = {brace}] (-8,-1.45) --  (-6,-1.45);
\draw [decorate, decoration = {brace}] (-5.9,-1.45) --  (-4,-1.45);

\draw [decorate, decoration = {brace}] (-2.95,-0.55) --  (-4.9,-0.55);
\draw [decorate, decoration = {brace}] (-0.9,-0.55) --  (-2.85,-0.55);

\node[below right] at (9,-4) {\fontsize{10pt}{10pt}$x_4$};
\node[above] at (0,-4) {\fontsize{10pt}{10pt}$x_4=0$};

\node[] (5dR) at (4.7,3.8) {\fontsize{10pt}{10pt}$\mathfrak{su}(2)+8F$};
\node[] (Rpar) at (4.7,3.1) {\fontsize{10pt}{10pt}$\phi,m_{{1\le i\le k}}<0,\, m_{k< i\le 8}=0$};
\node[] (Rpar) at (4.7,2.4) {\fontsize{10pt}{10pt}$\alpha_{1\le i< k}^{(R)} =\alpha_i^{\mathfrak{so}(16)}$,\, $\alpha_{k< i\le 8}^{(R)} =-\alpha_i^{\mathfrak{so}(16)}$};
\node[] (Lpar) at (4.7,1.7) {\fontsize{10pt}{10pt}$\alpha_k^{(R)} = \alpha_k^{\fso(16)} +2\sum_{i=k+1}^6 \alpha_i^{\fso(16)}$};
\node[] (Lpar) at (4.6,1) {\fontsize{10pt}{10pt}$ + \alpha_7^{\fso(16)} + \alpha_8^{\fso(16)}$};

\node[] (5dRyt) at (4.5,-1) {\ytableausetup{centertableaux,boxsize=12pt}
\begin{ytableau}
*(yellow) & *(yellow) & *(yellow) & *(yellow) & *(blue) & *(blue) & *(blue) & *(blue) \\
\none[] & \none[] & \none[] & \none[] & \none[] & \none[] & *(blue) & *(blue) & *(blue) & *(blue) & *(blue) & *(blue) & *(blue) & *(blue) \\ \none[] \\ \none[] \\
*(yellow) & *(yellow) & *(yellow) & *(yellow) & *(yellow) & *(yellow) & *(yellow) & *(yellow) \\
\none[] & \none[] & \none[] & \none[] & \none[] & \none[] & *(yellow) & *(yellow) & *(yellow) & *(yellow) & *(blue) & *(blue) & *(blue) & *(blue) \\
\end{ytableau}};
\end{tikzpicture}
\caption{The ECB phases associated with a domain wall of flux $\vec{\mathcal{F}}=\left(-1,...,-1,0,...,0\right)$ with even $k$ $(-1)$ entries and $8-k$ zeroes. Note that in this diagram $m_i$ denotes the mass of the $i$-th hypermultiplet. In the figure we show the example of $k=4$ for illustration.}
\label{F:GenBoxGraph}
\end{figure}
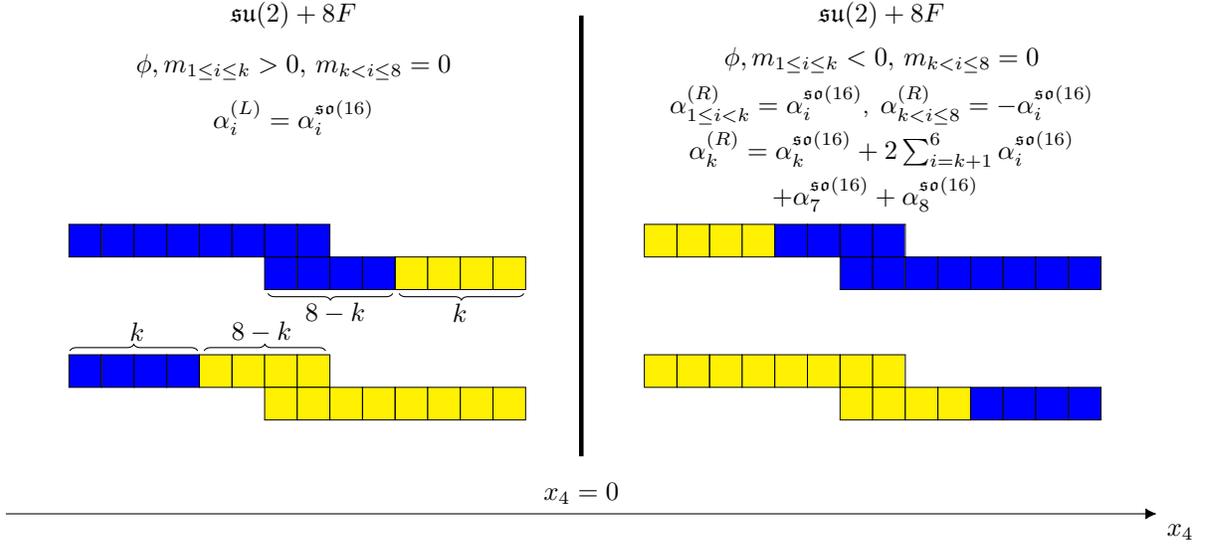

As we will see momentarily, in these more general configurations there are several mass parameters turned on and they all change as we move in the $x_4$ direction. Nevertheless, the change is such that eventually we have flux only for one specific $\mathfrak{u}(1)$, while there is no flux for the others. 
This is achieved by changing the mass parameter associated to several hypermultiplets that see the flux from positive to negative at each domain wall location, while the other mass parameters change continuously from positive to negative at the $(2i-1)$-th domain wall and back to positive at the $2i$-th domain wall, with $i=1,\cdots,\frac{\widetilde{n}}{2}$. This is possible thanks to an $m_{KK}$ shift at the concatenation interface of the two domain walls for the hyper multiplet masses related to the $\mathfrak{u}(1)$ with flux, while employing no such shift for the rest of the mass parameters, as we will show momentarily. 
It turns out that the only ECB phase consistent with the shift of the masses imposed by the gluing required to form the desired flux domain wall, is the one where $k$ hypers can be integrated out, which is shown on the left of Figure \ref{F:GenBoxGraph}. Accordingly on the other side of the domain wall we will need to flip the mass sign of $k$ hypermultiplet masses which would amount to flipping the sign of $k$ boxes on each side of the box graph. Nevertheless, as we showed in Figure \ref{basicFDWestring} for the case $k=8$, there is a non-trivial identification of the Weyl chamber used for the right and left sides of the domain wall, as specified in Figure \ref{F:GenBoxGraph}.\footnote{For odd $k$ the picture would be similar but the minimal flux is $(-2,\cdots,-2,0,\cdots,0)$ with $k$ $(-2)$ entries.} We will show how to determine the Weyl chamber of each side in the following examples.

We will now show how to construct these flux domain walls for generic $k$,  as well as the associated flux tori from the basic flux domain wall with $k=8$ shown in Figure \ref{basicFDWestring} using the above gluing procedure. This will be clarified first by using specific examples.

\subsubsection{Flux $\vec{\mathcal{F}}=\left(-n,-n,-n,-n,-n,-n,-n,-n\right)$}
\label{SS:5dPhigluing}

We start with the simplest example of higher flux domain wall and torus, where we simply add the fluxes of several $\vec{\mathcal{F}}=\left(-\half,\cdots,-\half\right)$ flux domain walls. In this case the mixing of the $\mathfrak{u}(1)$ we give flux to with $\mathfrak{u}(1)_{KK}$ is as described in the general case in \eqref{eq:mixKK}. Moreover, the Weyl transformation relating the right side of the left domain wall with the left side of the right domain wall is the trivial one since we only want to add the flux of the two domain walls. This means that the roots of the $\mathfrak{so}(16)$ global symmetry for the two 5d theories are trivially identified
\be
\alpha^{(R,0)}_i=\alpha^{(L,1)}_i\,,
\ee
where $\alpha^{(R,0)}_i$ are the roots of the theory on the right of the left domain wall, while $\alpha^{(L,1)}_i$ are the roots of the theory on the left of the right domain wall. This in particular means that the flavor symmetry is trivially identified between the two walls. 

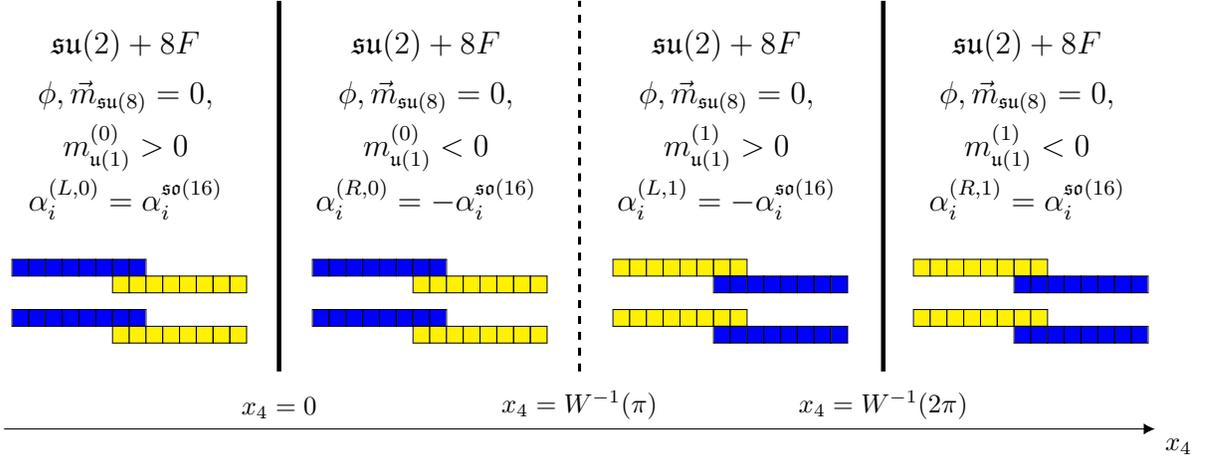
\begin{figure}[t]
\center
\begin{tikzpicture}[baseline,scale=0.85]
\tikzstyle{every node}=[font=\scriptsize]
\node[] (5dL) at (-7.1,2) {\fontsize{12pt}{12pt}$\mathfrak{su}(2)+8F$};
\node[] (Lpar) at (-7.1,1.2) {\fontsize{12pt}{12pt}$\phi,\vec{m}_{\mathfrak{su}(8)}=0,$};
\node[] (Lpar) at (-7.1,0.4) {\fontsize{12pt}{12pt}$m_{\mathfrak{u}(1)}^{(0)}> 0$};
\node[] (Lpar) at (-7.1,-0.4) {\fontsize{12pt}{12pt}$\alpha_i^{(L,0)} =\alpha_i^{\mathfrak{so}(16)}$};

\node[] (5dLyt) at (-7.1,-2) {\ytableausetup{centertableaux,boxsize=6pt}
\begin{ytableau}
*(blue) & *(blue) & *(blue) & *(blue) & *(blue) & *(blue) & *(blue) & *(blue) \\
\none[] & \none[] & \none[] & \none[] & \none[] & \none[] & *(yellow) & *(yellow) & *(yellow) & *(yellow) & *(yellow) & *(yellow) & *(yellow) & *(yellow) \\ \none[] \\
*(blue) & *(blue) & *(blue) & *(blue) & *(blue) & *(blue) & *(blue) & *(blue) \\
\none[] & \none[] & \none[] & \none[] & \none[] & \none[] & *(yellow) & *(yellow) & *(yellow) & *(yellow) & *(yellow) & *(yellow) & *(yellow) & *(yellow) \\
\end{ytableau}};

\draw[black,line width=0.6mm] (-4.7,-3.1) -- (-4.7,2.7);
\node[above] at (-4.7,-4) {\fontsize{10pt}{10pt}$x_4=0$};

\node[] (5dL) at (-2.4,2) {\fontsize{12pt}{12pt}$\mathfrak{su}(2)+8F$};
\node[] (Lpar) at (-2.4,1.2) {\fontsize{12pt}{12pt}$\phi,\vec{m}_{\mathfrak{su}(8)}=0,$};
\node[] (Lpar) at (-2.4,0.4) {\fontsize{12pt}{12pt}$m_{\mathfrak{u}(1)}^{(0)}<0$};
\node[] (Lpar) at (-2.4,-0.4) {\fontsize{12pt}{12pt}$\alpha_i^{(R,0)} =-\alpha_i^{\mathfrak{so}(16)}$};

\node[] (5dLyt) at (-2.4,-2) {\ytableausetup{centertableaux,boxsize=6pt}
\begin{ytableau}
*(blue) & *(blue) & *(blue) & *(blue) & *(blue) & *(blue) & *(blue) & *(blue) \\
\none[] & \none[] & \none[] & \none[] & \none[] & \none[] & *(yellow) & *(yellow) & *(yellow) & *(yellow) & *(yellow) & *(yellow) & *(yellow) & *(yellow) \\ \none[] \\
*(blue) & *(blue) & *(blue) & *(blue) & *(blue) & *(blue) & *(blue) & *(blue) \\
\none[] & \none[] & \none[] & \none[] & \none[] & \none[] & *(yellow) & *(yellow) & *(yellow) & *(yellow) & *(yellow) & *(yellow) & *(yellow) & *(yellow) \\
\end{ytableau}};

\draw[black,dashed,line width=0.4mm] (0,-3.1) -- (0,2.7);
\node[above] at (0,-4) {\fontsize{10pt}{10pt}$x_4=W^{-1}(\pi)$};

\draw[black,solid,->] (-9,-4) -- (9,-4);
\node[below right] at (9,-4) {\fontsize{10pt}{10pt}$x_4$};

\node[] (5dR) at (2.3,2) {\fontsize{12pt}{12pt}$\mathfrak{su}(2)+8F$};
\node[] (Rpar) at (2.3,1.2) {\fontsize{12pt}{12pt}$\phi,\vec{m}_{\mathfrak{su}(8)}=0,$};
\node[] (Rpar) at (2.3,0.4) {\fontsize{12pt}{12pt}$m_{\mathfrak{u}(1)}^{(1)}> 0$};
\node[] (Lpar) at (2.3,-0.4) {\fontsize{12pt}{12pt}$\alpha_i^{(L,1)} =-\alpha_i^{\mathfrak{so}(16)}$};

\node[] (5dRyt) at (2.3,-2) {\ytableausetup{centertableaux,boxsize=6pt}
\begin{ytableau}
*(yellow) & *(yellow) & *(yellow) & *(yellow) & *(yellow) & *(yellow) & *(yellow) & *(yellow) \\
\none[] & \none[] & \none[] & \none[] & \none[] & \none[] & *(blue) & *(blue) & *(blue) & *(blue) & *(blue) & *(blue) & *(blue) & *(blue) \\ \none[] \\
*(yellow) & *(yellow) & *(yellow) & *(yellow) & *(yellow) & *(yellow) & *(yellow) & *(yellow) \\
\none[] & \none[] & \none[] & \none[] & \none[] & \none[] & *(blue) & *(blue) & *(blue) & *(blue) & *(blue) & *(blue) & *(blue) & *(blue) \\
\end{ytableau}};

\draw[black,line width=0.6mm] (4.75,-3.1) -- (4.75,2.7);
\node[above] at (4.75,-4) {\fontsize{10pt}{10pt}$x_4=W^{-1}(2\pi)$};

\node[] (5dR) at (7,2) {\fontsize{12pt}{12pt}$\mathfrak{su}(2)+8F$};
\node[] (Rpar) at (7,1.2) {\fontsize{12pt}{12pt}$\phi,\vec{m}_{\mathfrak{su}(8)}=0,$};
\node[] (Rpar) at (7,0.4) {\fontsize{12pt}{12pt}$m_{\mathfrak{u}(1)}^{(1)}< 0$};
\node[] (Lpar) at (7,-0.4) {\fontsize{12pt}{12pt}$\alpha_i^{(R,1)} =\alpha_i^{\mathfrak{so}(16)}$};

\node[] (5dRyt) at (7,-2) {\ytableausetup{centertableaux,boxsize=6pt}
\begin{ytableau}
*(yellow) & *(yellow) & *(yellow) & *(yellow) & *(yellow) & *(yellow) & *(yellow) & *(yellow) \\
\none[] & \none[] & \none[] & \none[] & \none[] & \none[] & *(blue) & *(blue) & *(blue) & *(blue) & *(blue) & *(blue) & *(blue) & *(blue) \\ \none[] \\
*(yellow) & *(yellow) & *(yellow) & *(yellow) & *(yellow) & *(yellow) & *(yellow) & *(yellow) \\
\none[] & \none[] & \none[] & \none[] & \none[] & \none[] & *(blue) & *(blue) & *(blue) & *(blue) & *(blue) & *(blue) & *(blue) & *(blue) \\
\end{ytableau}};

\end{tikzpicture}
\caption{Structure of the ECB phases associated with the domain wall of flux $\vec{\mathcal{F}}=\left(-1,\cdots,-1\right)$. Here the box graphs on the two sides of each domain wall are identical since they are drawn using different conventions for the $\mathfrak{so}(16)$ roots as specified. Note that $m_{\mathfrak{u}(1)}^{(1)} = m_{\mathfrak{u}(1)}^{(0)}-m_{KK}$.}
\label{F:basicHFDWestring}
\end{figure}

Note that the mass parameter $m_{\mathfrak{u}(1)}^{(0)}$ associated to the left domain wall is negative in its right region, while the mass parameter $m_{\mathfrak{u}(1)}^{(1)}$ associated to the right domain wall is positive in its left side due to the shift by $m_{KK}$ we described in \eqref{eq:massmixKK}. Hence, since the roots are identified but the mass parameter changes sign, the box graph encoding the ECB phase of the theory on the left of the first domain wall and that encoding the ECB phase of the theory on the right of the second domain wall are oppositely coloured. We summarize this concatenation of two $\vec{\mathcal{F}}=\left(-\half,\cdots,-\half\right)$ flux domain walls in Figure \ref{F:basicHFDWestring}.

One can join $\widetilde{n}$ copies of the $\vec{\mathcal{F}}=\left(-\half,\cdots,-\half\right)$ flux domain wall to generate a $\vec{\mathcal{F}}=\left(-\frac{\widetilde{n}}{2},\cdots,-\frac{\widetilde{n}}{2}\right)$ flux domain wall in a similar manner. When trying to generate a flux torus, one needs to close the $x_4$ direction to a circle. We see that this can consistently be done provided that $\widetilde{n}$ is even due to Dirac's quantization condition \eqref{E:QuanCond}. Indeed, in this way the total flux for the $\mathfrak{u}(1)$ is $n=\frac{\widetilde{n}}{2}\in\mathbb{Z}$ in our normalization in which $q_{min}^{6d}=1$. We can also see that if we have an even number $\widetilde{n}$ of domain walls building the torus, then we can consistently identify the $\mathfrak{so}(16)$ roots of the theory on the very left with those of the theory on the very right
\be
\alpha^{(L,0)}_i=\alpha_i^{\mathfrak{so}(16)}=\alpha^{(R,2n)}_i\,,
\ee
as it can be seen explicitly in Figure \ref{F:basicHFDWestring} for $\widetilde{n}=2$. Moreover, the identification of the mass parameters 
\be
m_{\mathfrak{u}(1)}^{(0)}=m_{\mathfrak{u}(1)}^{(2n)}=m_{\mathfrak{u}(1)}^{(0)}+2n\,m_{KK}
\ee
implies that the charges of the KK symmetry should be $0$ modulo $2n$, that is it is broken to its discrete subgroup
\be\label{eq:breakKKk8}
\mathfrak{u}(1)_{KK}\,\,\,\to\,\,\, \mathbb{Z}_{2n}\,.
\ee

It is actually also possible to consider the situation in which the number of domain walls $\widetilde{n}$ is odd. As discussed in \cite{Kim:2017toz}, the resulting torus can be made consistent even if the flux for the $\mathfrak{u}(1)$ is half-integer by including a flux for the $\mathbb{Z}_2$ center of the $\mathfrak{e}_7$ symmetry preserved by the flux, which is enhanced from $\mathfrak{su}(8)$ at the SCFT point. This discrete flux nevertheless breaks the $\mathfrak{e}_7$ symmetry to a subgroup, such that under the branching rules the resulting representations won't be charged under the center $\Z_2$. The maximal subgroup of $\mathfrak{e}_7$ that can be preserved is $\mathfrak{f}_4$ (see Appendix C of \cite{Kim:2017toz}).\footnote{This can be seen by using the decomposition $\mathfrak{e}_8 \to \mathfrak{e}_7 \oplus \fsu(2) \to \mathfrak{f}_4 \oplus \fsu(2) \oplus \fsu(2) \to \mathfrak{f}_4 \oplus \fsu(2)_d \to \mathfrak{f}_4 \oplus \fu(1)$, where $\fsu(2)_d$ is the diagonal of both $\fsu(2)$.} The $\mathfrak{su}(8)$ symmetry of the 5d gauge theory is accordingly broken to $\fso(8)$. 
This can be directly seen in the configuration with stacked domain walls, since the roots of $\mathfrak{so}(16)$ of the theory on the left of the first domain wall are $\alpha_i^{L,0}=\alpha_i^{\mathfrak{so}(16)}$ while those on the right of the last domain wall are $\alpha_i^{R,\widetilde{n}}=-\alpha_i^{\mathfrak{so}(16)}$ for odd $\widetilde{n}$. Thus, in the gluing we should identify the $\textbf{8}$ of the residual $\mathfrak{su}(8)$ symmetry of one side of the domain wall with the $\overline{\textbf{8}}$ on the other side, breaking $\mathfrak{su}(8)\to \mathfrak{so}(8)$.\footnote{In \cite{Pasquetti:2019hxf} it was shown that this $\mathfrak{so}(8)$ symmetry can get enhanced to $\mathfrak{f}_4$ for the higher rank E-string.} Such a breaking instead doesn't occur for even $\widetilde{n}=2n$, as expected. The $\mathfrak{u}(1)_{KK}$ is still broken to $\mathbb{Z}_{\widetilde{n}}$, so for example in the case of the torus of flux $-\frac{1}{2}$ made of a single basic domain wall there is no residual symmetry.

\subsubsection{Flux $\vec{\mathcal{F}}=\left(-n,-n,-n,-n,-n,-n,0,0\right)$}

Now we move to a more involved example where we add the flux of the first six entries of the neighbouring $\vec{\mathcal{F}}=\left(-\half,\cdots,-\half\right)$ domain walls and subtract the last two entries to generate a $\vec{\mathcal{F}}=\left(-1,\cdots,-1,0,0\right)$ domain wall. This is achieved by using the Weyl element that flips the sign of the last two entries of the flux vector when performing the gluing.

In order to describe this gluing, we need to decompose $\mathfrak{su}(8)\to \mathfrak{su}(6) \oplus \mathfrak{su}(2)\oplus \mathfrak{u}(1)_b$ according to the flux. The $\mathfrak{u}(1)$ symmetry associated to the flux will be a combination of $\mathfrak{u}(1)_b$ and the $\mathfrak{u}(1)$ that had flux in the $\vec{\mathcal{F}}=\left(-\half,-\half,-\half,-\half,-\half,-\half,-\half,-\half\right)$ domain wall, which we will denote as $\mathfrak{u}(1)_a$. Accordingly the $8$ hypers transforming in the 5d theory as the $\textbf{16}$ of $\mathfrak{so}(16)$ will transform under $\mathfrak{su}(6) \oplus \mathfrak{su}(2)\oplus \mathfrak{u}(1)_a \oplus \mathfrak{u}(1)_b$ as
\be
\label{E:branching62}
\textbf{16} \to \textbf{8}^{+1} \oplus \overline{\textbf{8}}^{-1} \to (\textbf{6},\textbf{1})^{+1,+1} \oplus (\textbf{1},\textbf{2})^{+1,-3} \oplus (\overline{\textbf{6}},\textbf{1})^{-1,-1} \oplus (\textbf{1},\textbf{2})^{-1,+3}\,.
\ee 
We can see that the parametrization of the hypermultiplet masses is now
\be
\vec{m} = (m_{a}+m_{b}+m_{\mathfrak{su}(6)}^1,...,m_{a}+m_{b}+m_{\mathfrak{su}(6)}^6, m_{a}-3m_{b}+m_{\mathfrak{su}(2)},m_{a}-3m_{b}-m_{\mathfrak{su}(2)}) \,,\nonumber\\
\ee
with the constraint $\sum_{a=1}^6 m_{\mathfrak{su}(6)}^a = 0$. 

Before figuring out how the mass parameters of the two glued domain walls are identified, let us discuss the action on the roots. The first $5$ simple roots of $\mathfrak{so}(16)$ are the $\mathfrak{su}(6)$ roots which relate to the first six flux entries. Since we want to add the flux related to these first six entries, we should map these in a trivial way between the 5d theory on the right of the first domain wall and that on the left of the second domain wall. The roots $\alpha_7,\alpha_8$ of $\mathfrak{so}(16)$ are the $\mathfrak{so}(4)$ roots related to the last two flux entries. Since for these we want to deduct the flux, we should reflect these roots. Finally the root $\alpha_6$ allows us to combine the $\mathfrak{so}(4)$ and $\mathfrak{su}(6)$ symmetries to get an $\mathfrak{so}(16)$ box graph. Thus, there is only one sensible choice taking $\alpha_6^{(L,1)}= \alpha_6^{(R,0)} +\alpha_7^{(R,0)} +\alpha_8^{(R,0)}$. One can check that this choice indeed leads to a root of $\mathfrak{so}(16)$. Summarizing, the identifications of the roots is
\be\label{eq:rootidk8}
\alpha_{1\le i \le 5}^{(L,1)} = \alpha_{i}^{(R,0)}\,, \quad \alpha_6^{(L,1)} = \alpha_6^{(R,0)} + \alpha_7^{(R,0)} + \alpha_8^{(R,0)}\,, \quad \alpha_{7\le i \le 8}^{(L,1)} = -\alpha_{i}^{(R,0)}\,.
\ee 

\begin{figure}[t]
\center
\includegraphics[width=1\textwidth]{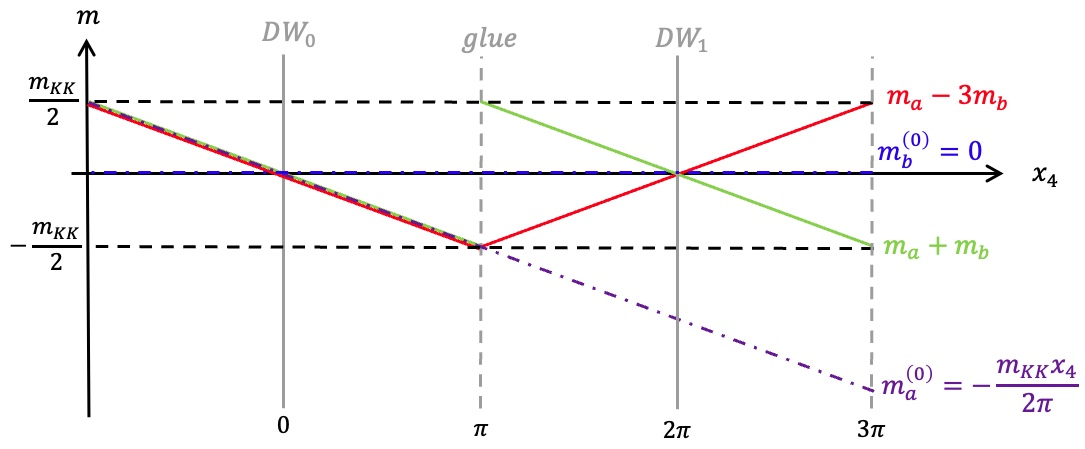} 
\caption{The mass variation of different mass parameters across two concatenated domain walls. The dashed purple line shows the mass parameter $m_a^{(0)}$ associated to $\mathfrak{u}(1)_a$ in the left domain wall and its extension to the right domain wall. For simplicity we chose a linear dependence of $m_a^{(0)}$ in $x_4$. The dashed blue line shows the mass parameter $m_b^{(0)}$ which is similarly extended and is chosen to be vanishing here for simplicity. In green we have the mass parameter $m_a+m_b$ for the first six hypermultiplets associated with the flux. This mass parameter is given as $m_a^{(0)}+m_b^{(0)}$ in the range $x_4\in (-\pi,\pi)$, and as $m_a^{(1)}+m_b^{(1)}$ in the range $x_4\in (\pi,3 \pi)$. One can see it passes through zero twice in the same direction, and in the gluing interface it jumps; this amounts to a $-1$ flux. We similarly show in red the mass parameter $m_a-3m_b$ of the other two hypers which is continuous and passes through zero twice but in opposite directions, which amounts to a $0$ flux as required.}
\label{F:MassVariation}
\end{figure}

This identification of the flavor symmetry translates into a non-trivial identifications of the $\mathfrak{u}(1)_a$ and $\mathfrak{u}(1)_b$ symmetries of the first and the second domain wall, which we shall denote by $\mathfrak{u}(1)_a^{(p)}$ and $\mathfrak{u}(1)_b^{(p)}$ for $p=0,1$. 
These symmetries also mix with $\mathfrak{u}(1)_{KK}$ in a way that shifts the first six mass parameters $m_a+m_b$, while leaving the other two mass parameters $m_a-3m_b$ unchanged. In addition, remember that the only non-vanishing mass parameter for the $\vec{\mathcal{F}}=\left(-\half,\cdots,-\half\right)$ flux domain wall was $m_{a}$, which flips its sign from positive to negative as we cross the wall. Thus, we demand $m_{b}^{(0)}=C$ with constant $C$, while $m_{a}^{(0)}$ varies from $\frac{m_{KK}}{2}-2C$ at $x_4=W^{-1}(-\pi)$ to $-\frac{m_{KK}}{2}$ at $x_4=W^{-1}(\pi)$. In general we can consider $m_{a}^{(0)}$ as a monotonically decreasing function of $x_4$ dropping by an amount $\frac{m_{KK}}{2}-C$ between $x_4=W^{-1}((n-1)\pi)$ and $x_4=W^{-1}(n\pi)$, see Figure \ref{F:MassVariation} for an example with $C=0$. Considering these changes of the mass parameters we would expect the following relations between the mass parameters:
\be
m_a^{(1)}+m_b^{(1)} = m_a^{(0)}-m_b^{(0)} +m_{KK}\,,\qquad m_a^{(1)}-3m_b^{(1)} = -(m_a^{(0)}+3m_b^{(0)}) -m_{KK}\,,
\ee
where at the transition point we flip the sign for the mass parameter $m_a+m_b$, from $-\frac{m_{KK}}{2}+C$ to $\frac{m_{KK}}{2}-C$, but keep it monotonically decreasing (the sign of the coefficient of $m_a^{(0)}$ is positive), while for the mass parameter $m_a-3m_b$ we do the opposite (the sign of the coefficient of $m_a^{(0)}$ is negative). This can be translated into the following identification of the mass parameters
\be\label{eq:massidk6}
m_a^{(1)}=\frac{1}{2}m_a^{(0)}-\frac{3}{2}m_b^{(0)}+\frac{1}{2}m_{KK}\,,\qquad m_b^{(1)}=\frac{1}{2}m_a^{(0)}+\frac{1}{2}m_b^{(0)}+\frac{1}{2}m_{KK}\,.
\ee

Now, we can understand the combination of $\mathfrak{u}(1)$'s that gets flux denoted $\mathfrak{u}(1)_\CF$ and the orthogonal combination which doesn't denoted $\mathfrak{u}(1)_\CO$. For $\mathfrak{u}(1)_\CF$ we demand that the charges of all the representations in the decomposition \eqref{E:branching62} on the left side of the first domain wall will be mapped to the same representations with an opposite charge on the right side of the second domain wall. Under $\mathfrak{u}(1)_\CO$ we demand the representations in the same respective locations will have the same charges. Before determining the above $\mathfrak{u}(1)$'s, first note that between the two sides of each domain wall the $\mathfrak{u}(1)_a$ charges are flipped. In addition, we need to use \eqref{eq:massidk6} to map the charges of the second domain wall to the notation of the first domain wall. All in all, the representations and charges under $\mathfrak{u}(1)_a^{(0)}$ and $\mathfrak{u}(1)_b^{(0)}$ of the right side of the second domain wall are
\be
(\textbf{6},\textbf{1})^{0,+2} \oplus (\textbf{1},\textbf{2})^{-2,0} \oplus (\overline{\textbf{6}},\textbf{1})^{0,-2} \oplus (\textbf{1},\textbf{2})^{+2,0}\,.
\ee
Considering the above constraints we find that
\be
\mathfrak{u}(1)_\CF = \frac1{4} \left(-3\mathfrak{u}(1)_a^{(0)}+\mathfrak{u}(1)_b^{(0)}\right)\,,\qquad \mathfrak{u}(1)_\CO = \frac1{4} \left(\mathfrak{u}(1)_a^{(0)}+\mathfrak{u}(1)_b^{(0)}\right)\,,
\ee
where the charges are in the convention of the left side of the first domain wall. In addition, the normalization was chosen to match the flux quantization. Under this choice of charges we have the following $\mathfrak{u}(1)_\CF$ and $\mathfrak{u}(1)_\CO$ charges on the two sides:
\be
(L,0) & : & (\textbf{6},\textbf{1})^{-\half,+\half} \oplus (\textbf{1},\textbf{2})^{-\frac{3}{2},-\half} \oplus (\overline{\textbf{6}},\textbf{1})^{+\half,-\half} \oplus (\textbf{1},\textbf{2})^{+\frac{3}{2},+\half}\,, \nonumber \\
(R,1) & : & (\textbf{6},\textbf{1})^{+\half,+\half} \oplus (\textbf{1},\textbf{2})^{+\frac{3}{2},-\half} \oplus (\overline{\textbf{6}},\textbf{1})^{-\half,-\half} \oplus (\textbf{1},\textbf{2})^{-\frac{3}{2},+\half}\,.
\ee

\begin{figure}[t]
\center
\begin{tikzpicture}[baseline,scale=0.85]
\tikzstyle{every node}=[font=\scriptsize]
\node[] (Lpar) at (-7.1,2.2) {\fontsize{10pt}{10pt}$m_a^{(0)} - 3m_b^{(0)}=0,$};
\node[] (Lpar) at (-7.1,1.5) {\fontsize{10pt}{10pt}$m_a^{(0)} + m_b^{(0)}> 0$};
\node[] (Lpar) at (-7.1,0.8) {\fontsize{10pt}{10pt}$\alpha_i^{(L,0)} =\alpha_i^{\mathfrak{so}(16)}$};

\node[] (5dLyt) at (-7.1,-2) {\ytableausetup{centertableaux,boxsize=6pt}
\begin{ytableau}
*(blue) & *(blue) & *(blue) & *(blue) & *(blue) & *(blue) & *(blue) & *(blue) \\
\none[] & \none[] & \none[] & \none[] & \none[] & \none[] & *(blue) & *(blue) & *(yellow) & *(yellow) & *(yellow) & *(yellow) & *(yellow) & *(yellow) \\ \none[] \\
*(blue) & *(blue) & *(blue) & *(blue) & *(blue) & *(blue) & *(yellow) & *(yellow) \\
\none[] & \none[] & \none[] & \none[] & \none[] & \none[] & *(yellow) & *(yellow) & *(yellow) & *(yellow) & *(yellow) & *(yellow) & *(yellow) & *(yellow) \\
\end{ytableau}};

\draw[black,line width=0.6mm] (-4.7,-3.1) -- (-4.7,2.7);
\node[above] at (-4.7,-4) {\fontsize{10pt}{10pt}$x_4=0$};

\node[] (Lpar) at (-2.4,2.2) {\fontsize{10pt}{10pt}$m_a^{(0)} - 3m_b^{(0)}<0,$};
\node[] (Lpar) at (-2.4,1.5) {\fontsize{10pt}{10pt}$m_a^{(0)} + m_b^{(0)}<0$};
\node[] (Lpar) at (-2.4,0.8) {\fontsize{10pt}{10pt}$\alpha_i^{(R,0)} =-\alpha_i^{\mathfrak{so}(16)}$};

\node[] (5dLyt) at (-2.4,-2) {\ytableausetup{centertableaux,boxsize=6pt}
\begin{ytableau}
*(blue) & *(blue) & *(blue) & *(blue) & *(blue) & *(blue) & *(blue) & *(blue) \\
\none[] & \none[] & \none[] & \none[] & \none[] & \none[] & *(yellow) & *(yellow) & *(yellow) & *(yellow) & *(yellow) & *(yellow) & *(yellow) & *(yellow) \\ \none[] \\
*(blue) & *(blue) & *(blue) & *(blue) & *(blue) & *(blue) & *(blue) & *(blue) \\
\none[] & \none[] & \none[] & \none[] & \none[] & \none[] & *(yellow) & *(yellow) & *(yellow) & *(yellow) & *(yellow) & *(yellow) & *(yellow) & *(yellow) \\
\end{ytableau}};

\draw[black,dashed,line width=0.4mm] (0,-3.1) -- (0,2.7);
\node[above] at (0,-4) {\fontsize{10pt}{10pt}$x_4=W^{-1}(\pi)$};

\draw[black,solid,->] (-9,-4) -- (9,-4);
\node[below right] at (9,-4) {\fontsize{10pt}{10pt}$x_4$};

\node[] (5dR) at (2.3,2.2) {\fontsize{10pt}{10pt}$m_a^{(1)} - 3m_b^{(1)}<0,$};
\node[] (Rpar) at (2.3,1.5) {\fontsize{10pt}{10pt}$m_a^{(1)} + m_b^{(1)}>0$};
\node[] (Rpar) at (2.3,0.8) {\fontsize{10pt}{10pt}$\alpha_{1\le i\le 5}^{(L,1)} =-\alpha_i^{\mathfrak{so}(16)}$};
\node[] (Lpar) at (2.3,0.1) {\fontsize{10pt}{10pt}$\alpha_6^{(L,1)} =-\sum_{j=6}^8\alpha_j^{\mathfrak{so}(16)}$};
\node[] (Lpar) at (2.3,-0.6) {\fontsize{10pt}{10pt}$\alpha_{7\le i\le 8}^{(L,1)} =\alpha_i^{\mathfrak{so}(16)}$};

\node[] (5dRyt) at (2.3,-2) {\ytableausetup{centertableaux,boxsize=6pt}
\begin{ytableau}
*(yellow) & *(yellow) & *(yellow) & *(yellow) & *(yellow) & *(yellow) & *(yellow) & *(yellow) \\
\none[] & \none[] & \none[] & \none[] & \none[] & \none[] & *(blue) & *(blue) & *(blue) & *(blue) & *(blue) & *(blue) & *(blue) & *(blue) \\ \none[] \\
*(yellow) & *(yellow) & *(yellow) & *(yellow) & *(yellow) & *(yellow) & *(yellow) & *(yellow) \\
\none[] & \none[] & \none[] & \none[] & \none[] & \none[] & *(blue) & *(blue) & *(blue) & *(blue) & *(blue) & *(blue) & *(blue) & *(blue) \\
\end{ytableau}};

\draw[black,line width=0.6mm] (4.75,-3.1) -- (4.75,2.7);
\node[above] at (4.75,-4) {\fontsize{10pt}{10pt}$x_4=W^{-1}(2\pi)$};

\node[] (5dR) at (7,2.2) {\fontsize{10pt}{10pt}$m_a^{(1)} - 3m_b^{(1)}=0,$};
\node[] (Rpar) at (7,1.5) {\fontsize{10pt}{10pt}$m_a^{(1)} + m_b^{(1)}<0$};
\node[] (Rpar) at (7,0.8) {\fontsize{10pt}{10pt}$\alpha_{1\le i\le 5}^{(R,1)} =\alpha_i^{\mathfrak{so}(16)}$};
\node[] (Lpar) at (7,0.1) {\fontsize{10pt}{10pt}$\alpha_6^{(R,1)} =\sum_{j=6}^8\alpha_j^{\mathfrak{so}(16)}$};
\node[] (Lpar) at (7,-0.6) {\fontsize{10pt}{10pt}$\alpha_{7\le i\le 8}^{(R,1)} =-\alpha_i^{\mathfrak{so}(16)}$};

\node[] (5dRyt) at (7,-2) {\ytableausetup{centertableaux,boxsize=6pt}
\begin{ytableau}
*(yellow) & *(yellow) & *(yellow) & *(yellow) & *(yellow) & *(yellow) & *(blue) & *(blue) \\
\none[] & \none[] & \none[] & \none[] & \none[] & \none[] & *(blue) & *(blue) & *(blue) & *(blue) & *(blue) & *(blue) & *(blue) & *(blue) \\ \none[] \\
*(yellow) & *(yellow) & *(yellow) & *(yellow) & *(yellow) & *(yellow) & *(yellow) & *(yellow) \\
\none[] & \none[] & \none[] & \none[] & \none[] & \none[] & *(yellow) & *(yellow) & *(blue) & *(blue) & *(blue) & *(blue) & *(blue) & *(blue) \\
\end{ytableau}};

\end{tikzpicture}
\caption{Structure of the $\mathfrak{su}(2)+8F$ ECB phases associated with the domain wall of flux $\vec{\mathcal{F}}=\left(-1,\cdots,-1,0,0\right)$. For brevity sake we didn't include that the vector of scalars as well as all the other mass parameters are set to zero throughout this diagram. Note that $m_a^{(1)} = \frac{1}{2}m_a^{(0)}-\frac{3}{2}m_b^{(0)}+\frac{1}{2}m_{KK}$ and $m_b^{(1)} = \frac{1}{2}m_a^{(0)}+\frac{1}{2}m_b^{(0)}+\frac{1}{2}m_{KK}$. For simplicity we assume a step function profile for $m_a^{(0)}$ dropping from $\frac{3m_{KK}}{10}$ to $-\frac{m_{KK}}{2}$ at $x_4=0$ and again from $-\frac{m_{KK}}{2}$ to $-\frac{13m_{KK}}{10}$ at $x_4=W^{-1}(2\pi)$, while keeping constant $m_b^{(0)}=\frac{m_{KK}}{10}$.}
\label{F:involvedDWestring}
\end{figure}
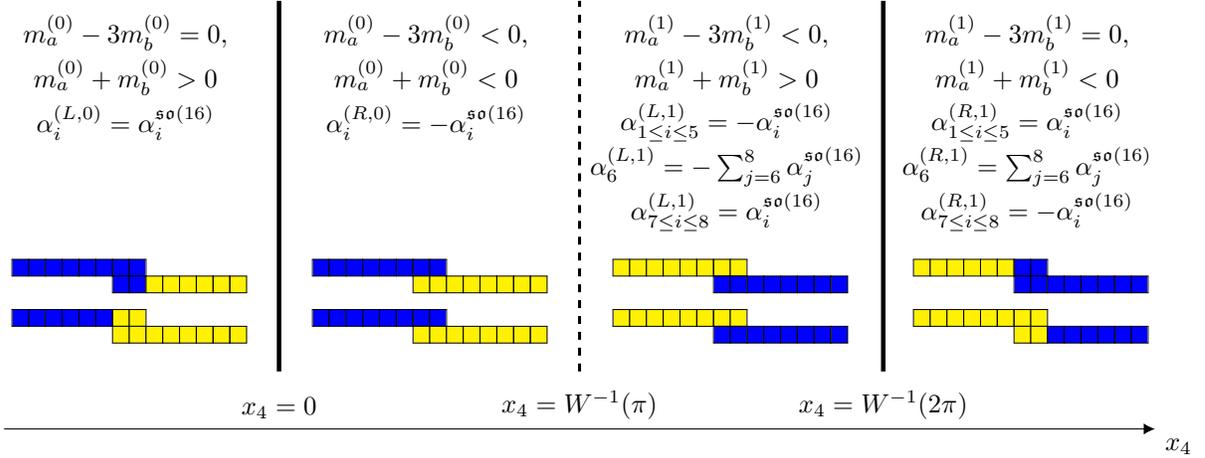

The last comment is that if we take the ECB phases of the first domain wall to be as in Figure \ref{basicFDWestring}, then after the identifications \eqref{eq:rootidk8}-\eqref{eq:massidk6} we would get non-sensible ECB phases for the second domain wall. In order to get a consistent picture, we need to start from the ECB phase depicted on the left of Figure \ref{F:GenBoxGraph} with $k=6$. Then the previous identifications are as summarized in Figure \ref{F:involvedDWestring}, so that the final ECB phase on the very right coincides with the one we drew on the right of Figure \ref{F:GenBoxGraph}.

With the $\vec{\mathcal{F}}=\left(-1,\cdots,-1,0,0\right)$ flux domain wall at hand we can now construct $\vec{\mathcal{F}}=\left(-n,\cdots,-n,0,0\right)$ flux domain walls and tori by gluing $n$ copies of it, or equivalently $2n$ copies of the most basic domain wall with half-integer flux. The gluing of two $\vec{\mathcal{F}}=\left(-1,\cdots,-1,0,0\right)$ domain walls is done by identifying the roots of the right side of the left domain wall with the roots of the left side of the right domain wall with the trivial Weyl element, as we did in the previous example of Subsubsection \ref{SS:5dPhigluing}. Recall that in this case there is no non-trivial identification of the flavor symmetries and the masses of all the hypers are shifted by $m_{KK}$ according to \eqref{eq:massmixKK}. Note that in the general case of non-vanishing $m_b^{(0)}=C$ we need to identify in addition $m_b^{(p+1)}=-m_b^{(p)}$ for odd $p$, since as in the previous example one needs the sign of all the masses to flip sign at the gluing interface. As in the former example, generating a flux torus with odd $n$ will require a flux for a $\mathbb{Z}_2$ subgroup of the center and this will break the flavor symmetry, while an even $n$  will correspond only to flux for the $\mathfrak{u}(1)$ symmetry. 

We can determine how the KK symmetry is broken when considering a torus as follows. Let us focus on the case of $n=2$ for concreteness. The configuration then involves $2n=4$ copies of the basic domain wall with flux $(-\frac{1}{2},\cdots,-\half)$. We shall denote by $\mathfrak{u}(1)_a^{(p)}$ and $\mathfrak{u}(1)_b^{(p)}$ for $p=0,1,2,3$ the abelian symmetries around each of the domain walls. From our previous discussion, we see that these are identified according to \eqref{eq:massidk6} for the $0/1$ and the $2/3$ domain walls, that is
\be
m_a^{(p+1)}=\frac{1}{2}m_a^{(p)}-\frac{3}{2}m_b^{(p)}+\frac{1}{2}m_{KK}\,,\qquad m_b^{(p+1)}=\frac{1}{2}m_a^{(p)}+\frac{1}{2}m_b^{(p)}+\frac{1}{2}m_{KK}
\ee
for $p=0,2$, while they are identified according to \eqref{eq:massmixKK} for the $1/2$ and the $4/1$ domain walls, that is
\be
m_a^{(p+1)}=m_a^{(p)}+m_{KK}\,,\qquad m_b^{(p+1)}=-m_b^{(p)}
\ee
for $p=1,3$, where the inversion of the $\mathfrak{u}(1)_b$ mass parameter is needed since this gluing increases the flux under $\mathfrak{u}(1)_a$ and not under $\mathfrak{u}(1)_b$.
Consistency of the torus then requires that the charges of the 5d hypermultiplets go back to their original value, which at the level of the mass parameters means
\be
&&m_a^{(0)}+m_b^{(0)}=m_a^{(3)}+m_b^{(3)}=m_a^{(0)}+m_b^{(0)}+2m_{KK}\,,\nn\\
&&m_a^{(0)}-3m_b^{(0)}=m_a^{(3)}-3m_b^{(3)}=m_a^{(0)}-3m_b^{(0)}+6m_{KK}\,,
\ee
The stronger condition comes from the first line and it translates to the following condition on charges:
\be
\mathfrak{u}(1)_a^{(0)}+\mathfrak{u}(1)_b^{(0)} = \mathfrak{u}(1)_a^{(0)}+\mathfrak{u}(1)_b^{(0)} +2\mathfrak{u}(1)_{KK}\,,
\ee
which tells us that the charges under the KK symmetry should be 0 modulo 2, hence the $\fu(1)_{KK}$ symmetry is broken to a discrete subgroup
\be
\mathfrak{u}(1)_{KK}\,\,\,\to\,\,\, \mathbb{Z}_{2}\,.
\ee
For generic flux $(-2n,-2n,-2n,-2n,-2n,-2n,0,0)$ the symmetry is instead broken to $\mathbb{Z}_{2n}$.

\subsubsection{Flux $\vec{\mathcal{F}}=\left(-n,...,-n,0,...,0\right)$}

In this final example we will give a general prescription generalizing the former examples to give a $\vec{\mathcal{F}}=\left(-n,...,-n,0,...,0\right)$ flux domain wall and torus with $k$ entries equal to $-n$ and $8-k$ zeroes. The prescription will be dependent on the parity of $k$ due to the fact that the $\mathfrak{e}_8$ roots of the form $(\pm \half,\cdots,\pm \half)$ corresponding to the basic flux domain walls come only with even number of minus signs.

We start with the case of even $k$ for which we have already covered the cases of $k=6,8$ in the former examples. The prescription here relies as before on gluing two $\vec{\mathcal{F}}=\left(-\half,\cdots,-\half\right)$ flux domain walls in a way that adds the flux of the first $k$ entries and deducts the flux of the last $8-k$ entries. This is achieved by identifying the domain walls up to the Weyl element that flips the sign of the last $8-k$ entries of the flux vector.
The mapping of the roots of $\mathfrak{so}(16)$ between the right side of the left domain wall and the left side of the right domain wall is given in this general case by
\be
\alpha_{1\le i \le k-1}^{(L,1)} = \alpha_{i}^{(R,0)}\,, \quad \alpha_k^{(L,1)} = \alpha_k^{(R,0)} +2\sum_{i=k+1}^6 \alpha_i^{(R,0)} + \alpha_7^{(R,0)} + \alpha_8^{(R,0)}\,, \quad \alpha_{k+1\le i \le 8}^{(L,1)} = -\alpha_{i}^{(R,0)}\,.\nonumber\\
\ee 

As we did before, we first look at the branching rules of the $\textbf{16}$ under the decomposition $\mathfrak{so}(16)\to \mathfrak{su}(8)\oplus \mathfrak{u}(1)_a\to \mathfrak{su}(k) \oplus \mathfrak{su}(8-k) \oplus \mathfrak{u}(1)_a \oplus \mathfrak{u}(1)_b$
\be\label{eq:decompgenk}
\textbf{16} \to \textbf{8}^{+1} \oplus \overline{\textbf{8}}^{-1} \to (\textbf{k},\textbf{1})^{+1,+(4-\frac{k}{2})} \oplus (\textbf{1},\textbf{8-k})^{+1,-\frac{k}{2}} \oplus (\overline{\textbf{k}},\textbf{1})^{-1,-(4-\frac{k}{2})} \oplus (\textbf{1},\overline{\textbf{8-k}})^{-1,+\frac{k}{2}}\,.\nn\\
\ee
From this decomposition we find that the hypermultiplet masses are given by
\be
\vec{m} & = &\Bigl(m_a + \bigl(4-\frac{k}{2}\bigr)m_b+m_{\mathfrak{su}(k)}^1,...,m_a + \bigl(4-\frac{k}{2}\bigr)m_b+m_{\mathfrak{su}(k)}^k,\nonumber\\
& & \quad m_a -\frac{k}{2}m_b+m_{\mathfrak{su}(8-k)}^1,...,m_a -\frac{k}{2}m_b+m_{\mathfrak{su}(8-k)}^{8-k}\Bigr)\,.
\ee
with the constraint $\sum_{a=1}^n m_{\mathfrak{su}(n)}^a = 0$.

In a similar way to the former example we can figure out what are the mapping of the mass parameters between the first and second domain wall from the hypermultiplet masses
\be\label{eq:massidgenk}
&&m_a^{(1)}=\left(\frac{k}{4}-1\right)m_a^{(0)}+k\left(\frac{k}{8}-1\right)m_b^{(0)}+\left(\frac{k}{4}-1\right)m_{KK}\,,\nn\\
&&m_b^{(1)}=\frac{1}{2}m_a^{(0)}+\left(\frac{k}{4}-1\right)m_b^{(0)}+\frac{1}{2}m_{KK}\,.
\ee
Note that in the case of $k=8$ the second line of the mapping is irrelevant since there is no $m_b$ parameter, and the identification of the $m_a$ parameter is given only by the first line.
This again guarantees that as we move in the $x_4$ directions the masses of the first $k$ hypers change sign from positive to negative value twice thanks to the jump by $m_{KK}$ at the junction of the two walls, while those of the other $(8-k)$ hypers change from positive to negative and eventually back to the original positive value. Again similarly to the former example these mass mappings allow us to find the $\mathfrak{u}(1)$ we give flux to and the orthogonal combination
\be
\mathfrak{u}(1)_\CF = \frac1{4} \left(-\frac{k}{2}\mathfrak{u}(1)_a^{(0)}+\mathfrak{u}(1)_b^{(0)}\right)\,,\qquad \mathfrak{u}(1)_\CO = \frac1{4} \left(\frac{8-k}{2}\mathfrak{u}(1)_a^{(0)}+\mathfrak{u}(1)_b^{(0)}\right)\,,
\ee
where this is again only relevant for $k\ne8$.

One can use the above prescription to generate a $\vec{\mathcal{F}}=\left(-1,...,-1,0,...,0\right)$ flux domain wall with an even number $k$ of $-1$ entries. Moreover, with this domain wall at hand we can concatenate $n$ copies of it with the trivial Weyl element. As we explained in Example \ref{SS:5dPhigluing}, this further gluing will trivially identify the abelian symmetries and shift the masses of all the hypers by $m_{KK}$. This will generate a $\vec{\mathcal{F}}=\left(-n,...,-n,0,...,0\right)$ flux domain wall or torus. Generating a flux torus with odd $n$  will correspond to an additional flux for a $\mathbb{Z}_2$  subgroup of the center of the flavor symmetry that breaks it to the invariant subgroup, while an even $n$ will lead to flux only to the chosen $\mathfrak{u}(1)$. 

Similarly to what we did for $k=6,8$, we can determine how the KK symmetry is broken when considering a torus. Again let us focus on the case of $n=2$ for simplicity. The configuration then involves $2n=4$ copies of the basic domain wall with flux $-\frac{1}{2}$. The identification of the $\mathfrak{u}(1)_a^{(p)}$ and $\mathfrak{u}(1)_b^{(p)}$ for $p=0,1,2,3$ between each domain wall is
\be
&&m_a^{(p+1)}=\left(\frac{k}{4}-1\right)m_a^{(p)}+k\left(\frac{k}{8}-1\right)m_b^{(p)}+\left(\frac{k}{4}-1\right)m_{KK}\,,\nn\\
&&m_b^{(p+1)}=\frac{1}{2}m_a^{(p)}+\left(\frac{k}{4}-1\right)m_b^{(p)}+\frac{1}{2}m_{KK}
\ee
for $p=0,2$, while it is
\be
m_a^{(p+1)}=m_a^{(p)}+m_{KK}\,,\qquad m_b^{(p+1)}=-m_b^{(p)}
\ee
for $p=1,3$. Consistency of the torus then requires that the charges of the 5d hypermultiplets go back to their original value
\be\label{eq:breakKKgenk}
&&m_a^{(0)}+\left(4-\frac{k}{4}\right)m_b^{(0)}=m_a^{(3)}+\left(4-\frac{k}{4}\right)m_b^{(3)}=m_a^{(0)}+\left(4-\frac{k}{4}\right)m_b^{(0)}+(k-4)m_{KK}\,,\nn\\
&&m_a^{(0)}-\frac{k}{2}m_b^{(0)}=m_a^{(3)}-\frac{k}{2}m_b^{(3)}=m_a^{(0)}-\frac{k}{2}m_b^{(0)}+km_{KK}\,,
\ee
meaning that $\mathfrak{u}(1)_{KK}$ is broken to the smallest group between $\mathbb{Z}_{|k-4|}$ and $\mathbb{Z}_k$.
For generic $(-2n,\dots,-2n,0,\dots,0)$ flux the symmetry is instead broken to the smallest group between $\mathbb{Z}_{n|k-4|}$ and $\mathbb{Z}_{nk}$.\footnote{For $k=4$ the residual group is $\mathbb{Z}_{4n}$, since the constraint from the first line in \eqref{eq:breakKKgenk} is trivial.}

The case of odd $k$ is more involved, as the former prescription will not work for it. The reason for this is that the former prescription of gluing two $\vec{\mathcal{F}}=\left(-\half,\cdots,-\half\right)$ domain walls actually shifts the flux of the right domain wall such that it corresponds to a $\vec{\mathcal{F}}=(-\half,...,-\half,\half,...,\half)$ flux domain wall, with $k$ entries equal to $-\half$. This domain wall can only be generated for even $k$, as it needs to correspond to a root of $\mathfrak{e}_8$.\footnote{Remember the $\mathfrak{e}_8$ roots corresponding to $\vec{\mathcal{F}}=\left(\pm\half,\pm\half,\pm\half,\pm\half,\pm\half,\pm\half,\pm\half,\pm\half\right)$ only come with even number of minus signs.} Thus, the minimal domain wall flux we can generate with odd $k$ will be for $n=2$, and its generation will require at least four $\vec{\mathcal{F}}=\left(-\half,\cdots,-\half\right)$ domain walls. For example for $k=1,3,5$ we can generate such a domain wall by the following concatenation of domain walls
\be
(\underbrace{-2,...,-2}_{k},\underbrace{0,...,0}_{8-k}) & = & \Bigl(\underbrace{-\half,...,-\half}_{k+1},\underbrace{\half,...,\half}_{7-k}\Bigr) + \Bigl(\underbrace{-\half,...,-\half}_{k},\half,-\half,\underbrace{\half,...,\half}_{6-k}\Bigr)\nonumber\\
 & &  + \Bigl(\underbrace{-\half,...,-\half}_{k},\half,\half,\underbrace{-\half,...,-\half}_{6-k}\Bigr) +\Bigl(-\half,...,-\half\Bigr)\,,
\ee
where the relative signs between the flux domain walls imply the required gluing. We will not explicitly show the building of such domain walls as they are quiet involved and just require several gluings as the ones we have shown before.

\section{The 4d domain wall theory}
\label{sec:4dDWtheory}

In this section we focus on the 4d theory living on a \emph{fundamental} flux domain wall. We define a \emph{fundamental} flux domain wall as a domain wall whose flux is associated with a single root of the 6d flavor symmetry, rather than a combination of roots. We will in particular provide some reasoning for determining such theory from the 5d field theory point of view. Since at the domain wall location the gauge coupling of a general KK theory goes to infinity, the theory on the domain wall will be determined by strong coupling dynamics. This means that in the general case field theory arguments alone can provide very little insights, and one needs stronger tools such as geometric engineering from M-theory. Nevertheless, one can still extract some information and clues about the 4d domain wall theory, which in many cases allows us to guess a Lagrangian for it.

\subsection{Determining the domain wall Lagrangian}

We start by considering the half-BPS boundary conditions one can give at the position of the 4d domain wall. A 5d vector multiplet gives a 4d $\CN=1$ vector multiplet and a 4d $\CN=1$ adjoint chiral multiplet. Consider choosing Neumann boundary conditions for the 4d vector multiplet, while the 4d adjoint chiral multiplet gets Dirichlet boundary conditions. This choice is similar to the boundary conditions an NS5-brane enforces on the $\mathfrak{u}(N)$ $\CN=2$ vector multiplet coming from a stack of $N$ D4-branes. In this more supersymmetric analogue, the 5d $\CN=2$ vector multiplet gives a 4d $\CN=2$ vector multiplet and a 4d $\CN=2$ adjoint hypermultiplet, which get Neumann and Dirichlet boundary conditions, respectively. The resulting theory from the 5d perspective has two $\mathfrak{u}(N)$ gauge nodes coming from the two half-infinite stacks of $N$ D4-branes, and a bifundamental 4d hypermultiplet coming from $\mathfrak{u}(N)$ adjoint strings split between the two gauge copies by the NS5-brane. Considering lowering the amount of supersymmetry, one can think of F-theory on Calabi-Yau four manifolds, where a complex codimension one locus in the base with $I_N$ fiber on top of it gives an $\mathfrak{su}(N)$ symmetry, which is a gauge symmetry if the locus surface is compact. When one collides two such codimension one loci in the base we find bifundamental matter of the two gauge symmetries. Going back to our case, we can expect a 5d unitary gauge symmetry to split to two copies, one for each half-infinite side, in accordance with our discussion from the previous sections. In addition, each 4d adjoint chiral coming from each 5d unitary gauge node vector multiplet can be conjectured to give a bifundamental chiral operator of the two copies. Specifically in the case of an $\mathfrak{su}(N)$ gauge node one needs to remove the singlet generated from $\textbf{F} \otimes \overline{\textbf{F}}= \textbf{Adj} \oplus \textbf{1}$. This can be done by the superpotential term $\delta W= \CF \CO^N$ where $\CF$ is the flipping field for the $\CO^N$ operator giving a singlet under the two gauge nodes. This matches all the examples given in \cite{Kim:2017toz,Kim:2018lfo} of rank 1 E-string and general $(G_{ADE},G_{ADE})$ conformal matter flux domain walls. In these examples the operator is a simple chiral field, but in general this doesn't have to be the case even for unitary gauge nodes.\footnote{For example, this kind of a naive guess doesn't work for the $\mathfrak{u}(1)_t$ symmetry of the general 6d $(A,A)$ conformal matter SCFT, which is the symmetry under which each pair of bifunadmental half-hypers are oppositely charged in the tensor branch quiver description of the theory. Specifically, this naive guess only works for the next-to-minimal conformal matter SCFT related to two M5-branes probing a $\Z_k$ singularity. In this family one can also consider the case of $k=1$ where we find by compactification to 4d models of class $\CS$ where again such a flux domain wall and subsequently $\CN=1$ class $\CS$ flux tube is only possible for the theory described by $N=2$ M5-branes \cite{Benini:2009mz,Bah:2012dg,Bah:2011je,Razamat:2019sea,Hwang:2021xyw}.}

For non unitary 5d gauge symmetries one can still expect two copies of the gauge symmetry for each half-infinite side, or one can in some cases couple using a duality domain wall two UV dual 5d theories with different gauge symmetry as in \cite{Kim:2018bpg}. In such cases the matter connecting the two gauge nodes is unknown in general and requires a string/M-theoretic understanding of this construction as we have in the higher supersymmetry cases. From the few known examples \cite{Pasquetti:2019hxf,Kim:2018bpg} of higher rank E-string and minimal $(D,D)$ conformal matter flux domain walls it seems one still gets a bifundamental operator between the gauge nodes on the two sides.

Next, we consider the half-BPS boundary conditions we can impose on the 5d hypermultiplets. Each 5d $\CN=1$ hypermultiplet reduced to 4d gives two 4d $\CN=1$ chiral multiplets in complex conjugate representations. We have two options for half-BPS boundary conditions in which we need to give a Neumann boundary condition to one of the two chirals while giving a Dirichlet boundary condition to the other. Thus, on the 4d domain wall only the chiral with the Neumann boundary condition will remain. Note that for each hypermultiplet of the 5d gauge theory we need to make two such choices, one for the left 5d theory hyper and one for the right 5d theory hyper.

Thus far we went over the gauge and the matter content, but for 4d $\CN=1$ supersymmetry one also needs to specify the superpotential among the chiral fields. We already mentioned the superpotential term in which we flip some of the operators of the theory when we discussed the vector multiplet boundary conditions. As noted before reducing the 5d theory to 4d, the 5d vector multiplet contributes a 4d adjoint chiral multiplet $\Phi$ while the 5d hypermultiplet contributes two 4d chiral multiplets $Q$ and $\widetilde{Q}$. If the 5d hypermultiplet transforms under a gauge symmetry associated to the 5d vector multiplet, their 4d matter contributions will be related by a superpotential term 
\be
\delta W = Q \Phi \widetilde{Q}\,,
\ee 
with the implied gauge indices contractions. From the above discussion we expect to have on the 4d domain wall an operator $\CO$ transforming in the bifundamental representation of two copies of each 5d gauge node with a non-trivial boundary condition coming from its associated 5d vector multiplet and the adjoint 4d chiral it contributes. In addition, we expect from each hypermultiplet with half BPS boundary conditions on the domain wall to get two chiral multiplets $q,\widetilde{q}$ one from each side of the domain wall. Thus we expect to have on the domain wall a superpotential term relating these chiral multiplets as 
\be
\label{E:DWSP}
\delta W = q \CO \widetilde{q}\,.
\ee
Note that the chirals of the trivial reduction of the 5d hypermultiplet to 4d have opposite charges and conjugate representations under the 5d flavor symmetry, but on the two sides of the domain wall we flip the sign of the $\mathfrak{u}(1)$ that gets flux; thus, the aforementioned two chirals will in fact have the same charge under this $\mathfrak{u}(1)$. Considering the superpotential relating the charges of these two chirals to the bifundamental chiral going between the two sides, one finds this bifundamental chiral is charged only under the $\mathfrak{u}(1)$ which we give flux to out of the full 6d flavor symmetry. This in turn fixes the charge of the flipping field mentioned before. In addition note that the superpotential implies that for a fundamental flux domain wall one can't freely choose the half-BPS boundary conditions for the hypermultiplets on the two sides and we need to choose the opposite half BPS boundary conditions on the two sides to get chirals related by the superpotential \eqref{E:DWSP}. One additional comment is that in the case where the hypermultiplet is not charged under the $\mathfrak{u}(1)$ we give flux to, the two chirals coming from the two sides will remain with opposite charges and the bifundamental chiral going between the two sides will have no charges. This will identify the two gauge nodes on the two sides leaving a single one and will leave the superpotential with a mass term for the two chirals coming from the two sides effectively removing them in the IR.

Before giving some examples let us discuss the boundary conditions required on the two sides of the domain wall in order to cut the infinite tube to a finite tube. We start from the 5d $\CN=1$ vector multiplet giving in 4d an $\CN=1$ vector and adjoint chiral multiplets. On both edges of the tube we choose Dirichlet boundary conditions for the 4d $\CN=1$ vector multiplet and Neumann boundary condition for the adjoint chiral multiplet. This effectively freezes the gauge symmetry to become non-dynamical and gives a 4d global symmetry on each side of the tube associated to each of the punctures of the tube, while the adjoint chiral doesn't survive due to the Dirichlet boundary conditions on the domain wall. The 5d $\CN=1$ hypermultiplet gives in 4d two $\CN=1$ chiral multiplets in conjugate representations, for which we can choose Neumann boundary conditions for one of the chirals and Dirichlet for the other. If the hyper boundary conditions on the edge of the tube match the boundary conditions on the domain wall in the same side, the chiral with Neumann boundary conditions will survive, while if the boundary conditions are opposite both chirals won't survive.

\subsection{Example: rank 1 E-string fundamental domain wall 4d theories}
\label{subsec:4dDWEstring}

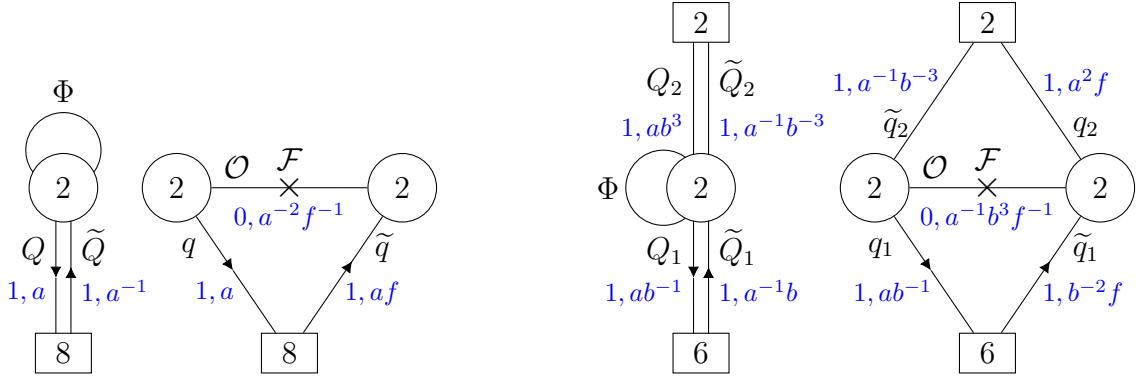
\begin{figure}[t]
    \centering
    \begin{subfigure}[b]{0.1\textwidth}
       \centering
       \begin{tikzpicture}[baseline,scale=1]
\tikzstyle{every node}=[font=\scriptsize]
	   \draw (0,0.5) circle (0.5cm);
       \node[draw, circle,fill=white] (p1) at (0,0) {\fontsize{12pt}{12pt}\selectfont $\,\,2\,\,$};
       \node[draw, rectangle] (p2) at (0,-2.2) {\fontsize{12pt}{12pt}\selectfont $\,\,8\,\,$};
       \draw[->] (-0.1,-0.44) to      (-0.1,-1.2);
       \draw[-] (-0.1,-1.2) to      (-0.1,-1.94);
       \draw[-<] (0.1,-0.44) to      (0.1,-1.2);
       \draw[-] (0.1,-1.94) to      (0.1,-1.2);
       \node[right] at (0.1,-0.8)  {\fontsize{12pt}{12pt}\selectfont $\widetilde{Q}$};
       \node[left] at (-0.1,-0.85)  {\fontsize{12pt}{12pt}\selectfont $Q$};
       \node[above] at (0,1)  {\fontsize{12pt}{12pt}\selectfont $\Phi$};
       \node[blue,right] at (0.1,-1.4)  {\fontsize{10pt}{10pt}\selectfont $1,a^{-1}$};
       \node[blue,left] at (-0.1,-1.4)  {\fontsize{10pt}{10pt}\selectfont $1,a$};
       \end{tikzpicture}
    \end{subfigure}%
    \begin{subfigure}[b]{0.3\textwidth}
       \centering
       \begin{tikzpicture}[baseline,scale=1]
\tikzstyle{every node}=[font=\scriptsize]
       \node[draw, circle] (p1) at (0,0) {\fontsize{12pt}{12pt}\selectfont $\,\,2\,\,$};
       \node[draw, circle] (p2) at (3,0) {\fontsize{12pt}{12pt}\selectfont $\,\,2\,\,$};
       \node[draw, rectangle] (p3) at (1.5,-2.2) {\fontsize{12pt}{12pt}\selectfont $\,\,8\,\,$};
       \draw[-] (p1) to  node[rotate=0]{\Large $\times$}    (p2);
       \draw[->] (p1) to      (0.75,-1.1);
       \draw[-] (0.75,-1.1) to      (p3);
       \draw[-<] (p2) to      (2.25,-1.1);
       \draw[-] (p3) to      (2.25,-1.1);
       \node[left] at (0.4,-0.8)  {\fontsize{12pt}{12pt}\selectfont $q$};
       \node[right] at (2.5,-0.8)  {\fontsize{12pt}{12pt}\selectfont $\widetilde{q}$};
       \node[above] at (0.8,0)  {\fontsize{12pt}{12pt}\selectfont $\CO$};
       \node[above] at (1.5,0.1)  {\fontsize{12pt}{12pt}\selectfont $\CF$};
       \node[blue,left] at (0.9,-1.4)  {\fontsize{10pt}{10pt}\selectfont $1,a$};
       \node[blue,right] at (2.1,-1.4)  {\fontsize{10pt}{10pt}\selectfont $1,af$};
       \node[blue,above] at (1.5,-0.7)  {\fontsize{10pt}{10pt}\selectfont $0,a^{-2}f^{-1}$};
       \end{tikzpicture}
    \end{subfigure}%
    \hspace{1.5cm}
    \begin{subfigure}[b]{0.2\textwidth}
       \centering
       \begin{tikzpicture}[baseline,scale=1]
\tikzstyle{every node}=[font=\scriptsize]
	   \draw (-0.5,0) circle (0.5cm);
       \node[draw, circle,fill=white] (p1) at (0,0) {\fontsize{12pt}{12pt}\selectfont $\,\,2\,\,$};
       \node[draw, rectangle] (p2) at (0,-2.2) {\fontsize{12pt}{12pt}\selectfont $\,\,6\,\,$};
       \node[draw, rectangle] (p3) at (0,2.2) {\fontsize{12pt}{12pt}\selectfont $\,\,2\,\,$};
       \draw[->] (-0.1,-0.44) to      (-0.1,-1.2);
       \draw[-] (-0.1,-1.2) to      (-0.1,-1.94);
       \draw[-<] (0.1,-0.44) to      (0.1,-1.2);
       \draw[-] (0.1,-1.94) to      (0.1,-1.2);
       \draw[-] (-0.1,0.44) to      (-0.1,1.94);
       \draw[-] (0.1,0.44) to      (0.1,1.94);
       \node[right] at (0.1,-0.8)  {\fontsize{12pt}{12pt}\selectfont $\widetilde{Q}_1$};
       \node[left] at (-0.1,-0.85)  {\fontsize{12pt}{12pt}\selectfont $Q_1$};
       \node[right] at (0.1,1.45)  {\fontsize{12pt}{12pt}\selectfont $\widetilde{Q}_2$};
       \node[left] at (-0.1,1.4)  {\fontsize{12pt}{12pt}\selectfont $Q_2$};
       \node[left] at (-0.95,0)  {\fontsize{12pt}{12pt}\selectfont $\Phi$};
       \node[blue,right] at (0.1,-1.4)  {\fontsize{10pt}{10pt}\selectfont $1,a^{-1}b$};
       \node[blue,left] at (-0.1,-1.4)  {\fontsize{10pt}{10pt}\selectfont $1,ab^{-1}$};
       \node[blue,right] at (0.1,0.8)  {\fontsize{10pt}{10pt}\selectfont $1,a^{-1}b^{-3}$};
       \node[blue,left] at (-0.1,0.8)  {\fontsize{10pt}{10pt}\selectfont $1,ab^{3}$};
       \end{tikzpicture}
    \end{subfigure}%
    \begin{subfigure}[b]{0.27\textwidth}
       \centering
       \begin{tikzpicture}[baseline,scale=1]
\tikzstyle{every node}=[font=\scriptsize]
       \node[draw, circle] (p1) at (0,0) {\fontsize{12pt}{12pt}\selectfont $\,\,2\,\,$};
       \node[draw, circle] (p2) at (3,0) {\fontsize{12pt}{12pt}\selectfont $\,\,2\,\,$};
       \node[draw, rectangle] (p3) at (1.5,-2.2) {\fontsize{12pt}{12pt}\selectfont $\,\,6\,\,$};
       \node[draw, rectangle] (p4) at (1.5,2.2) {\fontsize{12pt}{12pt}\selectfont $\,\,2\,\,$};
       \draw[-] (p1) to  node[rotate=0]{\Large $\times$}    (p2);
       \draw[->] (p1) to      (0.75,-1.1);
       \draw[-] (0.75,-1.1) to      (p3);
       \draw[-<] (p2) to      (2.25,-1.1);
       \draw[-] (p3) to      (2.25,-1.1);
       \draw[-] (p1) to      (p4);
       \draw[-] (p2) to      (p4);
       \node[left] at (0.4,-0.8)  {\fontsize{12pt}{12pt}\selectfont $q_1$};
       \node[right] at (2.5,-0.8)  {\fontsize{12pt}{12pt}\selectfont $\widetilde{q}_1$};
       \node[left] at (0.6,0.85)  {\fontsize{12pt}{12pt}\selectfont $\widetilde{q}_2$};
       \node[right] at (2.5,0.8)  {\fontsize{12pt}{12pt}\selectfont $q_2$};
       \node[above] at (0.8,0)  {\fontsize{12pt}{12pt}\selectfont $\CO$};
       \node[above] at (1.5,0.1)  {\fontsize{12pt}{12pt}\selectfont $\CF$};
       \node[blue,left] at (0.9,-1.4)  {\fontsize{10pt}{10pt}\selectfont $1,ab^{-1}$};
       \node[blue,right] at (2.1,-1.4)  {\fontsize{10pt}{10pt}\selectfont $1,b^{-2}f$};
       \node[blue,above] at (1.5,-0.7)  {\fontsize{10pt}{10pt}\selectfont $0,a^{-1}b^3f^{-1}$};
       \node[blue,left] at (1,1.4)  {\fontsize{10pt}{10pt}\selectfont $1,a^{-1}b^{-3}$};
       \node[blue,right] at (2.1,1.4)  {\fontsize{10pt}{10pt}\selectfont $1,a^{2}f$};
       \end{tikzpicture}
    \end{subfigure}
    \caption{On the right of each pair we have the Lagrangians for the half-BPS 5d/4d coupled system of the fundamental rank 1 E-string theory flux domain walls with fluxes $(-\half ,-\half ,-\half ,-\half ,-\half ,-\half ,-\half ,-\half)$ and $(-1 ,-1 ,0 ,0 ,0 ,0 ,0 ,0)$, respectively. On the left of each pair we display the 5d theory where we place the domain wall, where we display the fields as the chirals they contribute when compactified to 4d. Square and circle nodes denote special unitary global and gauge symmetries, respectively, while lines connecting them denote chiral fields, with the outgoing arrows standing for fundamental and ingoing arrows for anti-fundamental representation. The cross denotes a flipping field coupled by a superpotential term for which we don't write the charges. We denote in black the field names while in blue we denote the natural R-charge coming from 5d and the charges under the abelian symmetries, encoded in the powers of the fugacities $a,b$ and $f$. Specifically, $f$ denotes the fugacity for the residual $\mathfrak{u}(1)_{KK}$ coming from 5d. In the left pair we specify the Lagrangian for the $(-\half ,-\half ,-\half ,-\half ,-\half ,-\half ,-\half ,-\half)$ flux domain wall, with the superpotential $W=q \CO \widetilde{q} + \CF \CO^2$. In the right pair we specify the Lagrangian for the $(-1 ,-1 ,0 ,0 ,0 ,0 ,0 ,0)$ flux domain wall, with the superpotential $W=q_1 \CO \widetilde{q}_1 + q_2 \CO \widetilde{q}_2 + \CF \CO^2$. Note that in this case the superpotential forces us to choose matching boundary conditions for the lower and upper hypermultiplets.}
    \label{F:Estring4dFundDW}
\end{figure}

In the former section we mentioned two fundamental flux domain walls for the rank 1 E-string theory associated with fluxes $(-\half ,-\half ,-\half ,-\half ,-\half ,-\half ,-\half ,-\half)$ and $(-1 ,-1 ,0 ,0 ,0 ,0 ,0 ,0)$. The associated 4d flux domain wall theories are known from \cite{Kim:2017toz}, and follow the above matter content and superpotential to give WZ models as shown in Figure \ref{F:Estring4dFundDW}. These are the theories before cutting the edges of the tube, so they actually describe a coupled 5d/4d system which is why we still have gauge symmetries.\footnote{Notice that the charge assignements in Figure \ref{F:Estring4dFundDW} are also compatible with the cancellation of gauge anomalies on the 4d interface. To check this, remember that the bifundamental field is a genuine 4d field, while the diagonal fundamental fields descent from 5d fields that were given Neumann boundary conditions, so their contribution to the anomalies should be divided by 2. Similar anomaly checks were done in \cite{Gaiotto:2015una,Kim:2017toz}.} For the $(-\half ,... ,-\half)$ flux domain wall we give flux to $\mathfrak{u}(1)_a$ and therefore flip its sign on the right side of the quiver. In this Lagrangian one gets the expected superpotential terms discussed above. In the $(-1 ,-1 ,0 ,0 ,0 ,0 ,0 ,0)$ flux domain wall we give flux to $\mathfrak{u}(1)_\CF=\frac1{4}(-\mathfrak{u}(1)_a+ \mathfrak{u}(1)_b)$ and not to $\mathfrak{u}(1)_\CO=\frac1{4}(3\mathfrak{u}(1)_a+ \mathfrak{u}(1)_b)$. This translates to flipping the sign of the combination $a^{-1}b^3$ and keeping the sign of $ab$ on the right side of the quiver. This together with the required superpotential forces us to choose the pairs $Q_1,\widetilde{Q}_2$ or $\widetilde{Q}_1,Q_2$ on the left side of the quiver, where the former will give the wanted flux and the latter a flipped sign flux. Note that in both cases we add the $\mathfrak{u}(1)_{KK}$ charge on the left denoted by the $f$ fugacity. This in turn forces the $\CO$ field to be charged under $\mathfrak{u}(1)_{KK}$ due to the superpotential.

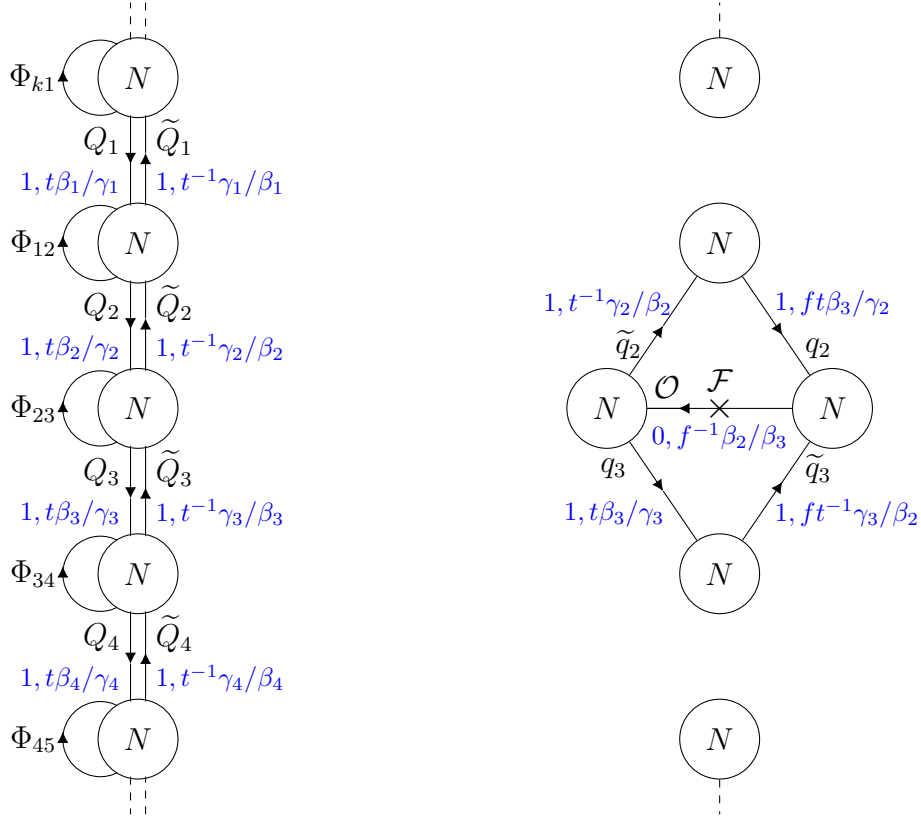
\begin{figure}[t]
    \centering
    \begin{subfigure}[b]{0.5\textwidth}
       \centering
       \begin{tikzpicture}[baseline,scale=1]
\tikzstyle{every node}=[font=\scriptsize]
	   \draw (-0.5,0) circle (0.5cm);
	   \draw (-0.5,-2.2) circle (0.5cm);
	   \draw (-0.5,2.2) circle (0.5cm);
	   \draw (-0.5,-4.4) circle (0.5cm);
	   \draw (-0.5,4.4) circle (0.5cm);
       \draw[->] (-1,0) to      (-1,0.1);
       \draw[->] (-1,-2.2) to      (-1,-2.1);
       \draw[->] (-1,-4.4) to      (-1,-4.3);
       \draw[->] (-1,2.2) to      (-1,2.3);
       \draw[->] (-1,4.4) to      (-1,4.5);
       \node[draw, circle,fill=white] (p1) at (0,4.4) {\fontsize{12pt}{12pt}\selectfont $\,\,N\,\,$};
       \node[draw, circle,fill=white] (p2) at (0,2.2) {\fontsize{12pt}{12pt}\selectfont $\,\,N\,\,$};
       \node[draw, circle,fill=white] (p3) at (0,0) {\fontsize{12pt}{12pt}\selectfont $\,\,N\,\,$};
       \node[draw, circle,fill=white] (p4) at (0,-2.2) {\fontsize{12pt}{12pt}\selectfont $\,\,N\,\,$};
       \node[draw, circle,fill=white] (p5) at (0,-4.4) {\fontsize{12pt}{12pt}\selectfont $\,\,N\,\,$};

       \draw[dashed] (-0.1,-4.9) to      (-0.1,-5.4);
       \draw[dashed] (0.1,-5.4) to      (0.1,-4.9);

       \draw[->] (-0.1,-2.7) to      (-0.1,-3.4);
       \draw[-] (-0.1,-3.4) to      (-0.1,-3.9);
       \draw[-<] (0.1,-2.7) to      (0.1,-3.4);
       \draw[-] (0.1,-3.9) to      (0.1,-3.4);

       \draw[->] (-0.1,-0.5) to      (-0.1,-1.2);
       \draw[-] (-0.1,-1.2) to      (-0.1,-1.7);
       \draw[-<] (0.1,-0.5) to      (0.1,-1.2);
       \draw[-] (0.1,-1.7) to      (0.1,-1.2);
       
       \draw[-<] (-0.1,0.5) to      (-0.1,1.2);
       \draw[-] (-0.1,1.2) to      (-0.1,1.7);
       \draw[->] (0.1,0.5) to      (0.1,1.2);
       \draw[-] (0.1,1.7) to      (0.1,1.2);
       
       \draw[-<] (-0.1,2.7) to      (-0.1,3.4);
       \draw[-] (-0.1,3.4) to      (-0.1,3.9);
       \draw[->] (0.1,2.7) to      (0.1,3.4);
       \draw[-] (0.1,3.9) to      (0.1,3.4);
       
       \draw[dashed] (-0.1,4.9) to      (-0.1,5.4);
       \draw[dashed] (0.1,5.4) to      (0.1,4.9);

       \node[left] at (-0.95,4.4)  {\fontsize{12pt}{12pt}\selectfont $\Phi_{k1}$};
       \node[left] at (-0.95,2.2)  {\fontsize{12pt}{12pt}\selectfont $\Phi_{12}$};
       \node[left] at (-0.95,0)  {\fontsize{12pt}{12pt}\selectfont $\Phi_{23}$};
       \node[left] at (-0.95,-2.2)  {\fontsize{12pt}{12pt}\selectfont $\Phi_{34}$};
       \node[left] at (-0.95,-4.4)  {\fontsize{12pt}{12pt}\selectfont $\Phi_{45}$};
       
       \node[right] at (0.1,3.6)  {\fontsize{12pt}{12pt}\selectfont $\widetilde{Q}_1$};
       \node[left] at (-0.1,3.55)  {\fontsize{12pt}{12pt}\selectfont $Q_1$};
       \node[right] at (0.1,1.4)  {\fontsize{12pt}{12pt}\selectfont $\widetilde{Q}_2$};
       \node[left] at (-0.1,1.35)  {\fontsize{12pt}{12pt}\selectfont $Q_2$};
       \node[right] at (0.1,-0.8)  {\fontsize{12pt}{12pt}\selectfont $\widetilde{Q}_3$};
       \node[left] at (-0.1,-0.85)  {\fontsize{12pt}{12pt}\selectfont $Q_3$};
       \node[right] at (0.1,-3)  {\fontsize{12pt}{12pt}\selectfont $\widetilde{Q}_4$};
       \node[left] at (-0.1,-3.05)  {\fontsize{12pt}{12pt}\selectfont $Q_4$};

       \node[blue,right] at (0.1,3)  {\fontsize{10pt}{10pt}\selectfont $1,t^{-1}\gamma_1/\beta_1$};
       \node[blue,left] at (-0.1,3)  {\fontsize{10pt}{10pt}\selectfont $1,t\beta_1/\gamma_1$};
       \node[blue,right] at (0.1,0.8)  {\fontsize{10pt}{10pt}\selectfont $1,t^{-1}\gamma_2/\beta_2$};
       \node[blue,left] at (-0.1,0.8)  {\fontsize{10pt}{10pt}\selectfont $1,t\beta_2/\gamma_2$};
       \node[blue,right] at (0.1,-1.4)  {\fontsize{10pt}{10pt}\selectfont $1,t^{-1}\gamma_3/\beta_3$};
       \node[blue,left] at (-0.1,-1.4)  {\fontsize{10pt}{10pt}\selectfont $1,t\beta_3/\gamma_3$};
       \node[blue,right] at (0.1,-3.6)  {\fontsize{10pt}{10pt}\selectfont $1,t^{-1}\gamma_4/\beta_4$};
       \node[blue,left] at (-0.1,-3.6)  {\fontsize{10pt}{10pt}\selectfont $1,t\beta_4/\gamma_4$};
       
       \end{tikzpicture}
    \end{subfigure}%
    \begin{subfigure}[b]{0.5\textwidth}
       \centering
       \begin{tikzpicture}[baseline,scale=1]
\tikzstyle{every node}=[font=\scriptsize]

       \node[draw, circle,fill=white] (p1) at (1.5,4.4) {\fontsize{12pt}{12pt}\selectfont $\,\,N\,\,$};
       \node[draw, circle,fill=white] (p2) at (1.5,2.2) {\fontsize{12pt}{12pt}\selectfont $\,\,N\,\,$};
       \node[draw, circle,fill=white] (p3_1) at (0,0) {\fontsize{12pt}{12pt}\selectfont $\,\,N\,\,$};
       \node[draw, circle,fill=white] (p3_2) at (3,0) {\fontsize{12pt}{12pt}\selectfont $\,\,N\,\,$};
       \node[draw, circle,fill=white] (p4) at (1.5,-2.2) {\fontsize{12pt}{12pt}\selectfont $\,\,N\,\,$};
       \node[draw, circle,fill=white] (p5) at (1.5,-4.4) {\fontsize{12pt}{12pt}\selectfont $\,\,N\,\,$};

       \draw[dashed] (p1) to      (1.5,5.4);

       \draw[->] (p3_1) to      (0.75,1.1);
       \draw[-] (0.75,1.1) to      (p2);
       \draw[-<] (p3_2) to      (2.25,1.1);
       \draw[-] (p2) to      (2.25,1.1); 

       \draw[-<] (p3_1) to      (1.1,0);
       \draw[-] (p3_1) to  node[rotate=0]{\Large $\times$}    (p3_2);
       
       \draw[->] (p3_1) to      (0.75,-1.1);
       \draw[-] (0.75,-1.1) to      (p4);
       \draw[-<] (p3_2) to      (2.25,-1.1);
       \draw[-] (p4) to      (2.25,-1.1);
       
       \draw[dashed] (p5) to      (1.5,-5.4);
       
       \node[left] at (0.6,0.85)  {\fontsize{12pt}{12pt}\selectfont $\widetilde{q}_2$};
       \node[right] at (2.5,0.8)  {\fontsize{12pt}{12pt}\selectfont $q_2$};
       \node[above] at (0.8,0)  {\fontsize{12pt}{12pt}\selectfont $\CO$};
       \node[above] at (1.5,0.1)  {\fontsize{12pt}{12pt}\selectfont $\CF$};
       \node[left] at (0.4,-0.8)  {\fontsize{12pt}{12pt}\selectfont $q_3$};
       \node[right] at (2.5,-0.8)  {\fontsize{12pt}{12pt}\selectfont $\widetilde{q}_3$};
     
       \node[blue,left] at (1,1.4)  {\fontsize{10pt}{10pt}\selectfont $1,t^{-1}\gamma_2/\beta_2$};
       \node[blue,right] at (2.1,1.4)  {\fontsize{10pt}{10pt}\selectfont $1,ft\beta_3/\gamma_2$};
       \node[blue,above] at (1.5,-0.7)  {\fontsize{10pt}{10pt}\selectfont $0,f^{-1}\beta_2/\beta_3$};
       \node[blue,left] at (0.9,-1.4)  {\fontsize{10pt}{10pt}\selectfont $1,t\beta_3/\gamma_3$};
       \node[blue,right] at (2.1,-1.4)  {\fontsize{10pt}{10pt}\selectfont $1,ft^{-1}\gamma_3/\beta_2$};
       
       \end{tikzpicture}
    \end{subfigure}
    \caption{On the right we have the Lagrangian for the half-BPS 5d/4d coupled system of the fundamental $(\mathfrak{su}(k),\mathfrak{su}(k))$ conformal matter theory flux domain wall with flux $(0 ,-\frac1{k} ,\frac1{k} ,0,...,0)$ for $\mathfrak{su}(k)_\beta$ and vanishing flux for the rest of the 6d flavor symmetry. On the left we display the 5d theory where we place the domain wall. As before we display the fields as the chirals they contribute when compactified to 4d. The dashed lines indicate the quivers continue on depending on the value of $k$ and eventually close off in a circular manner. We denote in black the field names while in blue we denote the natural R-charge coming from 5d and the charges under the abelian symmetries, encoded in the powers of the fugacities $t,\beta_i,\gamma_j$ and $f$. Specifically, $f$ denotes the fugacity for the residual $\mathfrak{u}(1)_{KK}$ coming from 5d. The superpotential for the 4d domain wall theory is $W=q_2 \CO \widetilde{q}_2 + q_3 \CO \widetilde{q}_3 + \CF \CO^N$. Note that in this case the superpotential forces us to choose matching boundary conditions for the lower and upper hypermultiplets.}
    \label{F:AA4dFundDW}
\end{figure}

\subsection{Example: $(A,A)$ conformal matter fundamental domain wall 4d theories}
Here we consider the fundamental flux domain walls of the $(A_{k-1},A_{k-1})$ conformal matter. Domain walls of $(G_{ADE},G_{ADE})$ conformal matter were studied in \cite{Kim:2018lfo}, where the $D$ and $E$ type follow a similar construction; therefore, we will exemplify the fundamental domain wall only for the $A$ case. In this case the 6d flavor symmetry is $\mathfrak{u}(1)_t \oplus \mathfrak{su}(k)_\beta \oplus \mathfrak{su}(k)_\gamma$ and we will only consider flux for $\mathfrak{u}(1)$ subgroups of $\mathfrak{su}(k)_\beta \oplus \mathfrak{su}(k)_\gamma$ as in \cite{Kim:2018lfo}. On the left of Figure \ref{F:AA4dFundDW} we give the Lagrangian for the 5d low energy effective field theory for the 5d KK theory of the $(A_{k-1},A_{k-1})$ conformal matter SCFT composed of $N$ minimal $(A_{k-1},A_{k-1})$ conformal matter SCFTs. This 5d Lagrangian is the affine $A_{k-1}$ quiver of $k$ $\mathfrak{su}(N)$ gauge nodes on a circle with bifundamental hypers between any two neighbouring gauge nodes. On the right of Figure \ref{F:AA4dFundDW} we give the fundamental flux domain wall with flux $(0 ,-\frac1{k} ,\frac1{k} ,0,...,0)$ for $\mathfrak{su}(k)_\beta$ and vanishing flux for the rest of the 6d flavor symmetry.\footnote{In this example one can also check that the charge assignments give vanishing gauge anomalies on the 4d interface. See \cite{Kim:2018lfo} for similar anomaly checks.} This domain wall corresponds to the root $(0 ,-1 ,1 ,0,...,0)$ of $\mathfrak{su}(k)_\beta$, and one can concatenate several such domain walls to find the non-fundamental domain walls described in \cite{Kim:2018lfo}. Specifically in this example the above flux implies we need to flip the sign of the fugacity combination $\beta_3/\beta_2$ and keep the sign of $\beta_2 \beta_3$ on the right side. This together with the implied superpotential requires us to choose the pairs $Q_2, \widetilde{Q}_3$ or $Q_3, \widetilde{Q}_2$ on the left side of the quiver, where the latter will give the chosen flux and the former a flipped sign flux. In this case we have in 5d many gauge nodes with hypermultiplets that are uncharged under the $\mathfrak{u}(1)$ for which we turn on flux, this means the associated superpotential on the domain wall of the inherited chirals will lead to a single gauge node and a mass term for the chirals coming from the two sides. Thus, these hypers will give no chirals on the domain wall and lead to unrelated gauge nodes. Note that this doesn't mean we can ignore these gauge nodes since chirals relating them can be added when cutting these infinite tubes to finite tubes and also when concatenating several such domain walls. As in the former example we also write down the $\mathfrak{u}(1)_{KK}$ charges according to the prescription given before.

\section{Explaining observations of the 4d theories}
\label{S:DW4dT}

In this section, we review some properties of the 4d flux domain wall theories that one can determine using the analysis of Section \ref{sec:4dDWtheory} and show how these match our predictions from the 5d perspective given in Sections \ref{S:5dFDW} and \ref{S:HFDW}. We will consider not only the fundamental domain walls, that is those associated with a flux that corresponds to a single root of the 6d flavor symmetry, but also more complicated flux domain walls. These can be obtained from the fundamental ones via a suitable gluing procedure, which is the 4d analogue of the one we described in 5d in Section \ref{S:HFDW} and which we are going to review. As before, our main example will be the rank 1 E-string theory.

\subsection{Example: 4d compactification of the rank 1 E-string theory}

\subsubsection{Flux $\vec{\mathcal{F}}=\left(-\half,-\half,-\half,-\half,-\half,-\half,-\half,-\half\right)$}

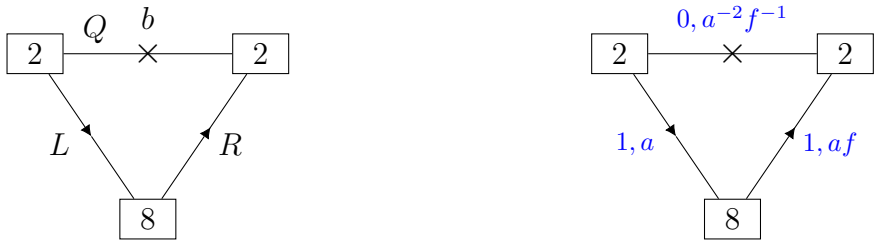
\begin{figure}[t]
    \centering
    \begin{subfigure}[b]{0.5\textwidth}
       \centering
       \begin{tikzpicture}[baseline,scale=1]
\tikzstyle{every node}=[font=\scriptsize]
       \node[draw, rectangle] (p1) at (0,0) {\fontsize{12pt}{12pt}\selectfont $\,\,2\,\,$};
       \node[draw, rectangle] (p2) at (3,0) {\fontsize{12pt}{12pt}\selectfont $\,\,2\,\,$};
       \node[draw, rectangle] (p3) at (1.5,-2.2) {\fontsize{12pt}{12pt}\selectfont $\,\,8\,\,$};
       \draw[-] (p1) to  node[rotate=0]{\Large $\times$}    (p2);
       \draw[->] (p1) to      (0.75,-1.1);
       \draw[-] (0.75,-1.1) to      (p3);
       \draw[-<] (p2) to      (2.25,-1.1);
       \draw[-] (p3) to      (2.25,-1.1);
       \node[left] at (0.6,-1.2)  {\fontsize{12pt}{12pt}\selectfont $L$};
       \node[right] at (2.3,-1.2)  {\fontsize{12pt}{12pt}\selectfont $R$};
       \node[above] at (0.8,0)  {\fontsize{12pt}{12pt}\selectfont $Q$};
       \node[above] at (1.5,0.2)  {\fontsize{12pt}{12pt}\selectfont $b$};
       \end{tikzpicture}
        \label{fig:Estring4dtubeA}
    \end{subfigure}%
    \begin{subfigure}[b]{0.5\textwidth}
       \centering
       \begin{tikzpicture}[baseline,scale=1]
\tikzstyle{every node}=[font=\scriptsize]
       \node[draw, rectangle] (p1) at (0,0) {\fontsize{12pt}{12pt}\selectfont $\,\,2\,\,$};
       \node[draw, rectangle] (p2) at (3,0) {\fontsize{12pt}{12pt}\selectfont $\,\,2\,\,$};
       \node[draw, rectangle] (p3) at (1.5,-2.2) {\fontsize{12pt}{12pt}\selectfont $\,\,8\,\,$};
       \draw[-] (p1) to  node[rotate=0]{\Large $\times$}    (p2);
       \draw[->] (p1) to      (0.75,-1.1);
       \draw[-] (0.75,-1.1) to      (p3);
       \draw[-<] (p2) to      (2.25,-1.1);
       \draw[-] (p3) to      (2.25,-1.1);
       \node[blue,left] at (0.6,-1.2)  {\fontsize{10pt}{10pt}\selectfont $1,a$};
       \node[blue,right] at (2.3,-1.2)  {\fontsize{10pt}{10pt}\selectfont $1,af$};
       \node[blue,above] at (1.5,0.15)  {\fontsize{10pt}{10pt}\selectfont $0,a^{-2}f^{-1}$};
       \end{tikzpicture}
        \label{fig:Estring4dtubeB}
    \end{subfigure}
    \caption{The 4d $\mathcal{N}=1$ Lagrangian for the compactification of the rank 1 E-string theory on a tube with flux $-\frac{1}{2}$ for a $\mathfrak{u}(1)$ whose commutant inside $\mathfrak{e}_8$ is $\mathfrak{e}_7$. Square nodes denote special unitary global symmetries, while lines connecting them denote chiral fields, with the outgoing arrows standing for fundamental and ingoing arrows for anti-fundamental representation. On the left we specify the names that we give to each fields. On the right we specify, in order, a possible assignement of R-charge and the charges under the abelian symmetries, encoded in the powers of the fugacities $a$ and $f$. In particular here we are using the R-symmetry coming from the Cartan of the 6d R-symmetry $\mathfrak{su}(2)_R$, but this will not in general be the 4d superconformal one since it can mix with the other abelian symmetries via $a$-maximization \cite{Intriligator:2003jj}.}
    \label{fig:Estring4dtube}
\end{figure}

We first consider the fundamental flux domain wall with $\vec{\mathcal{F}}=\left(-\half,\cdots,-\half\right)$ that was determined in \cite{Kim:2017toz} and which we reviewed in Subsection \ref{subsec:4dDWEstring}. Its construction from the 5d perspective is summarized in Figure \ref{basicFDWestring}. As we discussed in Section \ref{S:HFDW}, we can construct more general flux tubes and tori by gluing several such domain walls. After compactifying the $x_4$ direction on an interval and giving the boundary conditions, the $\mathfrak{su}(2)$ gauge symmetry of the two theories on the two sides of the domain wall are turned into $\mathfrak{su}(2)$ global symmetries associated to the two punctures. Moreover, we retain a set of chirals in the fundamental of $\mathfrak{su}(2)$ and in the $\bf 8$ of $\mathfrak{su}(8)$ from the theory on one side of the domain wall and a set of chirals in the fundamental of $\mathfrak{su}(2)$ and in the $\overline{\bf 8}$ of $\mathfrak{su}(8)$ from the theory on the opposite side of the domain wall. The last ingredient is the domain wall contribution to the 4d theory, which was found out in \cite{Kim:2017toz} building on previous results of \cite{Gaiotto:2015una}. In accordance with the reasoning we provided in the previous section, it is given by just an $\mathfrak{su}(2)\oplus \mathfrak{su}(2)$ bifundamental plus a signlet chiral field that flips the quadratic invariant made from the bifundamental. The matter content of the resulting theory is summarized in the quiver diagram in Figure \ref{fig:Estring4dtube} and the superpotential is
\be
\mathcal{W}=bQ^2+\sum_{i=1}^8L_iQR^i\,.
\ee

The manifest global symmetry of this theory is
\be
\mathfrak{su}(8)_u\oplus \mathfrak{u}(1)_a\oplus \mathfrak{u}(1)_f\oplus \mathfrak{su}(2)\oplus \mathfrak{su}(2)\,,
\ee
where $\mathfrak{u}(1)_a$ is the symmetry for which we turn on the flux, $\mathfrak{su}(8)_u$ is its commutant inside the 5d $\mathfrak{so}(16)$ global symmetry, the two $\mathfrak{su}(2)$ are the symmetries of the punctures and $\mathfrak{u}(1)_f$ is a symmetry that we can identify with the KK symmetry, as we will argue momentarily. This symmetry has been often neglected in the literature and it was identified with the isometry of the compactification manifold in \cite{Hwang:2021xyw}, where it was related to the $\mathfrak{su}(2)$ isometry in the case of sphere compactifications.

In \cite{Kim:2017toz} it was argued that the flux associated with this model is $-\frac{1}{2}$ by using anomaly matching arguments, that is comparing the anomalies of the theory with those predicted from 6d by compactifying the 8-form anomaly polynomial on the tube. This matches with the 5d prediction given in Section \ref{S:HFDW} using a top-down approach.

\subsubsection{Flux $\vec{\mathcal{F}}=\left(-n,-n,-n,-n,-n,-n,-n,-n\right)$}

\begin{figure}[t]
    \centering
       \begin{tikzpicture}[baseline,scale=1]
\tikzstyle{every node}=[font=\scriptsize]
       \node[draw, rectangle] (p1) at (0,0) {\fontsize{12pt}{12pt}\selectfont $\,\,2\,\,$};
       \node[draw, circle] (p2) at (3,0) {\fontsize{12pt}{12pt}\selectfont $\,\,2\,\,$};
       \node[draw, rectangle] (p3) at (6,0) {\fontsize{12pt}{12pt}\selectfont $\,\,2\,\,$};
       \node[draw, rectangle] (p4) at (3,-2.2) {\fontsize{12pt}{12pt}\selectfont $\,\,8\,\,$};
       \draw[-] (p1) to  node[rotate=0]{\Large $\times$}    (p2);
       \draw[-] (p2) to  node[rotate=0]{\Large $\times$}    (p3);
       \draw[->] (p1) to      (1.5,-1.1);
       \draw[-] (1.5,-1.1) to      (p4);
       \draw[->] (p3) to      (4.5,-1.1);
       \draw[-] (p4) to      (4.5,-1.1);
       \draw[-<] (p2) to      (3,-1.1);
       \draw[-] (p4) to      (3,-1.1);
       \node[blue,left] at (1.35,-1.2)  {\fontsize{10pt}{10pt}\selectfont $1,a$};
       \node[blue,left] at (3.1,-0.95)  {\fontsize{10pt}{10pt}\selectfont $1,af$};
       \node[blue,right] at (4.7,-1.2)  {\fontsize{10pt}{10pt}\selectfont $1,af^{2}$};
       \node[blue,above] at (1.5,0.15)  {\fontsize{10pt}{10pt}\selectfont $0,a^{-2}f^{-1}$};
       \node[blue,above] at (4.5,0.15)  {\fontsize{10pt}{10pt}\selectfont $0,a^{-2}f^{-3}$};
       \end{tikzpicture}
    \caption{The 4d $\mathcal{N}=1$ Lagrangian for the compactification of the rank 1 E-string theory on a tube with flux $-1$ for a $\mathfrak{u}(1)$ whose commutant inside $\mathfrak{e}_8$ is $\mathfrak{e}_7$. We focus here on the charges of the fields under the abelian symmetries.}
    \label{fig:Estring4dtubeglue}
\end{figure}
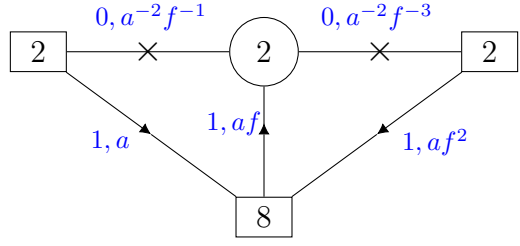

As we learnt from our discussion in Section \ref{S:HFDW}, we can now glue copies of this basic flux domain wall theory in different ways to generate tubes and tori with various values of flux. For example, we can concatenate two such flux domain walls by gluing with the trivial Weyl element. Remember that in 5d this amounted to identifying the $\mathfrak{su}(8)_u\oplus \mathfrak{u}(1)_a\subset \mathfrak{so}(16)$ symmetry of the theory on the right of the first domain wall with that of the theory on the left of the second domain wall. Denoting by $L$, $R$ and by $L'$, $R'$ the fields of the first and second domain wall theory respectively, this is achieved by introducing an $\mathfrak{su}(2)\oplus \mathfrak{su}(8)_u$ bifundamental field $\Phi$ and the superpotential interaction
\be\label{eq:gluesuperpot}
\delta\mathcal{W}=\sum_{i=1}^8\Phi_a\left(R^i+L^{'i}\right)\,.
\ee
Such a gluing is usually referred to in the literature as \emph{$\Phi$-gluing}. It can also be understood as the re-introduction of those 5d fields that were given Dirichlet boundary conditions, which are needed since the gluing is effectively removing the boundary. In the same manner, we should re-introduce the dynamical $\mathfrak{su}(2)$ vector multiplet; thus, in the gluing we also gauge the diagonal combination of the two $\mathfrak{su}(2)$ puncture symmetries that are glued, which is the combination preserved by \eqref{eq:gluesuperpot}. The superpotential \eqref{eq:gluesuperpot} also makes $\Phi$ and a combination of $R$ and $L'$ massive, so that at low energies we are left with only one $\mathfrak{su}(2)\oplus \mathfrak{su}(8)_u$ bifundamental field. The result is the theory depicted in Figure \ref{fig:Estring4dtubeglue}. The middle part of the quiver can be understood as the 5d theory living between the two domain walls that we concatenated, where according to our previous discussion the vector multiplet is given Neumann boundary conditions at the location of both domain walls, while for each pair of chirals one receives Neumann boundary condition at both walls while the other receives Dirichlet boundary conditions.

Consider now the charge assignement of the fields after the gluing, which is summarized in Figure \ref{fig:Estring4dtubeglue}. Notice that due to the superpotential \eqref{eq:gluesuperpot} as well as the requirement of cancellation of gauge anomalies at the central $\mathfrak{su}(2)$ gauge node the charges of the fields under $\mathfrak{u}(1)_f$ have been shifted. In particular, comparing the fields of the left and right domain wall we see that $\mathfrak{u}(1)_a$ mixes with $\mathfrak{u}(1)_f$ in terms of the fugacities $a$ and $f$ as
\be
a\to af\,.
\ee
In terms of the charges under these abelian symmetries we have
\be
\mathfrak{u}(1)_a^{(1)}=\frac{1}{2}\left(\mathfrak{u}(1)_a^{(0)}+\mathfrak{u}(1)_f\right)\,,
\ee
where $\mathfrak{u}(1)_a^{(0)}$ and $\mathfrak{u}(1)_a^{(1)}$ denote the $\mathfrak{u}(1)_a$ symmetries of the left and right domain wall, respectively. This agrees with the 5d prediction from \eqref{eq:mixKK} and justifies why we can interpret the flavor symmetry $\mathfrak{u}(1)_f$ of the 4d model as coming from the KK symmetry in 5d.

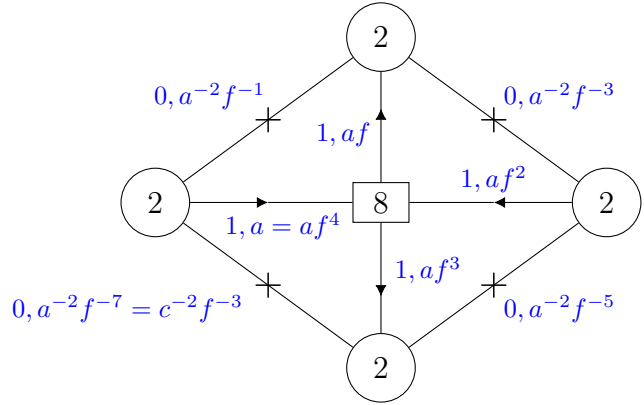
\begin{figure}[t]
    \centering
       \begin{tikzpicture}[baseline,scale=1]
\tikzstyle{every node}=[font=\scriptsize]
       \node[draw, circle] (p1) at (0,-2.2) {\fontsize{12pt}{12pt}\selectfont $\,\,2\,\,$};
       \node[draw, circle] (p2) at (3,0) {\fontsize{12pt}{12pt}\selectfont $\,\,2\,\,$};
       \node[draw, circle] (p3) at (6,-2.2) {\fontsize{12pt}{12pt}\selectfont $\,\,2\,\,$};
       \node[draw, rectangle] (p4) at (3,-2.2) {\fontsize{12pt}{12pt}\selectfont $\,\,8\,\,$};
       \node[draw, circle] (p5) at (3,-4.4) {\fontsize{12pt}{12pt}\selectfont $\,\,2\,\,$};
       \draw[-] (p1) to  node[rotate=45]{\Large $\times$}    (p2);
       \draw[-] (p2) to  node[rotate=45]{\Large $\times$}    (p3);
       \draw[-] (p3) to  node[rotate=45]{\Large $\times$}    (p5);
       \draw[-] (p5) to  node[rotate=45]{\Large $\times$}    (p1);
       \draw[->] (p1) to      (1.5,-2.2);
       \draw[-] (1.5,-2.2) to      (p4);
       \draw[->] (p3) to      (4.5,-2.2);
       \draw[-] (p4) to      (4.5,-2.2);
       \draw[-<] (p2) to      (3,-1.1);
       \draw[-] (p4) to      (3,-1.1);
       \draw[-<] (p5) to      (3,-3.3);
       \draw[-] (p4) to      (3,-3.3);
       \node[blue,below] at (1.7,-2.2)  {\fontsize{10pt}{10pt}\selectfont $1,a=af^{4}$};
       \node[blue,left] at (3,-1.3)  {\fontsize{10pt}{10pt}\selectfont $1,af$};
       \node[blue,above] at (4.5,-2.2)  {\fontsize{10pt}{10pt}\selectfont $1,af^{2}$};
       \node[blue,right] at (3.05,-3.1)  {\fontsize{10pt}{10pt}\selectfont $1,af^{3}$};
       \node[blue,above left] at (1.6,-1.1)  {\fontsize{10pt}{10pt}\selectfont $0,a^{-2}f^{-1}$};
       \node[blue,above right] at (4.5,-1.1)  {\fontsize{10pt}{10pt}\selectfont $0,a^{-2}f^{-3}$};
       \node[blue,below right] at (4.5,-3.3)  {\fontsize{10pt}{10pt}\selectfont $0,a^{-2}f^{-5}$};
       \node[blue,below left] at (1.3,-3.3)  {\fontsize{10pt}{10pt}\selectfont $0,a^{-2}f^{-7}=c^{-2}f^{-3}$};
       \end{tikzpicture}
    \caption{The 4d $\mathcal{N}=1$ Lagrangian for the compactification of the rank 1 E-string theory on a torus with flux $n=-\frac{\widetilde{n}}{2}=-2$ for a $\mathfrak{u}(1)$ whose commutant inside $\mathfrak{e}_8$ is $\mathfrak{e}_7$. The superpotential and the anomaly cancellation condition imply that the $\mathfrak{u}(1)_f$ symmetry is broken to $\mathbb{Z}_4$, which at the level of fugacities means that $f^4=1$.}
    \label{fig:Estring4dtoruseven}
\end{figure}

If we concatenate $\widetilde{n}$ such flux domain walls to build a flux $\vec{\mathcal{F}}=\left(-\frac{\widetilde{n}}{2},\cdots,-\frac{\widetilde{n}}{2}\right)$ domain wall, this shift will persist at each gluing, so that overall the shift of the $\mathfrak{u}(1)_a$ symmetry by $\mathfrak{u}(1)_f$ between the first and the last domain wall is
\be\label{eq:redeffugk8}
a\to af^{\widetilde{n}}\,,
\ee
which in terms of the charges gives
\be
\mathfrak{u}(1)_a^{(1)}=\frac{1}{2}\left(\mathfrak{u}(1)_a^{(0)}+\frac{1}{\widetilde{n}}\mathfrak{u}(1)_f\right)\,,
\ee
again in accordance with \eqref{eq:mixKK}.

We can eventually glue the two ends of the tube with flux $\vec{\mathcal{F}}=\left(-\frac{\widetilde{n}}{2},\cdots,-\frac{\widetilde{n}}{2}\right)$ to build a flux torus. As we saw in the previous section, the behaviour of this model changes depending on the parity of $\widetilde{n}$. Let us start from the case in which $\widetilde{n}=2n$ is even, so that the torus has a properly quantized integer flux $n$. The last gluing will necessarily force us to identify
\be
a=af^{2n}\quad\Longrightarrow\quad f^{2n}=1\,.
\ee
The fact that $f$ should be a $2n$-th root of unity implies that the $\mathfrak{u}(1)_f$ symmetry is broken to
\be
\mathfrak{u}(1)_f\to \mathbb{Z}_{2n}\,.
\ee
This again matches the 5d picture, see \eqref{eq:breakKKk8}. Let us stress that this behaviour of the $\mathfrak{u}(1)_f$ flavor symmetry of the 4d model was already pointed out in the literature \cite{Hwang:2021xyw}, but a top-down derivation as the one given in Section \ref{S:HFDW} was missing. In Figure \ref{fig:Estring4dtoruseven} we summarize the flux torus and the fields charge assignments for the case of gluing $\widetilde{n}=4$ flux domain walls.

\begin{figure}[t]
    \centering
       \begin{tikzpicture}[baseline,scale=1]
\tikzstyle{every node}=[font=\scriptsize]
       \node[draw, circle] (p1) at (0,0) {\fontsize{12pt}{12pt}\selectfont $\,\,2\,\,$};
       \node[draw, rectangle] (p2) at (3,0) {\fontsize{12pt}{12pt}\selectfont $\fso(8)$};
       \draw[-] (p1) edge [out=215,in=145,loop,looseness=5] node[rotate=0]{\Large $\times$}    (p1);
       \draw[-] (p1) to      (p2);
       \node[blue,above] at (1.5,0)  {\fontsize{11pt}{11pt}\selectfont $1,a$};
       \node[blue,left] at (-1.1,0)  {\fontsize{11pt}{11pt}\selectfont $0,a^{-2}$};
       \end{tikzpicture}
    \caption{The 4d $\mathcal{N}=1$ Lagrangian for the compactification of the rank 1 E-string theory on a torus with flux $n=-\frac{\widetilde{n}}{2}=-\frac{1}{2}$ for a $\mathfrak{u}(1)$ whose commutant inside $\mathfrak{e}_8$ is $\mathfrak{e}_7$. The $\mathfrak{su}(8)_u$ symmetry is broken to $\mathfrak{so}(8)$, while the $\mathfrak{u}(1)_f$ symmetry is completely broken.}
    \label{fig:Estring4dtorusodd}
\end{figure}
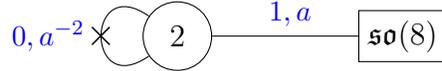

For $\widetilde{n}$ odd the gluing superpotential \eqref{eq:gluesuperpot} implies that the $\mathfrak{su}(8)$ symmetry is manifestly broken to $\mathfrak{so}(8)$, in accordance with the discussion from Section \ref{S:HFDW}. In particular, for $\widetilde{n}=1$ one gets the model of Figure \ref{fig:Estring4dtorusodd}, where the $\mathfrak{u}(1)_f$ is broken to $\mathbb{Z}_1$ so it is completely broken. Remember that in order to make sense of this model, one should also turn on a flux for the center $\mathbb{Z}_2$ subgroup of $E_7$, which can preserve at most its $F_4$ subgroup. In \cite{Kim:2017toz} it was found by index calculations that the conformal manifold of the 4d model is not large enough to accommodate a point where the enhancement $\mathfrak{so}(8)\to \mathfrak{f}_4$ actually occurs, but in \cite{Pasquetti:2019hxf} it was found that for the compactification of the higher rank E-string this is actually possible.

\subsubsection{Flux $\vec{\mathcal{F}}=\left(-n,...,-n,0,...,0\right)$}

As our last example, we consider gluing two $\vec{\mathcal{F}}=(-\frac{1}{2},\cdots,-\frac{1}{2})$ flux domain walls with the Weyl element that flips the sign of the last $8-k$ entries, such that we obtain a flux $\vec{\mathcal{F}}=(-1,\cdots,-1,0,\cdots,0)$ domain wall where the number of $-1$ entries is $k$. We explored this gluing from the 5d perspective in Section \ref{S:HFDW}, accordingly we will focus on the cases where $k$ is even. Let us review the construction of \cite{Kim:2017toz} for the implementation of this gluing in 4d.

\begin{figure}[t]
    \centering
       \begin{tikzpicture}[baseline,scale=1]
\tikzstyle{every node}=[font=\scriptsize]
       \node[draw, rectangle] (p1) at (-1,0) {\fontsize{12pt}{12pt}\selectfont $\,\,2\,\,$};
       \node[draw, circle] (p2) at (3,0) {\fontsize{12pt}{12pt}\selectfont $\,\,2\,\,$};
       \node[draw, rectangle] (p3) at (7,0) {\fontsize{12pt}{12pt}\selectfont $\,\,2\,\,$};
       \node[draw, rectangle] (p4) at (3,-2.2) {\fontsize{12pt}{12pt}\selectfont $\,\,k\,\,$};
       \node[draw, rectangle] (p5) at (3,2.2) {\fontsize{12pt}{12pt}\selectfont $\,8-k\,$};
       \draw[-] (p1) to  node[rotate=0]{\Large $\times$}    (p2);
       \draw[-] (p2) to  node[rotate=0]{\Large $\times$}    (p3);
       \draw[->] (p1) to      (1.5-0.5,-1.1);
       \draw[-] (1.5-0.5,-1.1) to      (p4);
       \draw[->] (p1) to      (1.5-0.5,1.1);
       \draw[-] (1.5-0.5,1.1) to      (p5);
       \draw[->] (p3) to      (4.5+0.5,-1.1);
       \draw[-] (p4) to      (4.5+0.5,-1.1);
       \draw[-<] (p3) to      (4.5+0.5,1.1);
       \draw[-] (p5) to      (4.5+0.5,1.1);
       \draw[-<] (p2) to      (3,-1.1);
       \draw[-] (p4) to      (3,-1.1);
       \node[blue,left] at (1.35-0.5,-1.2)  {\fontsize{10pt}{10pt}\selectfont $ab^{4-\frac{k}{2}}$};
       \node[blue,left] at (3.1,-0.95)  {\fontsize{10pt}{10pt}\selectfont $ab^{\frac{k}{2}-4}f$};
       \node[blue,right] at (4.7+0.5,-1.2)  {\fontsize{10pt}{10pt}\selectfont $a^{\frac{k}{2}-3}b^{\frac{(k-2)(k-8)}{4}}f^{\frac{k}{2}-2}$};
       \node[blue,above] at (1.5-0.5,0.1)  {\fontsize{10pt}{10pt}\selectfont $a^{-2}f^{-1}$};
       \node[blue,above] at (4.5+0.2,0.1)  {\fontsize{9pt}{9pt}\selectfont $a^{2-\frac{k}{2}}b^{\frac{k}{2}(4-\frac{k}{2})}f^{1-\frac{k}{2}}$};
       \node[blue,left] at (1.35-0.5,1.2)  {\fontsize{10pt}{10pt}\selectfont $ab^{-\frac{k}{2}}$};
       \node[blue,right] at (4.7+0.5,1.2)  {\fontsize{10pt}{10pt}\selectfont $a^{\frac{k}{2}-1}b^{\frac{k}{4}(k-6)}f^{\frac{k}{2}}$};
       \end{tikzpicture}
    \caption{The 4d $\mathcal{N}=1$ Lagrangian for the compactification of the rank 1 E-string theory on a tube with flux $\vec{\mathcal{F}}=(-1,\cdots,-1,0,\cdots,0)$, with $k$ $(-1)$ entries. We focus here on the charges of the fields under the abelian global symmetries, with the R-charges being as in the previous examples.}
    \label{fig:Estring4dtubegluegenk}
\end{figure}
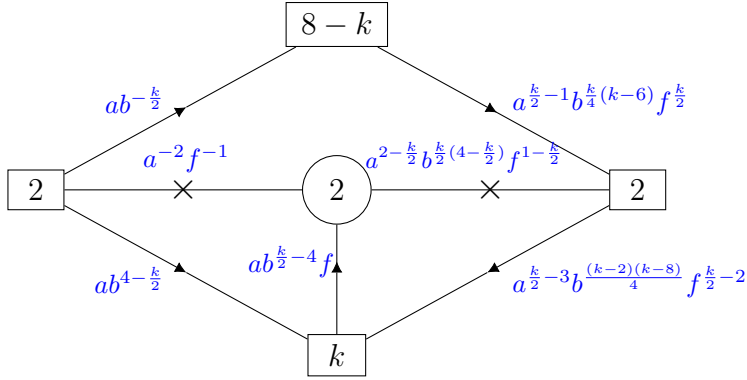

Remember that this operation breaks $\mathfrak{so}(16)\to \mathfrak{su}(8)\oplus \mathfrak{u}(1)_a\to \mathfrak{su}(k) \oplus \mathfrak{su}(8-k) \oplus \mathfrak{u}(1)_a \oplus \mathfrak{u}(1)_b$, where the embedding is as specified in \eqref{eq:decompgenk}. Each octet of fields $L$, $R$ in the basic flux domain wall theory of Figure \ref{fig:Estring4dtube} should be decomposed accordingly, so in particular the set of 8 chiral fields is split into $k$ plus $8-k$. Then, we perform the $\Phi$-gluing for the $k$ chirals, while we change the prescription for the $(8-k)$ chirals
\be
\label{eq:gluesuperpotgenk}
\delta\mathcal{W}=\sum_{i=1}^k\Phi_a\left(R^i+L^{'i}\right)+\sum_{i=k+1}^8R^iL'_i\,.
\ee
The second superpotential term corresponds to another type of gluing usually referred to as \emph{$S$-gluing} in the literature. Its effect is of giving mass to both $R^i$ and $L'_i$ for $i=k+1,\cdots,8$, such that the resulting model has only $k$ chirals transforming under the restored $\mathfrak{su}(2)$ gauge node. The result of the gluing is as depicted in Figure \ref{fig:Estring4dtubegluegenk}.

The prescription for $S$-gluing can be understood as follows. The Weyl reflection we are considering swaps the two 4d chirals inside a 5d hyper for $8-k$ of the hypers of the 5d gauge theory. This means that for $8-k$ of the chiral fields, the boundary conditions are Neumann at the location of the first domain wall and Dirichlet at the location of the second domain wall and oppositely for their partners. So these fields are expected not to survive the reduction to 4d, in agreement with the fact that the second term of the superpotential \eqref{eq:gluesuperpotgenk} kills any field transforming under the $\mathfrak{su}(2)$ gauge node and the $\mathfrak{su}(8-k)$ flavor node.

Notice that the flux domain wall model in Figure \ref{fig:Estring4dtubegluegenk} is dual to a Wess--Zumino (WZ) model with no gauge group only for $k=2$. This can be seen by Seiberg dualizing \cite{Seiberg:1994pq} the apparent $\mathfrak{su}(2)$ gauge node. For the case $k=8$ without gluings as shown in Figure \ref{fig:Estring4dtube} we also find a WZ model as it also corresponds to a single $\mathfrak{e}_8$ generator. This is related to the fact that the flux for the cases $k=2,8$ is realized by holonomies corresponding to a single $E_8$ generator and so they constitute fundamental flux domain walls as we discussed in Section \ref{sec:4dDWtheory}, while for the cases $k=4,6$ each is realized by holonomies corresponding to multiple $\mathfrak{e}_8$ generators, as was commented at the end of Section \ref{S:6dto5d}.

In Figure \ref{fig:Estring4dtubegluegenk} we also specified the charge assignments of all the fields. This was determined by taking the same assignment given in Figure \ref{fig:Estring4dtube} for the left domain wall theory, performing the decomposition \eqref{eq:decompgenk} and imposing the constraints coming from the gluing, which are due to the superpotential \eqref{eq:gluesuperpotgenk} and the gauge anomaly cancellation condition. We can see that as a consequence of the gluing the abelian symmetries are identified in a non-trivial way between the two domain walls. In terms of the fugacities we have
\be\label{eq:redeffuggenk}
a\to a^{\frac{k}{4}-1}b^{k\left(\frac{k}{8}-1\right)}f^{\frac{k}{4}-1}\,,\qquad b\to a^{\frac{1}{2}}b^{\frac{k}{4}-1}f^{\frac{1}{2}}\,.
\ee
This matches with the 5d prediction for such a mixing, see \eqref{eq:massidgenk}, where again we identify $\mathfrak{u}(1)_f$ with the KK symmetry.

Finally, we can consider stacking various copies of this new theory with flux $\vec{\mathcal{F}}=(-1,\cdots,-1,0,\cdots,0)$ using the trivial Weyl element, that is performing a $\Phi$-gluing. We can for example generate a flux $\vec{\mathcal{F}}=(-\widetilde{n},\cdots,-\widetilde{n},0,\cdots,0)$ by gluing $\widetilde{n}=2n$ of the theories in Figure \ref{fig:Estring4dtubegluegenk}, which is done by stacking $2\widetilde{n}=4n$ copies of the basic flux domain wall theory of Figure \ref{fig:Estring4dtube}. 

\begin{figure}[t]
    \centering
       \begin{tikzpicture}[baseline,scale=1.2]
\tikzstyle{every node}=[font=\scriptsize]
       \node[draw, circle] (p1) at (0,-2.2) {\fontsize{12pt}{12pt}\selectfont $\,\,2\,\,$};
       \node[draw, circle] (p2) at (3,0) {\fontsize{12pt}{12pt}\selectfont $\,\,2\,\,$};
       \node[draw, circle] (p3) at (6,-2.2) {\fontsize{12pt}{12pt}\selectfont $\,\,2\,\,$};
       \node[draw, rectangle] (p4) at (3,-2.2) {\fontsize{12pt}{12pt}\selectfont $\,\,k\,\,$};
       \node[draw, circle] (p5) at (3,-4.4) {\fontsize{12pt}{12pt}\selectfont $\,\,2\,\,$};
       \node[draw, rectangle] (p6) at (3,1.6) {\fontsize{12pt}{12pt}\selectfont $\,8-k\,$};
       \draw[-] (p1) to  node[rotate=45]{\Large $\times$}    (p2);
       \draw[-] (p2) to  node[rotate=45]{\Large $\times$}    (p3);
       \draw[-] (p3) to  node[rotate=45]{\Large $\times$}    (p5);
       \draw[-] (p5) to  node[rotate=45]{\Large $\times$}    (p1);
       \draw[->] (p1) to      (1.5,-2.2);
       \draw[-] (1.5,-2.2) to      (p4);
       \draw[->] (p3) to      (4.5,-2.2);
       \draw[-] (p4) to      (4.5,-2.2);
       \draw[-<] (p2) to      (3,-1.1);
       \draw[-] (p4) to      (3,-1.1);
       \draw[-<] (p5) to      (3,-3.3);
       \draw[-] (p4) to      (3,-3.3);
       \draw[-<] (p1) to      (0,0.2);
       \draw[-] (0,0.2) to      (0,1.6);
       \draw[-] (0,1.6) to      (p6);
       \draw[->] (p3) to      (6,0.2);
       \draw[-] (6,0.2) to      (6,1.6);
       \draw[-] (6,1.6) to      (p6);
       \node[blue,below] at (1.2,-2.2)  {\fontsize{8pt}{8pt}\selectfont $ab^{4-\frac{k}{2}}$};
       \node[blue,below] at (2,-2.5)  {\fontsize{8pt}{8pt}\selectfont $=ab^{4-\frac{k}{2}}f^{k-4}$};
       \node[blue,left] at (3,-1.4)  {\fontsize{8pt}{8pt}\selectfont $ab^{\frac{k}{2}-4}f$};
       \node[blue,above] at (4.1,-1.8)  {\fontsize{8pt}{8pt}\selectfont $a^{\frac{k}{2}-3}$};
       \node[blue,above] at (4.5,-2.2)  {\fontsize{8pt}{8pt}\selectfont $b^{\frac{(k-2)(k-8)}{4}}f^{\frac{k}{2}-2}$};
       \node[blue,right] at (3,-3)  {\fontsize{8pt}{8pt}\selectfont $ab^{\frac{k}{2}-4}f^{3}$};
       \node[blue,above left] at (1.6,-1.1)  {\fontsize{8pt}{8pt}\selectfont $a^{-2}f^{-1}$};
       \node[blue,above right] at (3.5,-0.6)  {\fontsize{8pt}{8pt}\selectfont $a^{2-\frac{k}{2}}b^{\frac{k}{2}(4-\frac{k}{2})}f^{1-\frac{k}{2}}$};
       \node[blue,below right] at (4.5,-3.3)  {\fontsize{8pt}{8pt}\selectfont $a^{2-\frac{k}{2}}b^{\frac{k}{2}(4-\frac{k}{2})}f^{-\frac{k}{2}-1}$};
       \node[blue,below left] at (1.3,-3.3)  {\fontsize{8pt}{8pt}\selectfont $a^{-2}f^{1-k}$};
       \node[blue,left] at (0,0.8)  {\fontsize{8pt}{8pt}\selectfont $ab^{-\frac{k}{2}}=ab^{-\frac{k}{2}}f^{k}$};
       \node[blue,right] at (6.1,0.8)  {\fontsize{8pt}{8pt}\selectfont $a^{\frac{k}{2}-1}b^{\frac{k}{4}(k-6)}f^{\frac{k}{2}}$};
       \end{tikzpicture}
    \caption{The 4d $\mathcal{N}=1$ Lagrangian for the compactification of the rank 1 E-string theory on a torus with flux $\vec{\mathcal{F}}=(-1,\cdots,-1,0,\cdots,0)$. The superpotentials and the anomaly cancellation condition imply that the $\mathfrak{u}(1)_f$ symmetry is broken to the smallest group between $\mathbb{Z}_{|k-4|}$ and $\mathbb{Z}_k$, which at the level of fugacities means that $f^{k-4}=1$ and $f^{-k}=1$.}
    \label{fig:Estring4dtorusgenk}
\end{figure}
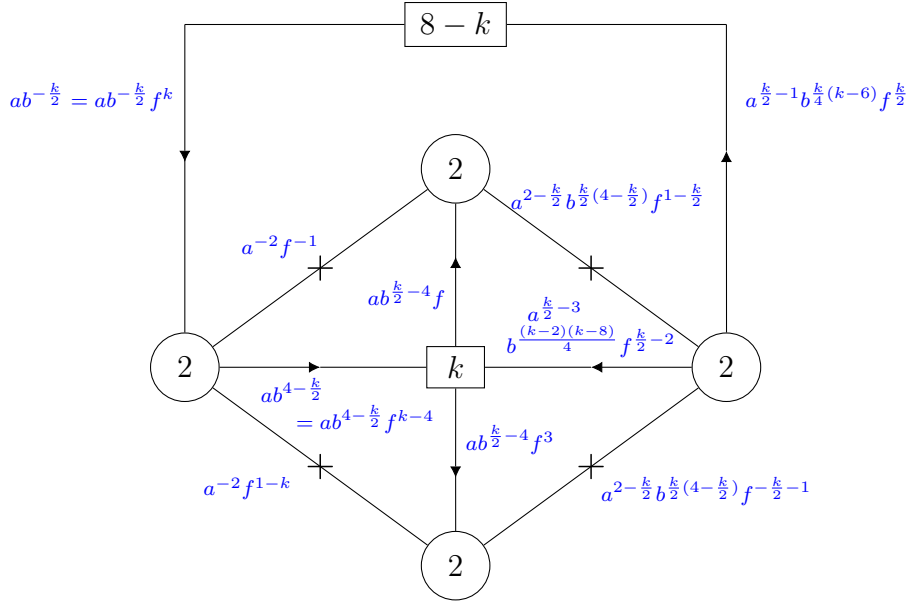

Let us consider as an example the case of a torus with flux $\vec{\mathcal{F}}=(-2,\cdots,-2,0,\cdots,0)$. The resulting model is depicted in Figure \ref{fig:Estring4dtorusgenk}. We can see that between the fields of each fundamental domain wall block, corresponding to each triangle of the quiver, there is a reparametrization of the abelian symmetries in accordance to what we previously described. Labelling each domain wall block with an index $p=0,\cdots,3$ where $p=0$ is the upper left triangle and increasing $p$ corresponds to moving clockwise in the quiver, we see that the fugacities $a_{(p)}$, $b_{(p)}$ for the abelian symmetries of each of these are redefined as in \eqref{eq:redeffuggenk} for $p=0,2$
\be
a_{(p+1)}= a_{(p)}^{\frac{k}{4}-1}b_{(p)}^{k\left(\frac{k}{8}-1\right)}f^{\frac{k}{4}-1}\,,\qquad b_{(p+1)}= a_{(p)}^{\frac{1}{2}}b_{(p)}^{\frac{k}{4}-1}f^{\frac{1}{2}}\,,
\ee
while they are redefined as in \eqref{eq:redeffugk8} for $p=1,3$
\be
a_{(p+1)}=a_{(p)}f\,,\qquad b_{(p+1)}=b_{(p)}^{-1}\,.
\ee
This is in accordance with the fact that the pairs first/second and third/fourth of domain walls are glued with a non-trivial Weyl element, while the pairs second/third and fourth/first are glued with the trivial Weyl element. 

After imposing the superpotentials and the anomaly cancellation constraints, we get
\be
f^{k-4}=1\,,\qquad f^{k}=1\,,
\ee
which means that $\mathfrak{u}(1)_f$ is broken to the smallest group between $\mathbb{Z}_{|k-4|}$ and $\mathbb{Z}_k$.\footnote{For $k=4$ the only non-trivial constraint is the second one $f^4=1$, so the symmetry is broken to $\mathbb{Z}_4$.} This is again compatible with our 5d expectation, see the discussion around \eqref{eq:breakKKgenk}, and the identification of $\mathfrak{u}(1)_f$ with $\mathfrak{u}(1)_{KK}$. For general even $n$, we expect the symmetry to instead be broken to the smallest group between $\mathbb{Z}_{n|k-4|}$ and $\mathbb{Z}_{nk}$.

\section{Conclusions and future directions}
\label{S:Conc}

In this paper we revisited the problem of compactifying 6d SCFTs on tubes and tori with flux to 4d $\mathcal{N}=1$ models and the connection with the study of duality domain walls in 5d KK theories. We focused in particular on the 5d perspective of the construction. This allowed us to give a top-down prediction for some of the known and lesser known properties of the resulting 4d $\mathcal{N}=1$ theories. In particular, the flux tube models generically possess a $\mathfrak{u}(1)$ flavor symmetry, which we can directly relate using the 5d analysis to the KK symmetry arising from the 6d to 5d circle compactification. Such a symmetry is generically broken to an abelian discrete symmetry when considering torus compactifications, which we again managed to understand in terms of consistency of the 5d construction. Throughout this paper we mainly focused on the example of the 6d rank 1 E-string theory, but this phenomenon should be a general feature of the 6d to 4d compactification and it would be interesting to investigate it also for the compactification of other 6d SCFTs. Another feature of the 4d theories that we managed to recover from the 5d perspective is the flux associated to a single domain wall, which was usually determined by anomaly matching arguments.

Our 5d analysis was performed in the language of the extended Coulomb branch phases and their associated decorated box graphs representations. This characterization is in direct correspondence with the geometric engineering picture of the 5d theories in terms of M-theory on non-compact Calabi--Yau three-folds with conical singularities, where each Coulomb branch phase is associated with a different resolution of the singularity. The hope is that the perspective given in this paper on the 6d to 4d compactifications would facilitate a geometrization of this field theory construction, allowing to possibly realize the same 4d $\mathcal{N}=1$ models but from geometric engineering of M-theory on new $G_2$-holonomy manifolds. Such an effort can possibly allow us to construct flux tube and torus 4d $\CN=1$ models for many other 6d SCFTs for which no such models are known.

The analysis of this paper can be extended in many directions. One such generalization would be to analyze flux tube and torus compactifications of 5d $\CN=1$ SCFTs to 3d $\CN=2$, as was suggested in \cite{Sacchi:2021afk,Sacchi:2021wvg}, using the 4d Coulomb branch and varying complex mass in a similar manner to what was done here. This could help build many other 3d $\CN=2$ models especially considering the fact that this analysis doesn't rely on 't Hooft anomalies  which are lacking for continuous symmetries in 5d and 3d. Moreover, it could also help in establishing a connection between the 5d to 3d compactifications and the geometric engineering of 3d $\CN=2$ models from M-theory on Calabi--Yau four-folds.

Another interesting direction to pursue that is related to this paper is to understand the RG flows considered in \cite{Razamat:2019mdt,Razamat:2019ukg,Sabag:2020elc} from the 5d KK-theory perspective. These RG flows are generated by vevs to operators charged under the 6d flavor symmetry and allow us in 4d to flow from flux tubes and tori to trinions with two maximal punctures and one minimal. From the 5d perspectives, such deformations should correspond to Higgs branch flows and it would be interesting to investigate how these affect the construction we provided in this paper. Translating this to geometry could also allow us to construct many new 4d models and gain a better understanding of the 4d $\CN=1$ SCFTs from geometry.

Finally, one can consider expanding this research to include discrete flux tubes and torus compactifications of 6d $(1,0)$ SCFTs. For example in \cite{Razamat:2018gro}  compactifications of Non-Higgsable clusters with algebras $\mathfrak{su}(3)$ and $\mathfrak{so}(8)$ were considered. Specifically 4d Lagrangians of compactifications on spheres with three and four punctures were found. These 6d SCFTs are special as they posses no flavor symmetry, and thus one cannot generate flux tubes and tori using the construction examined in this paper. Nonetheless, these theories do posses discrete symmetries which are the outer automorphism symmetries of their Dynkin diagram. One could study giving discrete flux to these symmetries in order to generate flux tubes and tori and check if they are consistent with the known Lagrangians under a closure of punctures. 

\section*{Acknowledgements}

We would like to thank Andreas P.~Braun and Sakura Sch\"afer-Nameki for many useful discussions and for collaboration at the early stages of this project. We are also grateful to Gabi Zafrir for useful discussions and comments. ES is supported by the European Union's Horizon 2020 Framework: ERC grant 682608 and the "Simons Collaboration on Special Holonomy in Geometry, Analysis and Physics". MS is partially supported by the ERC Consolidator Grant \#864828 “Algebraic Foundations of Supersymmetric Quantum Field Theory (SCFTAlg)” and by the Simons Collaboration for the Nonperturbative Bootstrap under grant \#494786 from the Simons Foundation.

\bibliographystyle{ytphys}
\bibliography{ref}

\end{document}